\documentclass{aa}
\usepackage[varg]{txfonts}
\graphicspath{{Png/}}
\usepackage{comment}
\usepackage[flushleft]{threeparttable}
\usepackage{xcolor}
\usepackage{commath}
\usepackage{soul}
\usepackage{CJKutf8}
\usepackage{soul}
\usepackage[normalem]{ulem}
\usepackage{natbib}

 \newcommand{\mi}{M_\mathrm{1,i}}
\newcommand{\qi}{q_\mathrm{i}}

\newcommand{\logpi}{\mathrm{log}\,P_\mathrm{orb,i}}
\newcommand{\logp}{\,\mathrm{log}\,P_\mathrm{orb}}

\newcommand{\dd}{\,\mathrm{d}}
\def\sfr{\,M_\odot\,\mathrm{yr}^{-1}}
\def\porb{\,P_\mathrm{orb}}

\def\ms{\,M_\odot}
\def\mob{\,M_\mathrm{OB}}
\def\kob{\,K_\mathrm{OB}}
\def\ls{\,L_\odot}
\def\rs{\,R_\odot}
\def\vrot{\,\upsilon_\mathrm{rot}}
\def\vc{\,\upsilon_\mathrm{crit}}

\def\kms{\,\mathrm{km\,s}^{-1}}

\def\mtr{\,M_\odot\,\mathrm{yr}^{-1}}
\def\domegas{\mathrm{ d\,}\Omega_\mathrm{ spin}}
\def\omegas{\Omega_\mathrm{ spin}}

\def\omegaorb{\Omega_\mathrm{ orb}}
\def\dd{\mathrm{ d\,}}
\def\mso{\,M_\odot}
 
 \def\lso{\,L_\odot}
 \def\kms{\, \mathrm{ km}\, \mathrm{ s}^{-1}}
 \def\egs{\, \mathrm{ erg}\, \mathrm{ s}^{-1}}
 \def\teff{\log\, T_\mathrm{ eff}\,}
 \def\llso{\log\, L/L_\odot \,}
 \def\simle{\mathrel{\hbox{\rlap{\hbox{\lower4pt\hbox{$\sim$}}}\hbox{$<$}}}}
 \def\simgr{\mathrel{\hbox{\rlap{\hbox{\lower4pt\hbox{$\sim$}}}\hbox{$>$}}}}
 \def\msoy{\, \mso~\mathrm{ yr}^{-1}}

\begin{document}

\title{Populations of evolved massive binary stars in the Small Magellanic Cloud I:
Predictions from detailed evolution models}
\titlerunning{}
\author{
\begin{CJK}{UTF8}{gkai}
X.-T. Xu (徐啸天)\inst{\ref{insit1}}
\thanks{email: xxu@astro.uni-bonn.de; xxu.astro@outlook.com}
\end{CJK}
\and 
C. Sch\"urmann\inst{\ref{insit1}} \thanks{email: chr-schuermann@uni-bonn.de}
\and 
N. Langer\inst{\ref{insit1},\ref{insit2}} \and 
C. Wang\inst{\ref{insit3}} \and
A. Schootemeijer\inst{\ref{insit1}} \and 
T. Shenar\inst{\ref{insit9}} \and
A. Ercolino\inst{\ref{insit1}} \and
F. Haberl\inst{\ref{insit6}} \and
B. Hastings\inst{\ref{insit1},\ref{insit2}} \and
H. Jin\inst{\ref{insit1}} \and
M. Kramer\inst{\ref{insit2}} \and
D. Lennon\inst{\ref{insit4},\ref{insit5}} \and 
P. Marchant\inst{\ref{insit8}}\and
K. Sen\inst{\ref{insit7}} \and 
T. M. Tauris\inst{\ref{insit11}} \and
S. E. de Mink\inst{\ref{insit3}}
}

\institute{
Argelander-Institut für Astronomie, Universität Bonn, Auf dem Hügel 71, 53121 Bonn, Germany\label{insit1}  
\and 
Max-Planck-Institut für Radioastronomie, Auf dem Hügel 69, 53121 Bonn, Germany\label{insit2} \and
Max-Planck-Institut f\"ur Astrophysik, Karl-Schwarzschild-Strasse 1, 85748 Garching, Germany\label{insit3}\and
Tel Aviv University, The School of Physics and Astronomy, Tel Aviv 6997801, Israel\label{insit9}\and
Instituto de Astrofísica de Canarias, 38200 La Laguna, Tenerife, Spain \label{insit4}\and
Dpto. Astrofísica, Universidad de La Laguna, 38205 La Laguna, Tenerife, Spain\label{insit5}\and
Max-Planck-Institut f\"ur Extraterrestrische Physik, Gie\ss enbachstraße 1, 85748 Garching, Germany\label{insit6} \and
Steward Observatory, Department of Astronomy, University of Arizona, 933 N. Cherry Ave., Tucson, AZ 85721, USA\label{insit7} \and
Department of Physics and Astronomy, Krijgslaan 281/S9 9000 Gent, Belgium\label{insit8} \and
Department of Materials and Production, Aalborg University, Fibigerstr{\ae}de 16, 9220 Aalborg, Denmark\label{insit11}
}

\def\egs{\, \mathrm{ erg}\, \mathrm{ s}^{-1}}
\def\egsu{$\, \mathrm{ erg}\, \mathrm{ s}^{-1}$}
\def\ms{\,M_\odot}
\def\msu{$\,M_\odot$}
\def\vorb{\upsilon_\mathrm{ orb}}
\def\ro{\,R_\mathrm{ O}}
\def\rd{\,R_\mathrm{ disc}}
\def\risco{\,R_\mathrm{ ISCO}}
\def\mbh{\,M_\mathrm{ BH}}
\def\vc{\upsilon_\mathrm{crit}}

\date{Submitted date / Received date / Accepted date }

\abstract
{The majority of massive stars are born with a close binary companion. How this affects their evolution and fate is still largely uncertain, especially at low metallicity.}
{We derive synthetic populations of massive post-interaction binary products and compare them with corresponding observed populations in the Small Magellanic Cloud (SMC).} 
{We analyse 53298 detailed binary evolutionary models computed with MESA.
Our models include the physics of rotation, mass and angular momentum transfer, magnetic internal angular momentum transport, and tidal spin-orbit coupling. They
cover initial primary masses of $5$---$100\mso$, initial mass ratios of 0.3---0.95, and  all initial periods for which interaction is expected, 1---3162 d.
They are evolved through the first mass transfer and the donor star death, 
a possible ensuing Be/X-ray binary phase, 
and they end when the mass gainer leaves the main sequence.
}
{
In our fiducial synthetic population, 8\% of the OB stars in the SMC are post-mass transfer systems,
and 7\% are merger products. In many of our models, the mass gainers are spun up and expected to form Oe/Be stars. 
While our model underpredicts the number of Be/X-ray binaries in the SMC, it reproduces the main features of their orbital period distribution and the observed number of SMC binary WR stars.
We further expect $\sim$50 OB+BH binaries below and $\sim$170 above 20\,d orbital period. The latter might produce merging double BHs. However, their progenitors, the predicted long-period WR+OB binaries, are not observed. 
}
{While the comparison with the observed SMC stars supports many physics assumptions in our high-mass binary models, a better match of the large number of observed OBe stars and Be/X-ray binaries likely requires a lower merger rate and/or a higher mass transfer efficiency during the first mass transfer. The fate of the initially wide O\, star binaries remains particularly uncertain.  
}

\titlerunning{Synthetic populations of evolved massive binary stars in the SMC -- Part I}
\authorrunning{X.-T. Xu et al.}

\maketitle 

\section{Introduction}

A new window to the Universe was opened by the first direct
detection of gravitational waves, emitted by the merging of two stellar-mass black holes \citep[BHs;][]{Abbott2016}. 
Up to now, more than one hundred compact object merger events were detected in this way \citep{Abbott2023}.
Several formation channels have been proposed \citep{Belczynski2016,Mapelli2020,Mandel2022}, 
including their formation in isolated massive binaries.
Due to the cosmological distance of these sources,
and considerable delay times between the compact object formation and the merger, many of
them will have formed in low-metallicity environment of high-redshift Universe \citep{Belczynski2016,Klencki2025arXiv250508860K}. 

An understanding of the formation of tight double-compact
binaries through binary evolution is intimately linked
to an understanding of massive stars, 
since most of them are born with a close companion with
which they will interact \citep{Sana2012}. Massive star
feedback, be it via emitting newly synthesized chemical elements, kinetic energy, or ionizing radiation, is strongly
affected by the presence of a companion \citep{Zapartas2017,Gotberg2019,Eldridge2022}, which is
therefore important for the evolution of star-forming galaxies \citep{Ma2016,Fichtner2022}.

While this holds for all redshifts, we can study metal-rich individual massive stars and binaries in detail. 
Galaxies are observed at redshifts beyond 14 \citep{Carniani2024,Helton2024},
and individual stars in these galaxies cannot be resolved. In this respect, the Small Magellanic Cloud (SMC) provides an ideal laboratory to investigate massive single-star and binary evolution at conditions prevalent in the early universe. Its metallicity of only 
$\sim$$1/5$th solar \citep{Hill1995,Korn2000,Davies2015} corresponds to the average metallicity of star forming galaxies at a redshift of $\approx 5$ \citep{Langer2006}. Yet, with a distance of only
$61.9\pm 0.6\,$kpc \citep{deGrijs2015}, and with a current star formation rate 
of $\sim$$0.05\msoy$ \citep{Harris2004,Rubele2015,Hagen2017,Rubele2018,Schootemeijer2021}, 
it shows us a rich population of massive stars.

The SMC is also an ideal testbed for massive star evolution for other reasons. Firstly, due to its low metallicity, the wind loss of hot stars are weak 
\citep{Mokiem2007}, such that the uncertainty in their mass and 
angular momentum loss is reduced. Secondly, the extinction towards most of the stars is very low \citep{Gorski2020}.
Except for potentially the youngest massive stars \citep{Schootemeijer2021},
it is thus generally assumed that we observe their complete populations, 
which makes the SMC well suited for population synthesis studies. The BLOeM survey will provide further information about the binarity of the SMC massive stars, offering critical tests for binary evolution models \citep{Shenar2024}.

\begin{figure}[t]
	\centering
	\includegraphics[width=0.8\linewidth]{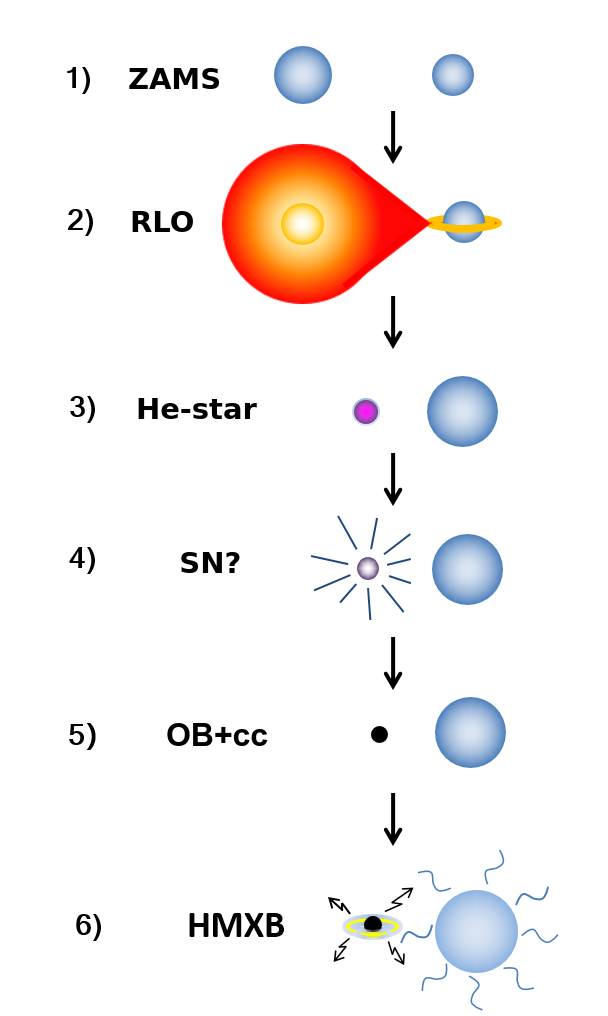}
	\caption{\label{intro}Schematic evolution of a massive
 binary system through the six evolutionary phases considered
 in this paper: pre-interaction, Roche lobe overflow (RLO),
 stripped envelope star (subdwarf, helium star or Wolf-Rayet star), compact object, referred to as the compact companion (cc) in a binary system, formation possibly 
 connected with a supernova (SN) explosion, X-ray silent compact object binary, and high-mass X-ray binary (HMXB). 
 After the 2nd stage, a fraction of the mass gainers may
 be fast rotating and appear as Oe or Be stars, and after
 NS formation as OBe/X-ray binary.
 Notably, many systems
 do not survive the 2nd, 4th and 6th stage as a binary,
 i.e., the number of systems evolving from top to bottom
 is being reduced at these stages. The figure is adapted from \citet{Kruckow2018}.}
\end{figure}

Figure\,\ref{intro} presents a schematic picture of the evolutionary 
phases of massive close binary systems, up to the time where the
initially less massive star leaves the main sequence. It demonstrates that 
binary evolution models may be compared with the properties of several
observed populations of evolved massive binaries in the SMC.
While so far, 14 radio pulsars are known in the SMC \citep{Carli2024} of which one has a B\,star companion, the SMC's massive X-ray binary population is particularly rich,
with about 150 objects, of which $\sim$100
are identified to be Be/X-ray binaries \citep[BeXBs,][]{Haberl2016}.
Furthermore, \citet{Schootemeijer2022} identified more than
1700 OBe stars brighter than $M_\mathrm{ V}\simeq -3\,$mag ($\simgr 9\mso$),
most of which are thought to represent spun-up mass gainers in binary systems \citep{Shao2014,Bodensteiner2020}. 
\citet{Drout2023}, by searching for UV excess, identified 9 hot stars in the SMC  with high surface gravities, consistent with the expectation for stars stripped in binaries.
The SMC also harbours 12 Wolf-Rayet (WR) stars, four of which show a close OB star companion \citep{Shenar2016,Neugent2018}.


While the predictions of population synthesis models can be 
compared with these observed samples of post-interaction binaries,
for several binary evolutionary stages we basically lack any observed
counterparts. This holds in particular for rich predicted populations of stripped mass donors which lack the strong winds to make them appear as WR stars \citep{Wellstein2001,Gotberg2018,Langer2020a,Shenar2020,Drout2023},
for wind accreting black holes with OB star companions \citep{Shao2019,Langer2020,Janssens2022},
which may lack observable X-ray emission \citep{Hirai2021,Sen2021,Sen2024}, and of which very few
are known in the Galaxy and Large Magellanic Cloud (LMC) \citep{Orosz2009,Orosz2011,Shenar2022,Mahy2022}, and none so far in the SMC.
Here, population synthesis predictions are useful to refine targeted searches for such objects.

We aim to provide predictions for the evolved phases of massive binary stars in the SMC based on detailed MESA binary evolution models. 
This work covers all the evolutionary stages presented in Fig.\,\ref{intro}. The later stages up to the merger of binary black holes will be investigated in our follow-up work by using the population derived in this work and our detailed model data.
Our work is closely related to
that of Sch\"urmann et al. (\citeyear{Schurmann2020}; hereafter Paper II), who undertake a similar effort using the rapid binary
evolution code ComBinE \citep{Kruckow2018}. Our paper is organized as follows.  
In Sect.\,\ref{method}, we introduce the grid of detailed binary evolution models used for our analysis, and detail the essential input physics. 
In Sect.\,\ref{results}, we present the results of our fiducial  population synthesis model, and we describe the results of parameter variations in Sect.\,\ref{initial_parameters_var}.
In Sect.\,\ref{s_obs}, we compare our results with observations. 
We discuss the key
uncertainties in our calculations in Sect.\,\ref{discuss}, before summarizing our results in Sect.\,\ref{conclusion}.

\section{Method\label{method}}

This work is based on a dense grid of detailed massive binary evolution models \citep{Wang2020}, 
which is calculated with the one-dimensional stellar evolution code MESA 
(Modules for Experiments in Stellar Astrophysics version 8845)  
using a tailored metallicity appropriate for the SMC, as in \citet{Brott2011}. The detailed description of this code can be found in
\citet{Paxton2011,Paxton2013,Paxton2015}. Using statistical weights depending on the
initial distributions and lifetimes allows us to perform population synthesis. 

In contrast to rapid binary population synthesis, 
where different binary model parameters can easily be explored,
we only have one fixed binary evolution grid to work with. 
However, in order to construct a synthetic population from that model
grid, parameters are introduced, which can be varied later on.
This concerns in particular the birth kicks of compact objects, the 
core mass ranges defining the emerging compact object type,
the threshold rotation for assuming an OBe nature, and the star formation history.
In the following subsections, we describe our method in detail.

\subsection{Input physics\label{MESA_input}}

Here we briefly summarize our input physics parameters for the MESA calculations. 
In order to model convection, the standard
mixing length theory is adopted with a mixing length parameter of
$\alpha_\mathrm{ MLT}=1.5$ \citep{Bohm1958}. We use the Ledoux criterion to identify convective layers, and adopt step-overshooting for the hydrogen-burning core
using a parameter $\alpha_\mathrm{ov}=0.335$ \citep{Brott2011}. We model semiconvection according to \citet{Langer1983} using a semiconvection efficiency parameter of
$\alpha_\mathrm{sc}=1$ \citep{Langer1991,Brott2011,Abel2019}. 
We follow \citet{Cantiello2010} to model thermohaline mixing with the 
efficiency parameter $\alpha_\mathrm{ th}=1$. Rotation-induced mixing is computed in the diffusion approximation as in
\citet{Heger2000}, including the dynamical \citep{Endal1978} and 
secular shear instability \citep{Maeder1997},  
Goldreich-Schubert-Fricke instability \citep{Goldreich1967,Fricke1968} 
and Eddington-Sweet circulation \citep{Eddington1929}. In addition, the Spruit-Taylor dynamo is included for the internal angular momentum transport \citep{Spruit2002}.

We adopt stellar wind mass loss following the treatment in \citet{Brott2011}, which includes the so-called bi-stability jump \citep{Vink1999}. The first jump is near an effective temperature of 22$\,$kK, below which the mass loss rate is computed according to
\citet{Nieuwenhuijzen1990} and \citet{Vink2000,Vink2001}. 
This mass-loss prescription may need revisions, as recent observations on galactic massive stars do not find evidence for the changes in mass-loss rates across the jump temperature \citep{deBurgos2024}.
For enhanced surface helium abundances, with helium mass fractions in the range 0.3---0.7, the mass-loss rate is 
smoothly increased towards empirical Wolf-Rayet rates \citep{Hamann1995}. In addition, 
enhanced mass loss by rotation is assumed as in \citet{Heger2000}. While rotation does not necessarily enhance mass loss \citep{Hastings2023}, the purpose of this assumption in our models is to prevent the accreting components in mass transferring models to achieve faster-than-critical rotation. Stellar wind also carries away angular momentum, braking stellar rotation. During each time step, our model (MESA ver. 8845) calculates the angular momentum loss by removing the angular momentum contained in the removed layer, which is done the same way as in \citet{Brott2011}. Assuming a constant surface specific angular momentum during one time step, which appears to be more physically correct and is done in the most recent MESA versions, would result in a stronger wind induced spin-down than our scheme, and would thus lead to a smaller number of OBe stars (cf., Paper\,II). However, as
\citep{Nathaniel2025} find that the spin-down of Galactic O\,stars appears to be better reproduced with the old angular momentum loss prescription, we consider this issue unresolved.

We consider only circular orbits.
The orbital evolution of our binary models is affected by mass transfer \citep{Tauris2023} and tides
\citep{Hut1981,Hurley2002,Sepinsky2007}. We use the $\tt{Hut\_rad}$ scheme 
coded in MESA to calculate radiative-damping dominated tidal interaction \citep{Detmers2008}, while recently \citet{Sciarini2024} argue that there could be inconsistency in this tide prescription. 
Mass transfer takes place when the radius of the primary star exceeds the 
radius of its Roche Lobe, where the Roche Lobe radius is calculated with 
the analytic fit derived by \citet{Eggleton1983}. In order to account
for long-term contact phases, the $\tt{contact}$ scheme in MESA is adopted 
\citep{Marchant2016,Menon2021}. During mass transfer, accretors 
can be spun up to critical rotation \citep{Packet1981,Petrovic2005}.
The accretor is assumed to accrete as much as possible before reaching critical rotation. When the expected mass transfer rate would cause it to exceed critical rotation, we assume the accretor can only accrete the amount that makes it remain below critical rotation \citep{Petrovic2005}, which sets a balance among tidal interaction, accretion, and structure adjustment that gives the mass transfer efficiency in a self-consistent way \citep{Paxton2015}. The non-accreted material is assume to be 
ejected from the binary system, carrying the specific orbital angular momentum of the mass gainer \citep[so called isotropic re-emission;][]{Soberman1997}. 
This leads to a rotation-dependent accretion efficiency. In wide-orbit binaries,
where tides are inefficient, the accretion efficiency is often below 10\%, 
while it can reach 60\% in narrow-orbit binaries. 

We set an upper limit on the 
mass loss rate from the binary system $\dot{M}_\mathrm{ limit}$ by 
comparing the photon luminosity of the stars with the energy input rate which is required to maintain the current mass loss rate from the binary, as \citet{pablothesis}:
\begin{equation}
    \mathrm{ log}\,\frac{\dot{M}_\mathrm{ limit}}{\ms\,\mathrm{ yr}^{-1}} = -7.19+\mathrm{ log}\frac{L_1+L_2}{\ls} - \mathrm{ log}\,\frac{M_2}{\ms} + \mathrm{ log}\frac{R_{\rm L,2}}{\rs}.
    \label{qmin}
\end{equation}
Here, $L_1$ and $L_2$ are the luminosities of the mass donor and the mass gainer, $M_2$ is the mass of 
mass gainer, and $R_{\rm L,2}$ is the Roche-lobe radius of the mass gainer, at which material is assumed to be ejected. If a larger mass loss rate from the binary system is required, we assume that the mass transfer is unstable and produces
a binary merger.With the above condition, the critical mass ratio for unstable mass transfer in our models is sensitive to the initial orbital period and the initial primary mass of the system (see Sect.\,\ref{binary_grid} and Fig.\,\ref{summary}).
The merger rate based on Eq.\,\eqref{qmin} is a lower limit, as the ejection of the same amount of material at other locations, like the surface of the mass gainer or the accretion disc around the mass gainer, would require higher radiation energies, resulting in a lower threshold for the mass loss rate from the system.
In addition, 
we also assume unstable mass transfer if the mass transfer rate exceeds 0.1$\mtr$, if inverse mass transfer occurs with a post-main-sequence (post-MS) 
donor, or if the second Lagrangian point (L2) overflow occurs during a contact phase. The survivors of unstable mass transfer are ignored, as their initial parameter space is narrow  \citep[e.g.,][]{Ercolino2024} and expected to take a small fraction of the population considered in this work.

Our binary models are evolved from the zero-age main-sequence (ZAMS) stage up to core carbon depletion of the primary star, except for donor stars with a helium core mass above 13$\ms$, for which we terminate the evolution at core 
helium depletion for numerical reasons. After the termination of the primary star, we further evolve the secondary star as a single star. This numerical setting allows to explore various post-supernova scenarios (cf. Sects. \ref{CC_formation} and \ref{natal_kick}). 
\subsection{Binary model grid\label{binary_grid}}

Our SMC binary model grid \citep{Wang2020} contains 53298 detailed evolution models.
The model data\footnote{https://wwwmpa.mpa-garching.mpg.de/stellgrid/} and necessary files\footnote{https://doi.org/10.5281/zenodo.10017597} to reproduce our models are available online.
We assume that the stars are initially 
tidally locked. Therefore, the initial binary parameters are 
initial primary mass $M_\mathrm{ 1,i}$, initial mass ratio $q_\mathrm{ i}$, and initial 
orbital period $P_\mathrm{ orb,i}$. Our model grid is computed with the following 
parameter space:
\begin{itemize}
	\item Initial primary masses $M_\mathrm{ 1,i}$ are set in the range of 5---100$\,\mathrm{ M}_\odot$, or
		log$\,(M_\mathrm{ 1,i}/\mathrm{ M}_\odot)$  between 0.7 and 2, with constant intervals of
		$\Delta\,$log$(M_\mathrm{ 1,i}/\mathrm{ M}_\odot) = 0.05$; 
	\item Initial mass ratios $q_\mathrm{ i}$ are considered between  0.3 and 0.95 with a constant interval of
	    $\Delta\,q_\mathrm{ i} = 0.05$;
	\item Initial orbital periods $P_\mathrm{ orb,i}$ are selected from 1\,d to 3162\,d, 
	    or log$(P_\mathrm{ orb,i}/\mathrm{ day}) = 0.0$---3.50, with constant intervals of
	$\Delta\,$log$\,(P_\mathrm{ orb,i}/\mathrm{ day}) = 0.025$.
\end{itemize}

The input parameter space covers all initial primary masses of BH/NS progenitors below $100\mso$ and all initial orbital periods for interacting systems. The lower limit on initial mass ratio is chosen to be 0.3, below which we expect mass transfer to be unstable \citep{Soberman1997}. While recent studies \citep{Ge2020,Marchant2021,Schurmann2024A&A...691A.174S}  find that stable mass transfer is possible for $\qi<0.3$ in the long-period regime, the corresponding parameter space is small and can be neglected for the purpose of this study.

\begin{figure}[t]
	\centering
    \includegraphics[width=\linewidth]{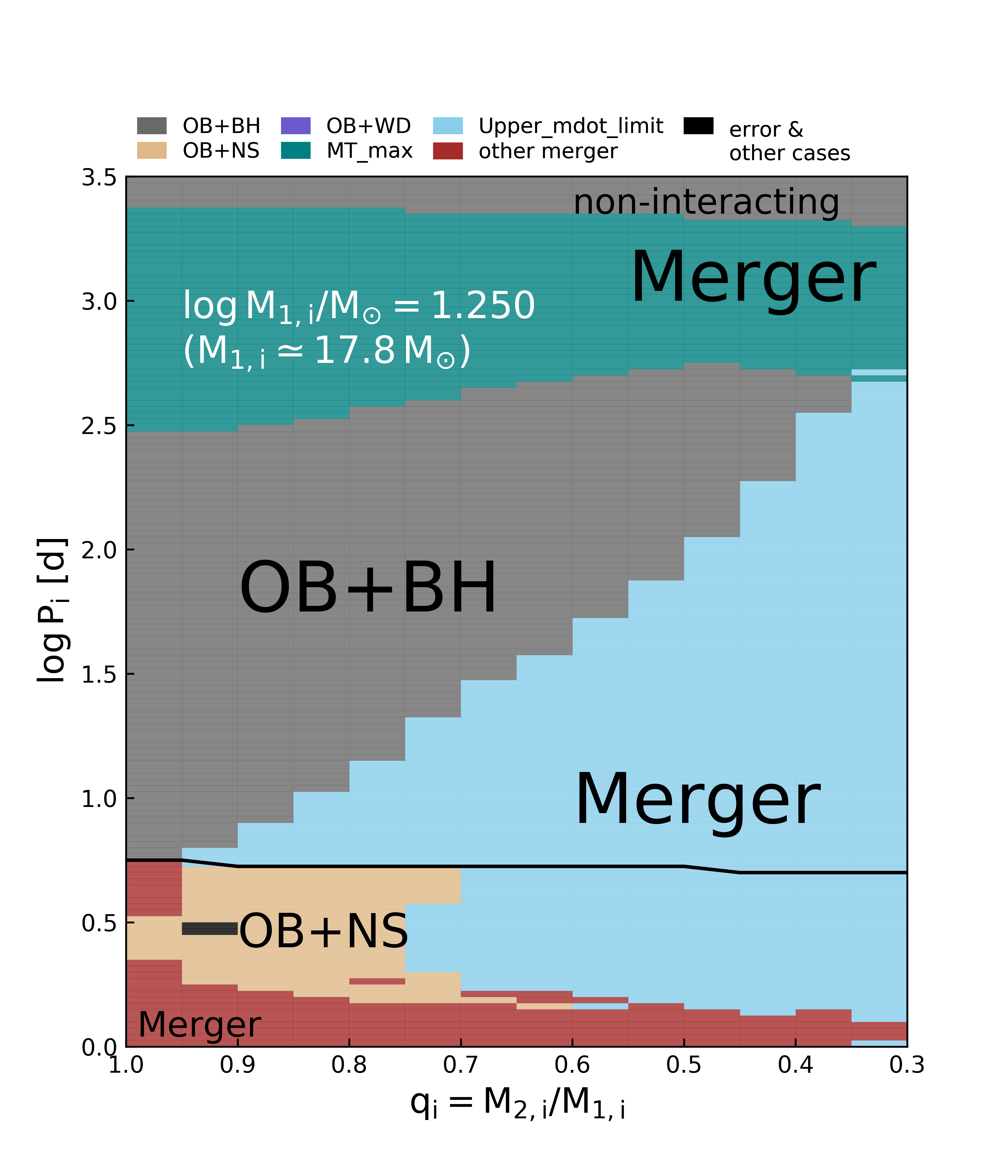}
	\caption{\label{summary}The outcome of the first mass transfer phase in our detailed binary evolution models 
	with an initial primary mass 17.8$\ms$. In this figure, each pixel represents 
	one detailed MESA binary evolution model, and the corresponding evolutionary outcome is colour-coded (see top legend). Here, "OB+cc" (cc= BH, NS, WD) implies that
    the corresponding model produced an OB+cc phase, with the compact companion type
    indicated by the corresponding colour. However, depending on the adopted
    compact object birth kicks, these systems may also break up. Systems indicated 
    as "Upper\_mdot\_limit" or "MT\_max" are terminated during their first mass transfer phase as the mass transfer rate exceeds
    limiting values (see text) and a merger is expected, and those indicated as "other merger" undergo L2-overflow 
    in a contact situation. 
	  Corresponding plots for different initial donor star masses can be found in App.\,\ref{Pq}.    \label{fig_2}
	}
\end{figure}

Figure \ref{summary} provides an overview of the models, and the outcome of the first mass transfer phase, for all models with an initial primary mass of M$_\mathrm{ 1,i} =17.8\,M_\odot$ (see App.\,\ref{Pq} for other initial primary masses). 
According to the definition of \citet{Kippenhahn1967}, our closest binaries undergo Case A mass transfer, where the primary star fills its Roche Lobe 
during its core hydrogen burning phase. Above an initial 
orbital period of $\sim$$5\,$d, the mass transfer takes place with a shell hydrogen burning donor (Case B mass transfer). 
At this initial primary mass, our Case A mass donors form neutron stars, and our Case B mass donors form black holes, according to our adopted maximum final He-core mass limit of 6.6$\mso$ for NS formation (cf. Sect. \ref{CC_formation}).
For initial orbital periods above $\sim$300\,d, the mass transfer rate violates our upper limit of 0.1$\mtr$.
Here, mass transfer is expected to become unstable due to the convective envelope of the donor star \citep{Soberman1997}. The widest binaries do not experience any binary interactions and serve as a grid of single star models. 

Based on our stability criterion from Eq. \eqref{qmin},
we find that a large fraction of models undergo unstable mass transfer
(labelled by "Upper\_mdot\_limit" in Fig. \ref{summary}).
For higher initial primary masses, fewer binary models experience unstable 
mass transfer (cf. Figs. \ref{summary}, \ref{A1k}, and \ref{A2k}). Also, the Case A/B boundary is shifted to 
larger initial orbital periods, even exceeding 1000\,d for the highest initial primary mass, due to envelope inflation near the Eddington limit 
\citep{Sanyal2015,Sanyal2017,Sen2023}.

If unstable mass transfer occurs during a Case\,A mass transfer, the merger product 
is expected to be a MS star, and we follow its evolution by using the merger models at 
the SMC metallicity computed by \citet{Wang2022}. In these models, the mass of the merger
product $M_{\rm merger}$ is determined by \citep{Glebbeek2013}
\begin{equation}
    M_{\rm merger} = \left[1-\frac{0.3 q}{(1+q)^2}\right](M_1+M_2),
    \label{Mass_merger}
\end{equation}
where $M_\mathrm{1}$ and $M_\mathrm{2}$ are the masses of the primary and secondary stars at the time of the merger, and $q=M_2/M_1$. The ejected mass during merger is below 7.5\% ($q=1$) of the mass of the pre-merger system.
This equation is based on head-on collision, and less mass ejection can be expected for a more gentle merger process \citep{Schneider2016}. The ejected material is assumed to be envelope material, which has the initial composition of the two stars.  

The angular momentum content of merger stars is uncertain. While the available orbital angular momentum exceeds the possible angular momentum content of the merger product, angular momentum can be lost efficiently along with the ejected fraction of the total mass. \citet{Schneider2019} find, in detailed 3D MHD and 1D stellar simulations, that the product of a merger between two massive main sequence stars is in fact slowly rotating due to angular momentum loss and  thermal relaxation. We therefore assume the same for our merger products, which implies in particular that they do not contribute to the predicted population of OBe stars.

\subsection{OBe stars} 

The Be phenomenon is produced by rapidly rotating massive main sequence stars which show the H$\alpha$ line in emission, and which have an spectral energy distribution showing an infrared excess. Both features are thought to originate from a circumstellar decretion disc which is fed from the layers near the stellar equator \citep{Reig2011,Rivinius2013}. 
While Be stars are fast rotators, it is unclear how close they are to their critical rotation.
\citet{Townsend2004} suggested that considering gravity darkening,
all Be stars spin near their critical rotation. However, the
concept of critical rotation is difficult to define \citep{Hastings2023}, and
observations seem to imply that some Be stars can rotate sub-critically, with rotational velocities as low as $\sim$60\% of the critical value \citep[e.g.,][]{Huang2010,Zorec2016,El-Badry2022,Dufton2022}.

In our fiducial model, we define Be stars as rotating faster than 0.95 of their critical rotational velocities ($\vc$). \citet{Golden2016} showed that Oe stars are the high-mass extension of Be stars, and therefore we adopt the same threshold value to define Oe stars. Since our spun-up accretors generally
reach critical rotation, and keep it up during their further main sequence evolution \citep[e.g.,][]{Hastings2020}, the threshold value only affects the
massive accretors that may slowly spin down due to their stellar wind \citep[above $\sim 40\mso$;][]{Langer1998}.
We quantify the effect of different threshold values in Sec.\,\ref{initial_parameters_var}.
We do not consider the detailed properties of the circumstellar discs \citep[cf.][]{Rubio2025}.  

\subsection{ Helium stars and Wolf-Rayet stars\label{He_WR}}

Following the common nomenclature, we define helium stars (He-stars) as 
stripped-envelope core helium burning stars \citep{Shenar2020,WangBo2021}, whose winds remain optically thin, such that they will not show emission lines in their spectra. Conversely, we speak of our model corresponding to a WR star when we can assume that it forms an optically thick stellar wind. \citet{Shenar2020} found a
metallicity-dependent luminosity threshold for the WR phenomenon, as
log$\,L/\ls \approx 4.9, 5.25, \mathrm{and\,} 5.6$ for the Galaxy, the LMC, and the SMC, respectively. 
\citet{Aguilera-Dena2022} and \citet{Sen2023}
showed that these luminosity limits are roughly reproduced by simple wind models, using the mass loss rates as adopted here.
Accordingly, we assume that our stripped stars with log$\,L/\ls > 5.6$ correspond to WR stars. This limit is exceeded by He-stars above $\sim$$15\mso$, which requires initial masses above $\sim$$40\mso$.

In addition, we define hydrogen-free (H-free) WR stars by a surface hydrogen mass fraction of less than 0.05, according to the typical uncertainties of the observationally derived hydrogen abundances in \citet{Shenar2016}.

\subsection{Formation of compact objects \label{CC_formation}}

We adopt the types of compact objects formed by our donor stars according to their final model properties.  \citet{Sukhbold2018} performed detailed simulations on the explodability of stars. They found a sudden increase in the compactness parameter of pre-supernova (pre-SN) stars at a final He core mass of 6.6$\ms$, which marks the formation of BHs. 
While stellar winds at different metallicities could alter the evolution of He stars \citep[e.g.,][]{WangBo2010,Doherty2015,Aguilera-Dena2022,Guo2024}, the relation between the compactness parameter and the final He core mass is insensitive to metallicity \citep{Sukhbold2016}. 
Accordingly, we assume that a BH forms if the final mass of a helium core $M_\mathrm{ He,c}$ reaches 6.6$\ms$. 
For simplicity, the non-monotonous behaviour of the compactness parameter \citep{OConnor2011,Sukhbold2016,Sukhbold2018,Schneider2023,Aguilera-Dena2023} is ignored, and this assumption may overestimate the number of low-mass BHs. 
Different assumptions on mixing process can affect the explodability of stars by changing the final core carbon abundance \citep{Patton2020,Schneider2021}, resulting in different values of the mass threshold. This  can affect the predicted number of low-mass BH but has a small effect otherwise \citep{Janssens2022}.
The mass of the BH is computed by using the same assumption as in the ComBinE code \citep[][and Paper II]{Kruckow2018}, which is that 20\% of the mass of the He-rich envelope of the core He depleted star is ejected, and after that  20\% of the remaining mass is lost due to the release of gravitational binding energy.

\begin{figure}[t]
    \centering
    \includegraphics[width=\linewidth]{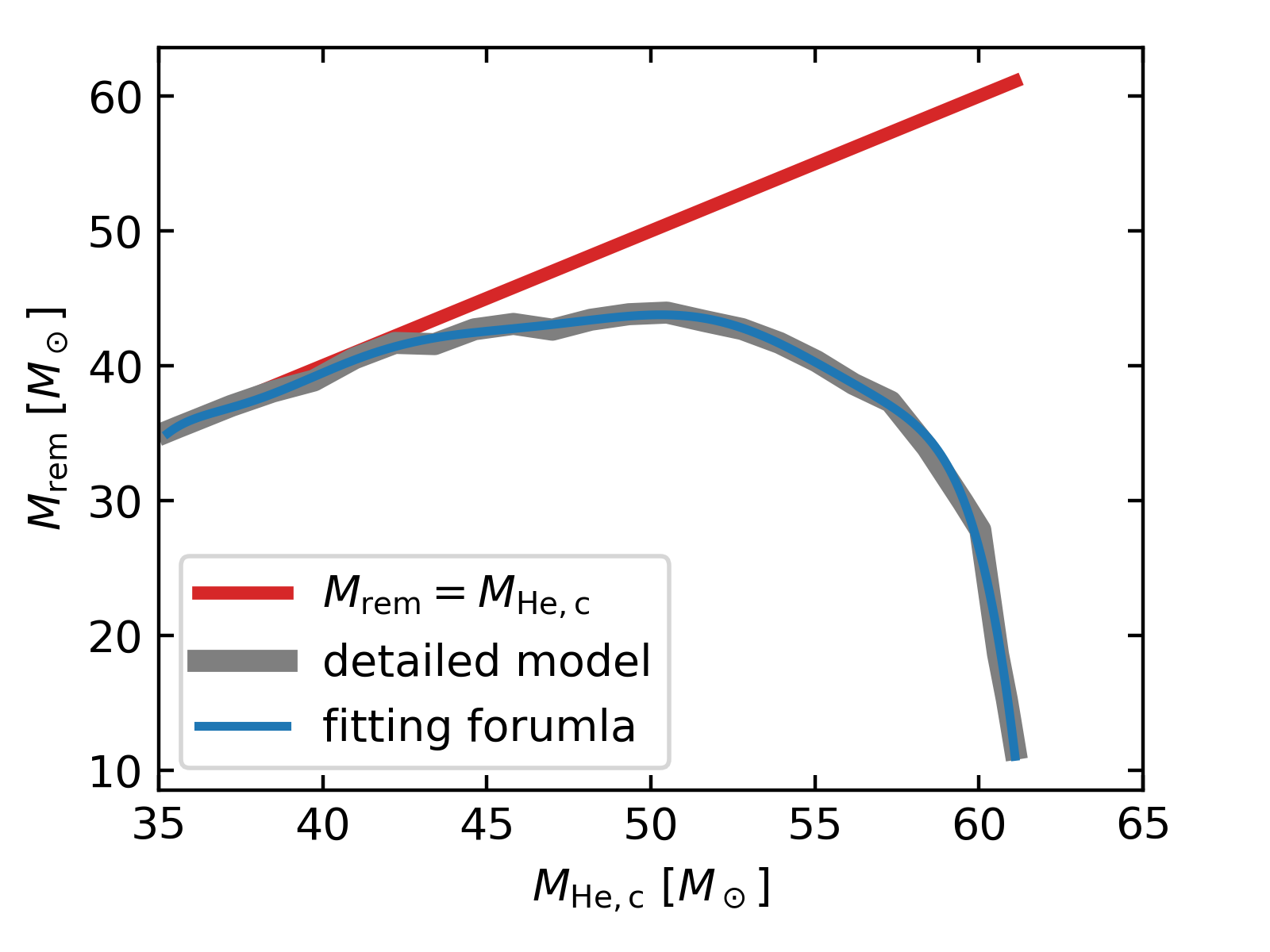}
    \caption{Remaining mass $M_\mathrm{rem}$ after the pulsations triggered by pair instability, as a function of the helium core mass $M_\mathrm{He,c}$ at core helium depletion. The gray and blue lines correspond to the model data in \citet{Marchant2019} and Eq\,\eqref{mrem_ppisn} respectively. The red line shows the relation $M_\mathrm{rem} = M_\mathrm{He,c}$.}
    \label{ppisn_fitting}
\end{figure}

Stars with very massive helium cores become unstable due to the production of electron–positron pairs. Those not massive enough to form a pair-production supernova \citep{Heger2002,Langer2007} undergo a series of energetic pulses accompanied by strong mass ejections \citep{Heger2002,Chatzopoulos2012,Woosley2017,Marchant2019}.
This process is known as pulsational pair-instability (PPI). 
\citet{Marchant2019} simulates these mass ejections during the PPI with the MESA code. Base on their result,
we perform a polynomial fitting to the helium core mass at core helium 
depletion and the remaining mass $M_\mathrm{rem}$ after the pulsations,
which is\footnote{In this formula, all digits are necessary.}
\newcommand{\xm}{M_\mathrm{He,c}}
\begin{equation}
\begin{split}
M_\mathrm{rem} =
& - 8.65828957\times 10^{-8}\,\xm^8\\
& + 3.25183895\times10^{-5}\, \xm^7\\
& - 5.31786630\times 10^{-3}\, \xm^6\\
& + 4.94552158\times10^{-1}\, \xm^5\\
& - 2.86053873\times10^{1}\, \xm^4\\
& + 1.05372343\times10^{3}\, \xm^3\\
& - 2.41399605\times10^{4}\, \xm^2\\
& + 3.14452667\times10^{5}\, \xm\\
& - 1.78318233\times10^6,
\label{mrem_ppisn}
\end{split}
\end{equation}
where $M_\mathrm{rem}$ and $\xm$ are in the unit of solar mass. Figure\,\ref{ppisn_fitting} provides a comparison between our fitting formula and the model data in \citet{Marchant2019}. 
This formula is only valid for $35.3\ms\leq\xm\leq 61.1\ms$. 
This mass range is somewhat sensitive to key nuclear reaction rates \citep{Takahashi2018,Farmer2020}, 
for which current best estimates lead to a slightly higher mass range \citep[e.g.,][]{Farag2022}.

We assume that stars with final helium core masses below $6.6\ms$ form NSs or, for stars with final carbon core masses below 1.37$\ms$ form WDs. The masses of our NSs are fixed to be 1.4$\ms$. 
Neutron star formation is accompanied by a SN explosion due to the collapsing core.  
Core collapse in the lowest-mass SN progenitors may be triggered by electron captures, producing a so called
electron-capture supernovae \citep[ECSNe,][]{Nomoto1984, Tauris2015}.
Since the low mass progenitors are expected to produce a weak momentum kick on the newborn NSs \citep{Janka2012},
they play an important role in the formation of Be/X-ray binaries \citep[BeXBs;][]{Podsiadlowski2004}.
However, our model grid cannot resolve the narrow mass range of ECSNe well (see Sect. \ref{SN_window}).
For consistency with the results obtained with the ComBinE code (Paper II), we apply the SN scheme adopted in \citet{Kruckow2018} (see App. \ref{SN_window} for details). 
This considers four types of SNe:
\begin{itemize}
    \item ordinary iron core collapse SNe (CCSNe), if the pre-SN star has a final carbon core mass larger than 1.435$\ms$ and a H-rich envelope;
    \item H-envelope-stripped iron-core collapse SNe (CCSN-He) if the H-rich envelope has been stripped by mass transfer;
    \item He-envelope-stripped iron-core collapse SNe (CCSN-C) if most of the He-rich envelope has been stripped by mass transfer;
    \item  ECSNe if the final carbon core mass is in the range 1.37 - 1.435$\ms$.
\end{itemize}
Different types of SNe are associated with different kick velocity distributions,
which are introduced in the next subsection.

\subsection{Natal kick\label{natal_kick}}

During the SN explosion, asymmetries in the neutrino emission or in the SN ejecta can generate a momentum kick in the new born NSs.
Empirical constraints on the kick velocities is still debated \citep[see the discussion in][]{Valli2025arXiv250508857V}, making this one of the major uncertainties for the formation of NS binaries.
The eccentricity of high-mass X-ray binaries \citep{Pfahl2002} and 
double pulsar binaries \citep{Tauris2017,Vigna-Gomez2018} imply that stripped
SNe produce weak momentum kick. \citet{Hobbs2005} find the spatial 
velocity of young pulsars can be described by the Maxwellian distribution 
with $\sigma=265\kms$. With a sample of 28 young pulsars having VLBI measurements,
\citet{Verbunt2017} suggest that there is a slow-moving group of young
pulsars \citep[also see][]{Igoshev2020}. Recently, \citet{Fortin2022} and \citet{ODoherty2023} argue that the kick velocities
derived from isolated pulsars overestimate the intrinsic NS kick velocities, since
the isolated ones biased towards strong kicks that unbind the binary. The authors
find much lower kick velocities by using NS low-mass X-ray binaries.

The more energetic the explosion, the stronger is the expected kick \citep{Janka2012}.
We use a Monte Carlo method to account for these dynamical effects of 
SNe. For each pre-SN binary, we determine the type of SN with the SN scheme explained above. 
Then we randomly draw a sample of kick 
velocities from the corresponding kick velocity distribution with random orientations. Here we adopt
the same kick distributions as ComBinE, which are summarized in Tab.\,\ref{tab2}. 

For normal CCSNe, the kick velocity $\upsilon_\mathrm{ K}$ distribution is assumed to be a Maxwell-Boltzmann 
distribution $f(\upsilon_\mathrm{ K},\,\sigma)$ with a root-mean-square velocity 
$\sigma = 265\kms$, which is based on the proper motion analysis of young pulsar
\citep{Hobbs2005}, where
\begin{equation}
	f(\upsilon_\mathrm{ K},\,\sigma)=\sqrt{\frac{2}{\pi}}\frac{\upsilon_\mathrm{ K}^2}{\sigma^3}\mathrm{ exp}\left(-\frac{\upsilon_\mathrm{ K}^2}{2\sigma^2}\right).
	\label{kick}
\end{equation}
The SNe from stripped stars are thought to be less energetic and therefore to generate weaker kicks \citep[cf.][]{Tauris2015}.
On the other hand, the observed double neutron star binaries imply that also rather large kicks 
may happen in close binaries \citep[][and references therein]{Tauris2017}.
Accordingly, for stripped SNe we adopt a bi-modal Maxwellian distribution 
$f_2(\upsilon_\mathrm{ K},\,\sigma_1,\,\sigma_2)$ with a 80\% low-kick component
and a 20\% high-kick component,
\begin{equation}
\begin{split}
	f_2(\upsilon_\mathrm{ K},\,\sigma_1,\,\sigma_2)=&0.8\sqrt{\frac{2}{\pi}}\frac{\upsilon_\mathrm{ K}^2}{\sigma_1^3}\mathrm{ exp}\left(-\frac{\upsilon_\mathrm{ K}^2}{2\sigma_1^2}\right)\\
    &+0.2\sqrt{\frac{2}{\pi}}\frac{\upsilon_\mathrm{ K}^2}{\sigma_2^3}\mathrm{ exp}\left(-\frac{\upsilon_\mathrm{ K}^2}{2\sigma_2^2}\right) .\label{kick2}
\end{split}
\end{equation}
Based on the observed migration of Galactic high-mass X-ray binaries
\citep{Coleiro2013},  the low-kick component $\sigma_1$ is taken 
to be $120\kms$ and $60\kms$ for CCSN-He and CCSN-C 
respectively, and the high-kick component $\sigma_2$ is taken to be 
200$\kms$\citep{Tauris2017}. For ECSNe, based on the observed low-eccentricity of X\,Persei-type X-ray binaries \citep{Pfahl2002,Podsiadlowski2004}, a flat distribution in the range $[0,\,50]\kms$ is adopted.

The momentum kick imparted on BHs is highly uncertain. It has been
proposed that the low-mass BHs can obtain a natal kick due to
fallback during the BH formation
\citep[e.g.,][]{Belczynski2008,Fryer2012,Janka2017}. \citet{Mirabel2017} and \citet{Mandel2020}
provided evidence that the BHs of $\sim$$10\mso$ and $\sim$$21\mso$ in the
high-mass BH binaries GRS 1915+105 and Cygnus X-1 formed
with essentially no kick. \citet{Shenar2022} and \citet{Vigna-Gomez2024} find that
the near-circular X-ray quiet BH binary VFTS\,243 in the LMC was born with a negligible kick. However, some BHs in LMXBs may be born with strong kicks \citep{Repetto2015}.
Recent evidence for and against kicks in the formation of BHs in LMXBs is discussed in \citet{Nagarajan2024}. 
For simplicity, we assume no BH formation kick in our fiducial population synthesis model.
We investigate the effect of this assumption in Sect.\,\ref{sect4}.

Besides the kinetic energy injected by a natal kick, mass loss during 
the SN weakens the gravitational binding 
of a binary \citep{Blaauw1961}. 
Whether a binary remains bound after a SN 
depends on whether the orbital energy of the post-SN system
is negative. When a model binary 
remains bound, we calculate the parameters of the post-SN orbit using the formulae given
in appendix\,A.1 of \citet{Hurley2002}, which are based on \citet{Hills1983}. The velocity of the center of  mass generated by the supernova explosion is not considered in this work. If a binary gets disrupted, we count the main-sequence companion as a runaway star regardless of its space velocity.

\begin{table}
\caption{Birth kick velocity distributions for neutron stars used in our synthetic populations.}
\label{tab2}
\begin{tabular}{c |c c}
\hline
	
	ordinary CCSN & \multicolumn{2}{c}{Maxwell-Boltzmann distribution}\\
	                             & \multicolumn{2}{c}{with $\sigma = 265\kms$\tablefootmark{a}}\\
	\hline
        stripped CCSN  & \multicolumn{2}{c}{bi-modal Maxwell-Boltzmann distribution} \\
           & $\sigma_1$\tablefootmark{a}  & $\sigma_2$\tablefootmark{a} \\
	$-$ CCSN-He\tablefootmark{b} & 120 km s$^{-1}$ & 200 km s$^{-1}$ \\
	$-$ CCSN-C\tablefootmark{b} & 60 km s$^{-1}$ & 200 km s$^{-1}$\\
	\hline
 ECSN & \multicolumn{2}{c}{flat distribution, 0 $-$ 50 km$\,$s$^{-1}$} \\
	\hline
\end{tabular}
\tablefoot{
\tablefoottext{a}{Parameters $\sigma$, $\sigma_{1/2}$ are the 1D root mean squares of the Maxwellian and bi-modal Maxwellian distributions, respectively.}
\tablefoottext{b}{CCSN-He/C stand for pre-SN stars having their hydrogen/helium envelope stripped.}
}
\end{table}

\subsection{Population synthesis}\label{PopSyn}

In contrast to the commonly adopted Monte Carlo method, we perform population synthesis based on a grid-approach. 
This is suitable as the number of dimensions that our grid span is not very large ($\mi$, $\qi$, and $\logpi$) and ensures a uniform coverage allowing us to capture the full variation in behaviour.
We assume an initial binary fraction of 100\%. We then assign a statistical weight to each binary model in a given cell of our initial parameter space which are in a specified evolutionary stage, given by the birth probability and the lifetime of the considered stage (see App.\,\ref{formulas} for detailed equations).

The birth probability is determined by the adopted star formation rate, the
initial mass function (IMF, $f_\mathrm{ IMF}$), and the initial distribution of mass ratios
$f_{\qi}$ and orbital periods,  $f_{\logpi}$, while the lifetime 
given by the evolutionary model in the considered grid cell.
For an OB+cc binary, 
we determine its lifetime by
the time between the compact object formation and the secondary star leaving the main sequence or filling 
its Roche Lobe. 
We adopt the IMF derived by \citet{Kroupa2001},
\begin{equation}
    f_\mathrm{ IMF} \propto \mi^{-\alpha},
\end{equation}
where
\begin{equation}
	\begin{cases}
	    \alpha=0.3~~~0.01\leq M_\mathrm{ 1,i}/M_\odot < 0.08\\
		\alpha=1.3~~~0.08\leq M_\mathrm{ 1,i}/M_\odot < 0.50\\
		\alpha=2.3~~~0.50\leq M_\mathrm{ 1,i}/M_\odot \\
	\end{cases}.
	\label{IMF}
\end{equation}
For our fiducial model, initial distributions of mass ratio and orbital period 
are taken from \citet{Sana2012}, which is derived from 
the O star population in open star clusters. It shows a preference 
for short orbital periods
and a near flat mass ratio distribution,
\begin{equation}
    f_{\qi} \propto \qi^{-0.1}
    \label{f_qi}
\end{equation}
and
\begin{equation}
    f_{\logpi} \propto (\logpi)^{-0.55}.
    \label{f_logpi}
\end{equation}
In addition, a constant star formation rate of  $0.05\sfr$ is adopted 
\citep{Harris2004,Rubele2015,Hagen2017,Rubele2018,Schootemeijer2021} in our fiducial model.
We explore different star formation histories in Sect.\,\ref{initial_parameters_var}.

During the OB+cc phase, the orbits may slowly expand or shrink due to the stellar wind 
from the main-sequence companions \citep{Quast2019,Mellah2020}. Considering 
the low mass-loss rate of OB stars in the SMC, we do not expect significant changes in orbital
separation during OB+cc phase. Therefore, we simply assume the orbital parameters
remain unchanged. Also the stellar parameters are assumed to remain constant during this phase. 
For stellar rotation,
we consider the tidal interaction during the OB+cc phase by calculating the 
synchronization timescale at the beginning of the OB+cc phase. We found that 
tides during the OB+cc phase are too weak to further spin down the OB star
(see App. \ref{sync_timescale} for details).

For He-stars, we determine their parameters at the middle of core helium burning, which we define as the moment when
when the core He mass fraction dropped to 0.5,
and the lifetime is determined by their core
He burning time. For WR stars, we go through the evolutionary
tracks of core He burning phase step by step to check whether the stars
are luminous enough to be considered as WR stars.

We count pre-interaction MS binaries by going through the evolutionary tracks 
of all double MS stars in our binary model grid. Only the primary stars are counted,
since they are visually brighter than the secondary stars in pre-interaction systems.
We define O stars as MS stars with effective temperatures
hotter than 31.6\,kK ($\teff /{\rm K} > 4.5$) . 
Here we adopt the relation between spectral type and 
temperature derived by \citet[][]{Schootemeijer2021}.
This work does not investigate interacting binaries \citep[see][for a detailed study on massive Algol systems]{Sen2022}

\section{Properties of our fiducial synthetic population\label{results}}

In this section we present the main properties of our synthetic
population based on our fiducial parameter choice, as explained above
(see App. \ref{further_details} for further model details). 
In Sect.\,\ref{initial_parameters_var}, we discuss the effects of variations of the birth kicks of compact objects, the core mass
ranges defining the emerging compact object type, the threshold rotation for assuming an OBe nature, and the star formation
history.
We compare our predictions with the observed numbers in Sect.\,\ref{s_obs}
and discuss the implication for mass transfer physics in Sect.\,\ref{discuss}.

\subsection{The number of O\,stars} \label{no}
Our adopted star formation rate of 0.05$\msoy$ determines the absolute number of predicted stars of any type. Our fiducial population contains  1070 O\,stars, which are defined as the main sequence stars with $T_{\rm eff} > 31.6\,$kK. This number is slightly larger than the estimated number of observed O\,stars in the SMC, which is consistent enough for the purposes of this study. For example, in the BLOeM survey of the SMC, \citet{Shenar2024} identify 159 O\,stars, with a completeness of about 35\%. While the survey covers less than half of the surface of the SMC, it focuses on the regions rich in massive stars. Considering the effect of crowding, the BLOeM estimates lead to a total number of about 500 O stars in the SMC.

A similar overprediction has already been noticed by \citet{Schootemeijer2021}, who
also obtained a total number of O stars in the SMC of just above 1000 with the same SFR and IMF as our fiducial model. However, the authors found that young O\,stars appear to be missing in the SMC by analysing  the Gaia DR2 catalogue \citep{Brown2018} and on the spectral type catalogue of \citet{Bonanos2010} (see also \citealt{Castro2018}, which is based on the RIOTS survey \citealt{Lamb2016}). This means that deep embedding may hide the youngest O\,stars from our view. 
Alternately, a star formation history with no star formation at the present time can explain this missing, but it fails to reproduce the observed number of the SMC WR star binaries (see Sect.\,\ref{initial_parameters_var}).

\subsection{Compact object binaries amongst main-sequence stars}\label{s_pie}
\begin{figure*}[!htbp]
    \centering
    \includegraphics[width=0.47\linewidth]{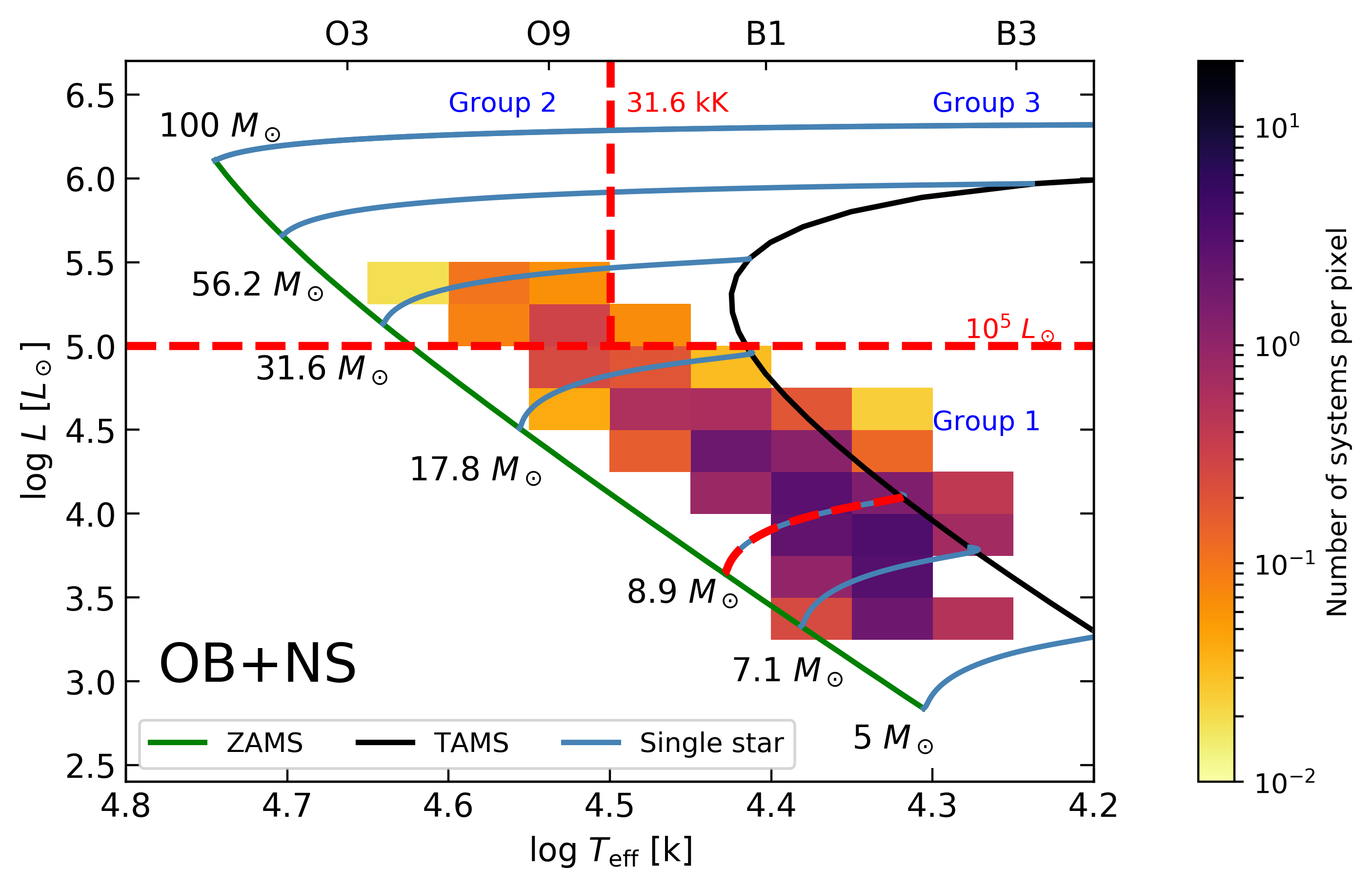}
    \includegraphics[width=0.47\linewidth]{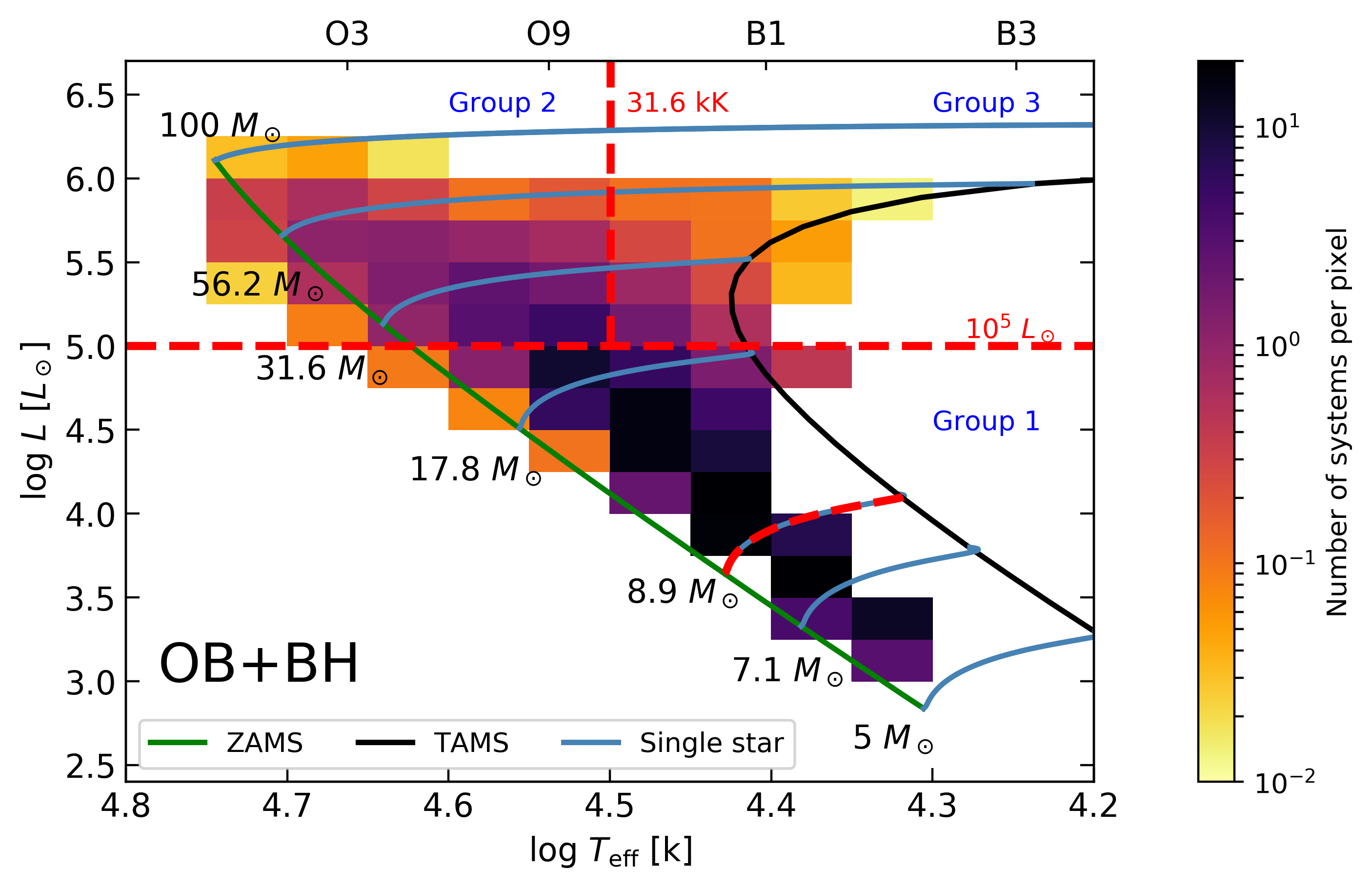}
    \caption{Predicted distribution of OB stars with a NS (left panel) and BH (right panel) companion in the Hertzsprung–Russell diagram. The green line is the zero-age main sequence (ZAMS), and the black line is the terminal-age main sequence (TAMS). The blue lines represent evolutionary tracks of non-rotating core hydrogen burning single stars, with the indicated initial masses. The red dashed lines show the boundaries of the three groups of stars defined in the main text, i.e., $T_{\rm eff}=31.6\,$kK, $\llso =5$, and the MS evolutionary track of a $8.9\ms$ single star. On top, we label several spectral types at their approximate effective temperature. \label{HRD_cc}}

    \includegraphics[trim=0.2cm 0 0 0,clip,width=0.87\linewidth]{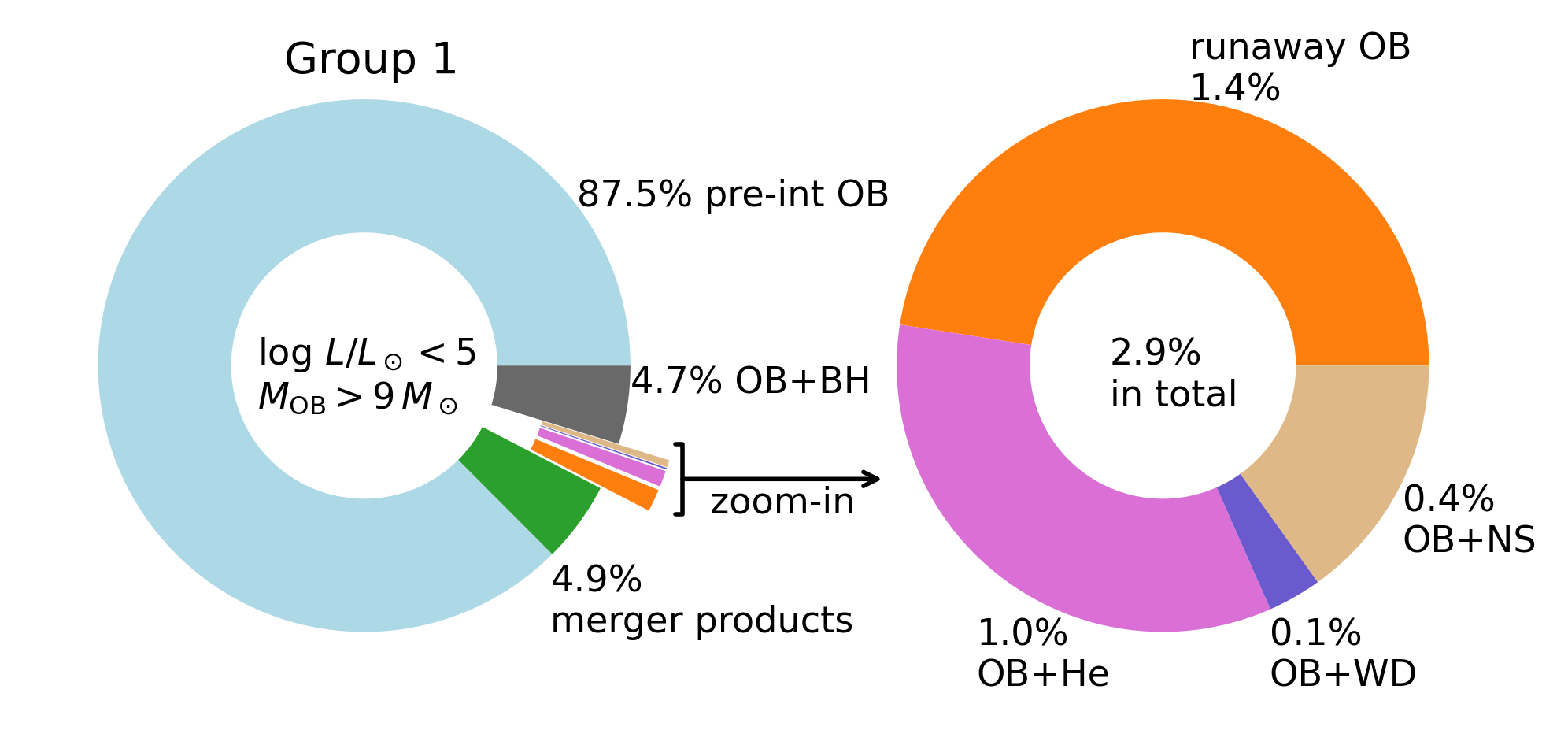}
    \vskip -0.4cm
    \includegraphics[trim=0.2cm 0 0 0,clip,width=0.87\linewidth]{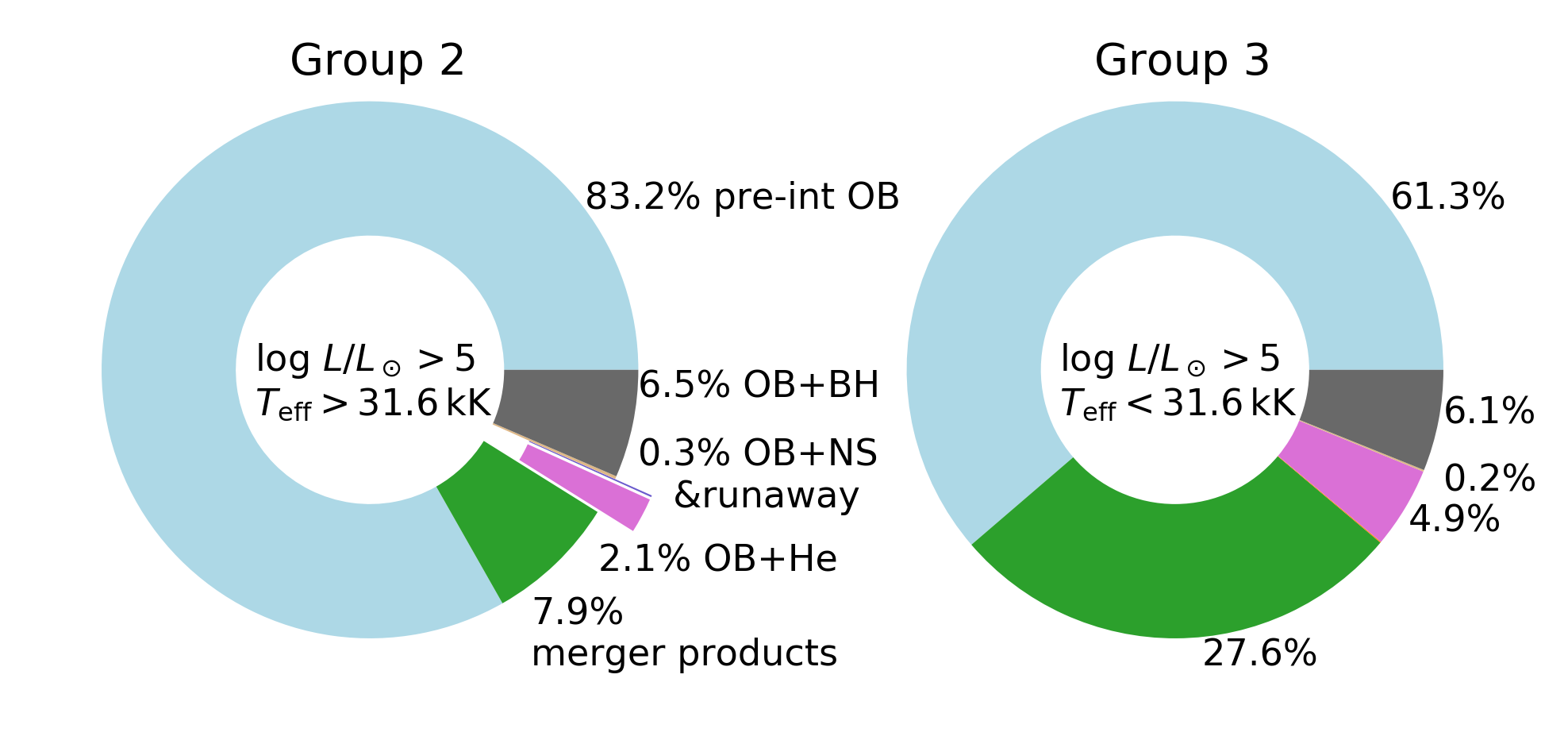}
    \caption{Pie charts for the predicted numbers of pre-interaction main-sequence OB stars primaries, and of OB stars in different types of post-interaction binaries, divided into three groups, Group 1 (top left; top right provides a zoom-in), Group 2 (lower left), and Group 3 (lower right), based on luminosities and effective temperatures. The considered types of binary stars and the corresponding fractions with respect to the total number within the considered group are indicated by text. The corresponding absolute numbers in each wedges are listed in Tab.\,\ref{number_Piechart}. \label{SMC_Pie}  }
\end{figure*}

\begin{table}[]
    \centering
    \caption{The predicted numbers of the pre- and post-interaction main-sequence OB stars in our fiducial synthetic SMC population. The numbers are divided into three groups, depending on the location on the Hertzsprung–Russell diagram, 1) Group 1: log$\,(L/\lso) < 5$ and $M_\mathrm{OB}>9\,\mso$, 2) Group 2: log$\,(L/\lso) > 5$ and $T_\mathrm{eff}> 31.6\,$kK , 3) Group 3: log$\,(L/\lso) > 5$ and $T_\mathrm{eff}< 31.6\,$kK.}
    \label{number_Piechart}
    \begin{tabular}{c|cc|c|c}
    \hline\hline
      Types   &\multicolumn{2}{c|}{Group 1} & Group 2 & Group 3\\
        &  O type & B type & O type  & B type \\\hline &&&&\\[-2ex]
    pre-int. &705 & 2041 &274 & 40.4\\
    merger products &17.0 & 138 &26.2& 17.8 \\
    runaway & 0.20 & 42.6 &0.34&0.04  \\
    OB+He & 2.09 & 28.6&6.96& 3.19 \\
    OB+WD & 0&2.94 & 0 &0  \\
    OB+NS &0.29 &13.3 &0.58& 0.07  \\
    OB+BH & 18.0& 131.1&21.4& 3.97 \\\hline &&&&\\[-2ex]
    Total & 742 & 2398&329& 65.5\\\hline
    \end{tabular}
\end{table}

To predict the fraction of massive MS stars expected to be in a binary with a compact object, we first determine the predicted number of OB+cc binaries. We further consider the MS stars with lower mass MS companions, with helium star companions, and those without companions, where the latter are MS mergers and runaways from disrupted systems.

Figure \ref{HRD_cc} presents the distribution of the MS stars of our OB+NS/BH binaries 
in the HR-diagram (left panel for NS companions, and right panel for BH companions). The MS components of our OB+NS systems are located between the single-star tracks of $\sim$6$\mso$ and $\sim$32$\mso$. The lower mass is set by the smallest initial mass ratio (~0.65) in the stable mass transferring binaries with the lowest initial primary mass to provide NSs ($\sim$$9\mso$), as can be seen from the overview diagrams in Fig.\,\ref{A1k}. The highest mass primaries to provide NSs in our grid are about 20$\mso$ stars in Case\,A binaries (Fig.\,\ref{A1k}). Those are short period systems which, especially for high initial mass ratios, evolve with mass transfer efficiencies of up to 60\%, such that the mass gainers can grow up to $\sim$$30\mso$. The peak of the mass distribution of the MS stars in NS-binaries is at about $9\mso$. Essentially all OB companions in our NS binary models are somewhat evolved. Hence they are not close to the ZAMS. The reason is that the NS forming binaries all have initial mass ratios larger than $\sim$0.5, with an average of about 0.8, which means that the MS lifetimes of the mass gainer and mass donor are comparable. Consequently, when the mass donor produces a NS, the mass gainer has burnt a significant amount of its core hydrogen. 

As we discuss in more detail below, the majority of the MS components below $\sim$$15\mso$ are expected to rotate close to critical rotation and display themselves as OBe stars. The remaining ones are also expected to rotate rapidly. As most NS forming binaries break up due to the NS birth kick (see below), we expect many more Be stars without NS companion than as part of a NS binary. 

The highest mass of a MS companion to a WD produced in our grid is $\sim$$12\mso$ (Fig.\,\ref{Companions}). As we do not expect newborn WDs to receive kicks like NSs, all WDs are predicted to remain in binary systems, mostly associated with fast-rotating companions, some of which may be observed as WD Be/X-ray binaries \citep[e.g.,][]{Gaudin2024MNRAS.534.1937G,Marino2025ApJ...980L..36M}. Since our model grid is highly incomplete for WD binaries, their detailed analysis is beyond our scope, and we focus on NS/BH-forming systems.

Since our model predicts more mergers for less massive systems,
we expect more BH than NS binaries in the SMC, as shown in the right panel of Fig.\,\ref{HRD_cc}. We find that the MS companions of BHs are stretching over a wider mass and and temperature range. While BHs are only formed from primaries above $\sim$$18\mso$, our BH forming binaries can have stable mass transfer for initial mass ratios as low as 0.3 (Fig.\,\ref{A1k}).
Due to the different accretion efficiencies of Case A/B binaries ($\simle$ 10\% for Case B and up to $\sim$ 60\% for Case A), we find the lowest mass MS companions ($6\mso$) to BHs in our models formed in Case B systems but the most massive ones (near $100\mso$) formed in Case\,A systems (Fig\,\ref{App_Mass}). The difference in accretion efficiencies also leads to a stronger surface abundance enrichment for the mass gainers in Case A systems (Fig.\,\ref{App_abundance}).  

Towards higher masses, our results become incomplete, but pair-instability will also reduce the number of BH binaries produced there (our most massive BHs are $\sim$$38\mso$; see below). As the our most massive main sequence star models undergo envelope inflation, we expect some early B\,supergiants companions to BHs \citep[cf.,][]{Quast2019}. The peak of the MS companion mass distribution is near $10\mso$, close to that of the NS companions. Most of the BH companions are expected to rotate very rapidly.

To obtain absolute numbers, we integrate the number densities for three groups of stars, mainly depending on the location of the MS component, or of the more massive star in case of pre-interaction binaries, in the HR diagram as follows (see Fig.\,\ref{HRD_cc} for the group borderlines).
\begin{itemize}
    \item Group 1: MS stars with masses above 9$\mso$ and luminosities below $10^{5}\lso$;
    \item Group 2: MS stars with luminosities above $10^{5}\lso$ and effective temperatures above 31.6 kK;
    \item Group 3: MS stars with luminosities above $10^{5}\lso$ and effective temperatures below 31.6 kK.
\end{itemize}

The chosen boundaries are motivated as follows. The lowest initial primary mass in our binary model grid is $5\ms$. The most massive merger product involving 5$\mso$ primaries occur in systems with an initial mass ratio of 0.95, and produce $9.02\ms$ merger products according to Eq. \eqref{Mass_merger}. We therefore chose a lower mass limit of $9M_\odot$ to ensure completeness of the merger products and mass gainers. Furthermore, the lowest initial primary mass for forming BHs in our Case\,B system is $17.8\ms$ (Fig.\,\ref{fig_2}), whose luminosity at terminal-age main sequence is about $10^5\ls$. Finally, we chose an effective temperature of 31.6\,kK do distinguish between O\,type and B\,type MS stars.

The predicted numbers and fractions of the pre- and post-interaction OB stars are presented in Tab.\,\ref{number_Piechart} and Fig.\,\ref{SMC_Pie} respectively. To obtain the proper fractions, we include the OB stars produced by the post-interaction single star channels, which are merger products and runaway stars.   

In Group\,1 we find 742 O\,type and 2398 B\,type stars, of which 2746 (87.5\%) reside in pre-interaction binaries, 155 (4.9\%) are merger products, and 42.6 (1.4\%) are runaway stars.
The remaining $\sim$$200$ OB stars have evolved companions.

Group\,1 contains 13.6 OB stars with NS companions, most of them B\,type stars, which implies that only 0.4\% of the stars in this group would have a NS companion. At the same time, we expect 4.9\% of the MS stars in Group\,1 to orbit a BH companion, which is 149 OB+BH binaries in Group\,1, 131 of them with a B\,star. The vast majority of them is expected to be X-ray silent \citep[cf.,][]{Sen2024}.

Group\,1 is further expected to contain 30.7 OB+He-star binaries, with none of them expected to produce dense enough winds to appear as Wolf-Rayet star. However, their winds may collide with the OBe disc or with the wind of their companion, which may turn them into observable X-ray sources \citep{Langer2020a}. We further find 3 WD binaries in Group\,1, representing the high mass end of the WD companion mass distribution. 

The high mass Groups\,2 and\,3 contain 329 O-type stars and 65.5 B\,supergiants, respectively, which corresponds to 9.3\% and 1.8\% of all considered OB stars, which means that Group\,1 contains 88.9\% of them. By design, Groups\,2 and\,3 contain essentially no NS binaries or runaway stars. However, the most massive BHs and the Wolf-Rayet binaries will be found in these groups. With about 10\%, the fraction of OB stars with evolved companions is almost twice that in Group\,1, with 21.4 (6.5\%) and 3.97 (6.1\%) of the MS stars forming O+BH and B+BH binaries in Groups\,2 and\,3, respectively. Also the MS merger fraction in Group\,3 is very high with 27.6\%.

These numbers are determined by several factors. 
The fraction of MS merger products goes up with mass due to the growth of the parameter space of Case A mergers (App.\,\ref{Pq}). The general decreasing of merger fraction for higher primary masses leads to the increasing of the expected number of post-interaction binaries.  
This is also relevant for the lower number of NS star binaries compared to BH binaries, which is aggravated by the adopted NS birth kicks, and the neglect of BH kicks.
Of course, a different lower mass limit or criterion for BH formation will also shift the relative numbers. The fraction of OB+He-star binaries is at just 10---20\% of the OB+cc binaries, because their lifetimes are determined by their core He burning time, while the lifetimes of the former is fixed by the remaining lifetime of the OB star after the formation of the compact object, which is a fair fraction of their MS lifetime.

\subsection{Number of post-interaction binaries \label{s_numbers}}
While the prediction of MS mergers from our grid is incomplete below $\sim$$9\mso$ (see above), our choice of the lowest initial primary mass of $5\mso$ guarantees completeness for the expected NS and BH binaries. In the previous subsection, we used the $9\mso$ limit for a meaningful comparison of the number of evolved binaries with that of MS stars.  
Here, we use our complete grid down to primary masses of $5\mso$, in order to derive absolute numbers of expected post-interaction binaries in the SMC, as well as the their mass dependence. 

\begin{table}
\caption{Number of post-main sequence companions of OB stars in our fiducial synthetic SMC population. Besides the total numbers, we give the numbers emerging from Case\,A mass transfer, from Case\,B mass transfer, and from chemically homogeneous evolution (CHE).
As OBe stars, we count OB mass gainers which rotate faster than 95\% of their critical rotation. For core-helium burning mass donors, we
distinguish three different initial mass ranges as indicated, as well as stars with $\log L/ \lso > 5.6$ as WR stars
\label{tab3}}
\centering
\setlength{\tabcolsep}{5pt}
\begin{tabular}{l | ll|ll|ll}
\hline\hline
	  & \multicolumn{2}{c|}{Total} &  \multicolumn{2}{c|}{Case A} & \multicolumn{2}{c}{Case B}\\
	  &   OBe & OB    &   OBe & OB & OBe & OB \\
	\hline\rule{0pt}{\dimexpr.7\normalbaselineskip+0.4mm}
	He-stars & 223    & 12.1  & 56.5  & 11.1 & 166 & 0.96\\
	- $\mi \leq 10\mso$ &  191 & 0.03   & 51.0 & 0 & 140 & 0.03 \\
	- $\mi, 10-17\mso$ &  18.4 &  2.52  & 4.86 & 2.52 & 13.6 &  0\\
    - $\mi \geq 17\mso$ &  13.6 & 9.53   & 0.61 & 8.60 &13.0 &  0.93\\
    \hline\rule{0pt}{\dimexpr.7\normalbaselineskip+0.4mm}
	 WR-stars &  3.97  &  2.75 &  0.83 & 2.30 & 3.14 & 0.45\\
    - H-free WR & 0.02 & 0.20  &  0.02 & 0.20 & 0 & 0\\
    - CHE WR & 0 & 0.04 & & & & \\ 
 	\hline\rule{0pt}{\dimexpr.7\normalbaselineskip+0.4mm}
    NSs & 92.5  & 6.59  & 7.98 & 6.59 & 84.6 & 0 \\
	 - bound & 21.0  &  3.74 & 5.41 & 3.74 & 15.6 & 0 \\
	- disrupted & 71.5& 2.85 & 2.57 & 2.85 & 69.0 & 0 \\
 	\hline\rule{0pt}{\dimexpr.7\normalbaselineskip+0.4mm}
    BHs & 170 & 40.3& 5.60 & 22.3 & 165 &18.0 \\
    - CHE BH & 0 & 0.32 & & & & \\
	\hline
\end{tabular}
\end{table}

\begin{figure*}
    \centering
    \includegraphics[width=\linewidth]{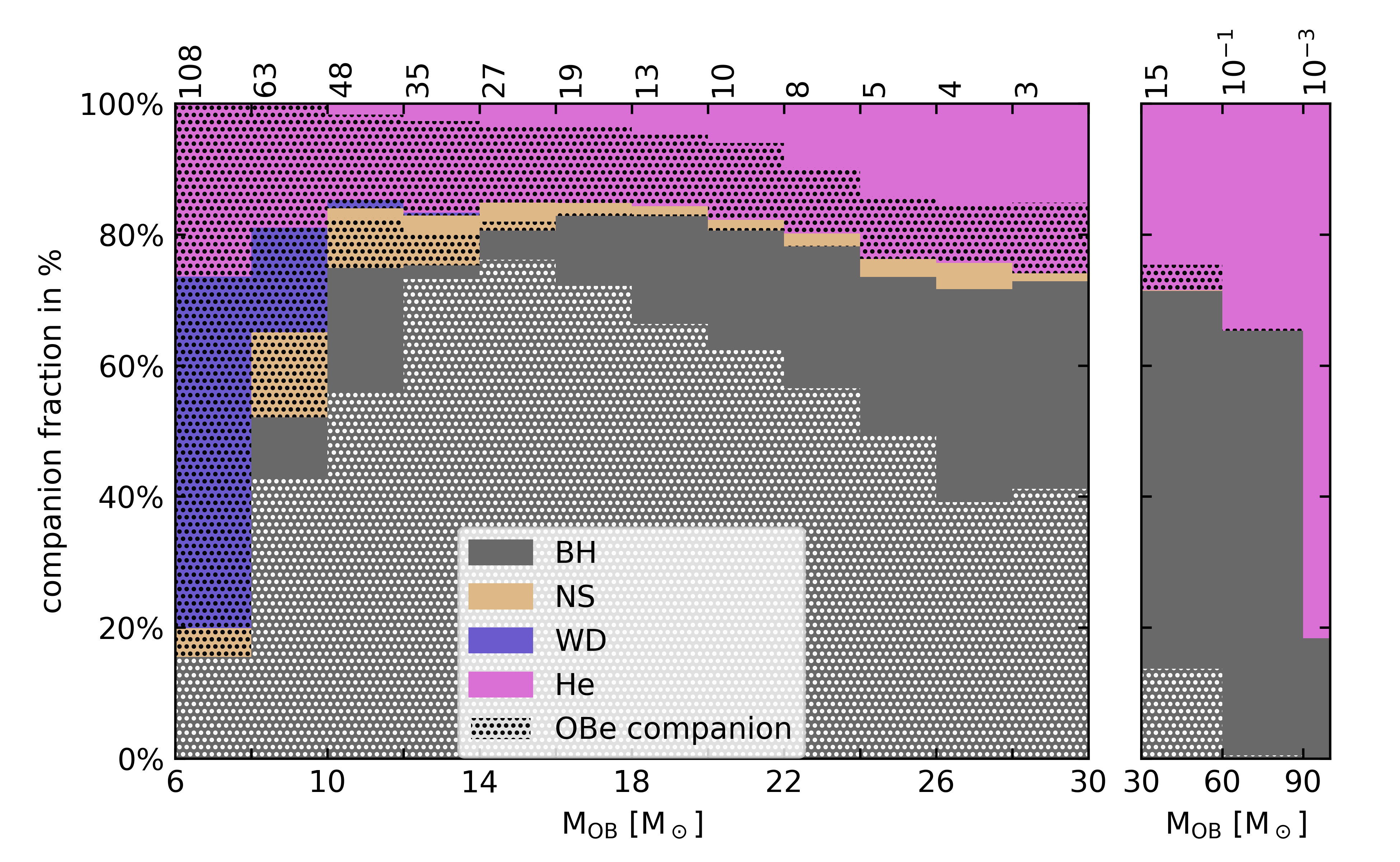}
    \caption{The fractions of different types of post-main sequence companions of OB stars in our fiducial synthetic binary population relative to all post-main sequence companions, as a function of the mass of the OB star. For the companions, we distinguish core-helium burning stars (magenta),
    white dwarfs (blue), neutron stars (yellow) and black holes (gray). Shading identifies those companions which are paired with an OB star that rotates  with more than 95\% of its critical rotational velocities. The absolute number of binaries with post-main sequence companions expected in the SMC for each mass bin is given on top of the bin. 
    The high-mass end (30 - 100 $M_\odot$) is presented with a wider bin width. 
    }
    \label{Companions}
\end{figure*}

The numbers of the various types of binary systems consisting of a main sequence star (the mass gainer) and a post-main sequence companion  predicted to exist in the SMC are shown in Tab.\,\ref{tab3} and Fig.\,\ref{Companions}. In Tab.\,\ref{tab3}, we distinguish core-helium burning stripped mass donors originating from three different initial donor mass ranges, which roughly correspond to those forming WDs, NSs, and BHs, respectively, according to our assumptions (Sect.\,\ref{CC_formation}). We expect that the numbers in Tab.\,\ref{tab3} and Fig.\,\ref{Companions} are complete, except for the WDs ($M_\mathrm{ i,1} \leq 10\mso$). Note, however, that while the initial mass limits for forming a certain compact object type are rather sharp for Case\,B binaries, the emerging helium star mass in Case\,A systems is not only a function of the initial donor mass but also of the initial orbital period \citep{Wellstein2001, Schurmann2024}. We can see the effect of this in Fig.\,\ref{fig_2}, which shows that $17.8\mso$ donors form BHs in Case\,B systems but NSs in Case\,A evolution (see also the plot for the initial donor mass of $10\mso$ in Fig.\,\ref{A1k}). 

Tab.\,\ref{tab3} further gives the predicted number of OB+WR binaries, and for the much smaller fraction of them in which the WR star is expected to be hydrogen-free. We also give the small number of WR stars produced in very massive and very close binaries in our sample undergoing chemically homogeneous evolution (CHE; see below) in Tab.\,\ref{tab3}, for completeness.
Finally, we list the predicted number of OB stars with NS and BH companions. We also count the number of binaries which were disrupted due to the adopted NS birth kick in Tab.\,\ref{tab3}, while Fig.\,\ref{Companions} shows only the remaining binaries. Since our fiducial model ignores BH birth kicks, none of our model binaries is broken up when a BH is formed.

In Tab.\,\ref{tab3} and Fig.\,\ref{Companions}, we also distinguish
systems in which the mass gainer is spun up to close-to-critical rotation
and designate them as OBe stars. Notably, we find that 
60---100\% of the post-interaction systems contain an OBe star.
Besides the high efficiency of the spin-up process, these numbers reflect
the generally low mass loss rates of our SMC main sequence star models, which effectively spins down star only above $\sim$$30\mso$.

Our overview plot for $10\mso$ donors (Fig.\,\ref{A1k}) shows that we expect the lowest-mass main sequence stars with a NS companion to have $\sim$$6.5\mso$. Similarly, Fig.\,\ref{fig_2} shows that the lowest mass OB+BH progenitors have initial mass ratios near $\sim$$0.35$, implying BH companion masses above $\sim$$6.2\mso$. Therefore, we do not expect NS or BH binaries in mass bins
to the left of the $6-8\mso$ bin in Fig.\,\ref{Companions}. The numbers on top of the bins in Fig.\,\ref{Companions} show that the number of systems with $M_\mathrm{ OB} > 100\mso$ which we might be missing is negligible.

As shown in Fig.\,\ref{A1k}, we expect Case\,C mass transfer (mass transfer with a core-He depleted mass donor) in a small range of initial orbital periods for initial donor masses $\simle 15.8\mso$.  Furthermore, 
about half of them may undergo stable mass transfer \citep{Ercolino2024}.
We ignore those in our statistics, because they have a negligible lifetime as post-interaction binary given that they have exhausted He in their core before the mass transfer begins. Their fates are uncertain and may be quite diverse \citep{Marchant2021,Ercolino2024}.

To interpret the numbers shown in Tab.\,\ref{tab3} and Fig.\,\ref{Companions},
it is helpful to consider some basic trends in our summary plots
(Fig.\,\ref{fig_2} and App.\,\ref{fig_g1}). For the
lowest initial donor masses in our model grid
(5$\ms$, Fig.\,\ref{A1k}), binaries that can avoid merging occupy a small corner in the Case\,A regime and
a somewhat larger triangular region in the high-mass ratio region of the
Case\,B regime,
and the models with the largest orbital periods are non-interacting systems.
This shows, in agreement with many previous detailed
binary evolution calculations
\citep{Pols1994,Wellstein2001,dM2007}, 
that only $\sim$10\% of all binary systems with donor masses of 5$\ms$
are expected to survive their first mass transfer, and appear in Tab.\,\ref{tab3} and Fig.\,\ref{Companions}. The other $\sim$90\% are expected to merge\footnote{Our employed initial orbital period and mass ratio distributions do not deviate strongly from flat distributions in $\log\,P_\mathrm{ i}$ and\,$q_\mathrm{ i}$, such that the area
in the summary plots are roughly representatives of the number of systems
born in these areas, at the considered initial donor mass. 
The low-$\qi$ binaries have higher weights in population synthesis because of their longer OB+cc lifetimes.}.
For larger initial donor masses, the surviving fraction increases and reaches about 60\% above
30$\mso$.

Our population synthesis model predicts a similar number of OB+He-star systems in the SMC formed from the initial mass ranges of NS ($M_\mathrm{ i} \simeq 10$---$17\mso$: 20.9) and of BH progenitors ($M_\mathrm{ i} \simgr 17\mso$: 23.1). This is so because the IMF predicts a similar number of systems in both mass ranges, and the shorter lifetime of the more massive binaries is compensated by a smaller merger fraction. Notably, our fiducial model also predicts 6.7 WR+O star binaries. Consistent with Fig\,\ref{SMC_Pie}, many more BH-binaries (210) than NS-binaries (24.7) are predicted to exist in the SMC, most of them with OBe type main-sequence components.



Besides the binaries emerging from stable mass transfer, 
there are 0.04 O+WR and 0.32 O+BH having close orbits but 
not undergoing any mass transfer. These binaries are formed 
from tidally-induced CHE 
\citep[][]{dM2009,Marchant2017}. The primary stars
are spun up by tides, and the enhanced rotational mixing 
prevents the establishment of the chemical gradient. 
Consequently, the stars evolve directly into helium stars
without significant expansion. 
In our model grid, this process occurs with 
initial primary mass above 70$\ms$, initial mass ratio below 0.4, and 
initial orbital period around 1.6 days. This small parameter space makes the expected number of CHE products in the SMC very small (see below).
In addition, CHE can also occur in massive near-equal-mass binaries, 
which does not
produce O+BH phases but only very brief WR+BH phases \citep{Marchant2016}.

Figure~\ref{Companions} shows the predicted relative fractions of compact 
and He-star companions to OB stars as the function of mass.
The total BH fraction is nearly constant at about 70\%.
The OBe fraction is a decreasing function of the OB stars mass.
Here, most MS star companions below $\sim$$15\mso$ are Be stars. 
For higher masses, the 
OBe+BH fraction drops, which reflects the growing fraction of Case A systems,
where stars are braked by tides. Above $40\ms$, stars are spun down through wind braking.
The NS fraction reaches a maximum near $\mob=8\ms$ with
a value about 20\% and then decreases to zero near 30$\ms$.
The most massive O+NS binaries form in Case A systems, 
which feature a strong tidal interaction and a high
accretion efficiency. The He-star fraction is nearly constant 
with $\sim$20\%,
which is in the range inferred by \citet{El-Badry2022}.
A distribution of OB star masses in NS and BH binaries is provided
in Fig.\,\ref{App_Mass}.

\subsection{Properties of OB+He/WR binary systems\label{SMC_He}}

Here, we first discuss the observable properties of the WR+O star binaries in our fiducial synthetic population. We compare those with the observed SMC WR stars in Sect.\,\ref{wrobs}. The total predicted number of WR stars in the SMC is 6.7. All of them are formed through binary interaction, including tide-induced CHE, as the adopted mass loss rate is not high enough to form WR stars through wind-stripping (see Fig.\,\ref{smc_single} for our non-rotating single star models).

\begin{figure}[!htbp]
    \centering
    \includegraphics[trim=1.3cm 0 0.1cm 0, clip,width=\linewidth]{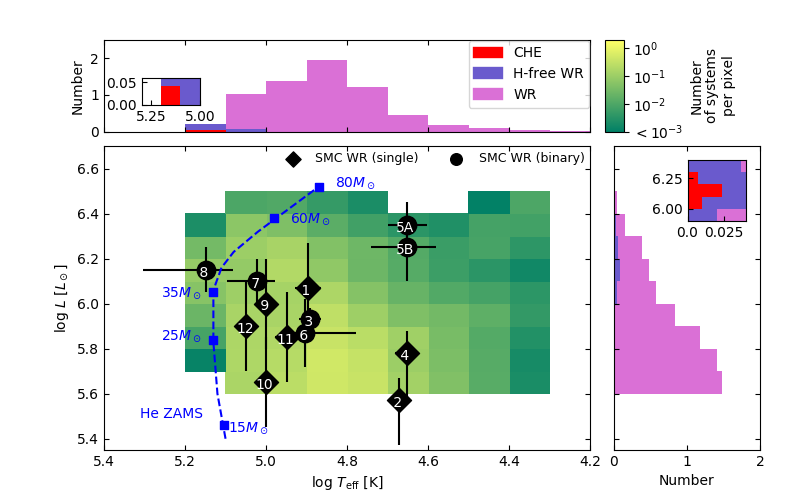}
    \includegraphics[trim=1.3cm 0 0.1cm 0, clip,width=\linewidth]{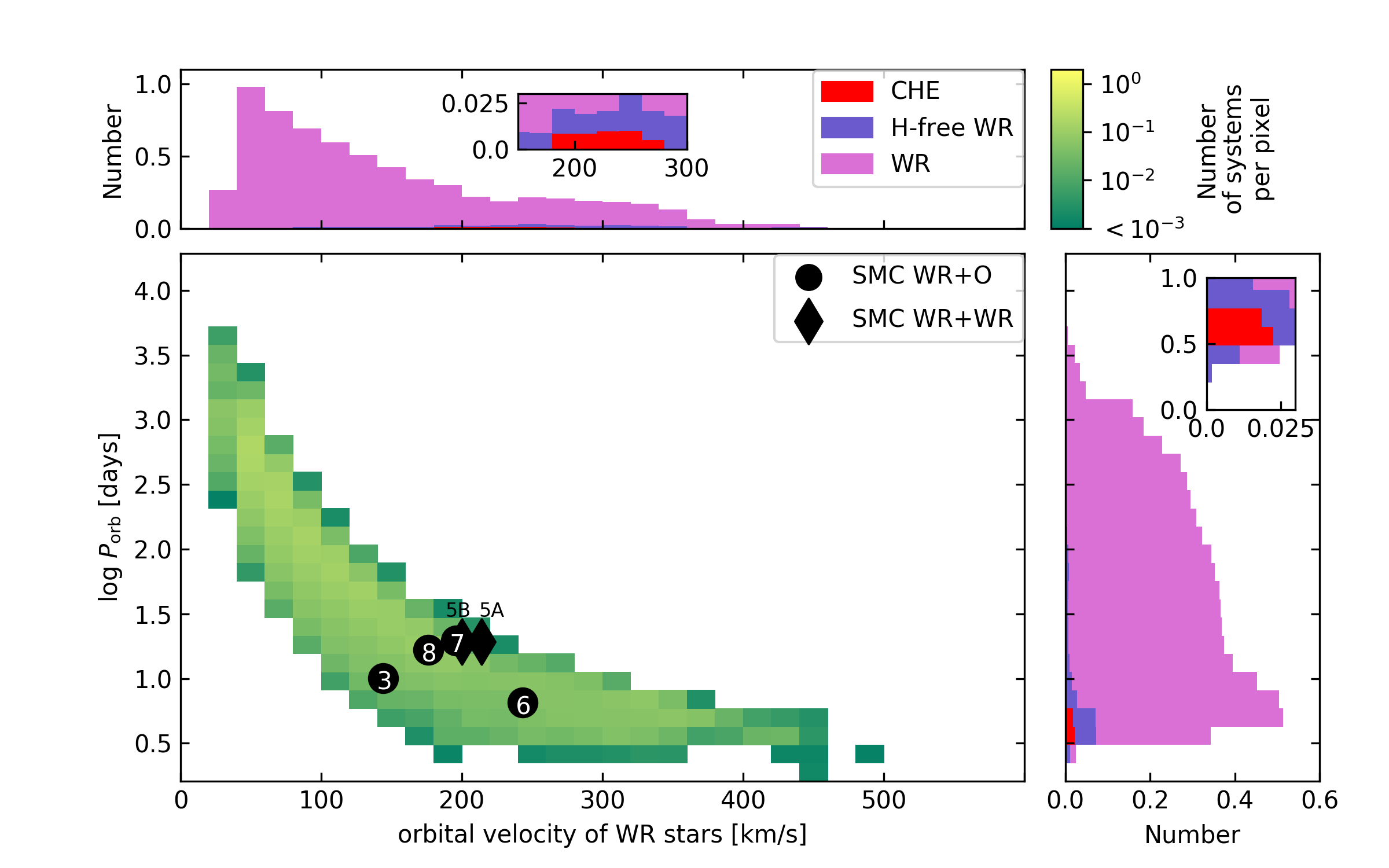}  
    \caption{Distributions of properties of the predicted SMC OB+WR binary population. The top panel shows the distribution of the predicted position of the WR components in the Hertzsprung Russell diagram (background colours), where the colour indicates the expected number per pixel (see colour bar on the top right). The total number of observed WR star components, or WR binaries, is four. The plots to the right and on top give the 1D-projections of the distribution, with the inset magnifying the WR binaries produced via chemical homogeneous evolution (CHE). The black circles and  diamonds represent the WR components of the observed WR binaries, and single WR stars, in the SMC 
    \citep{Foellmi2003,Foellmi2004,Koenigsberger2014,Hainich2015,Shenar2016,Shenar2018}, 
    where the numbers are related to the identifier, e.g., "1" for "SMC AB1". 
    The blue dashed line is the zero-age main sequence of helium stars models (He-ZAMS) with SMC metallicity  \citep{Kohler2015}, with the indicated helium star 
    masses.
    The bottom panel shows the distribution of the predicted WR binaries in the plane of orbital period versus the orbital velocity of the WR component. Here, the black symbols represent the projected orbital velocity semi-amplitudes of the observed WR+O star binaries (circles) and of the WR+WR binary SMC\,AB5 (diamond).}
    \label{SMC_WR}
\end{figure}

The top panel of Fig.\,\ref{SMC_WR} shows the predicted distribution of the WR components of OB+WR binaries in the HR-diagram, together with its 1D-projections in the top and right plots. The predicted effective temperatures of most of WR models are slightly cooler than those of pure He star models (blue line in Fig.\,\ref{SMC_WR}).  While the H-rich envelope of these donor stars gets nearly completely stripped by mass transfer, the remaining hydrogen leads to larger radii and smaller surface temperatures of stripped stars, compared to hydrogen-free models \citep{Schootemeijer2018,Gilkis2019,Laplace2020}. In the majority of our models, the hydrogen layer is not removed by the relatively weak winds in the WR phase at this metallicity. In our synthetic population, the WR models with log$\,T_\mathrm{eff}/{\rm K}=4.9$ have a surface hydrogen mass fraction of about 0.3. Models with a surface temperatures above log$\,T_\mathrm{eff}/{\rm K}=5.1$ are hydrogen-free WR. The predicted population is sharply cut off at log$\,L/\ls = 5.6$ 
due to the threshold luminosity for defining WR stars (see Sect. \ref{He_WR}). 
Towards the higher luminosities, the predicted number drops as a consequence of the adopted IMF.

The bottom panel of Fig.\,\ref{SMC_WR} shows the distribution of the orbital periods of our OB+WR binary models, and the orbital velocities of the WR components. The orbital periods span a wide range, from $\sim$$3\,$days to $\sim$$3\,$years, with about half of the population below 30\,days. We do not find shorter-period WR binaries because our ZAMS binaries with the shortest initial orbital periods evolve into contact and eventually merge due to L2-overflow (orange colour in Fig.\,\ref{A2k}). The relation between orbital velocity $\upsilon_{\rm orb,WR}$ and orbital period $P_{\rm orb}$ is determined by
\begin{equation}
    \upsilon_{\rm orb,WR} = \frac{M_{\rm OB}}{M_{\rm OB}+M_{\rm WR}}\left(2\pi G\frac{M_{\rm OB}+M_{\rm WR}}{P_{\rm orb}}\right)^{1/3},
    \label{vwr}
\end{equation}
where $M_{\rm WR}$ is the mass of WR star, and  $G$ is the gravitational constant.
The corresponding orbital velocities go beyond 400$\kms$ for the short period models, and go down to $\sim$$20\kms$ for the longest period binaries. The orbital velocity distribution between 200$\kms$ and 350$\kms$ is rather flat because the accretion efficiency in our Case\,A binaries is increasing towards shorter periods, increasing the scatter in stellar masses in Eq.\,\eqref{vwr}. In about 2/3rd of our WR binary models, the O\,star is more massive than the WR\,star (see Fig.\,\ref{WR_Pq}).

The insets in Fig.\,\ref{SMC_WR} magnify the small contributions from binary models which evolve through mass transfer and produce hydrogen-free WR\, stars,
and those which form WR stars through CHE, which avoid mass transfer (cf., Tab.\,\ref{tab3}). While CHE WR stars are expected to have a hydrogen-rich phase during late hydrogen burning and early helium burning \citep{Koenigsberger2014,Schootemeijer2018}, we do not include this short phase in our statistics (cf., Fig.\,\ref{CHE_example}).

Our fiducial SMC population also contains a large number of OB+He-star binaries,
in which the wind of the He-star is expected to be transparent such that WR-type emission lines would not be produced. These so called ``stripped star binaries'' largely lack observed counterparts \citep{Gotberg2023,Drout2023}. It has been suggested that, at least the more massive of such systems, may appear as hard X-ray sources, with X-rays being produced by the collision of the He-star wind with the OBe disc and the OBe star wind \citep{Langer2020a}. While their correspondence to the sub-class of Be/X-ray binaries called $\gamma\,Cas$-stars is debated \citep{Naze2022,Gunderson2025}, the nature of the companion to OBe stars in observed Be/X-ray binaries is often unclear. Therefore, the predicted OB+He-star binaries could contribute to the large Be/X-ray binary population of the SMC. This would reduce the discrepancy between our predicted number of OBe-NS systems and the observed number of Be/X-ray binaries (cf., Sects.\,\ref{OBNSfid},\,\ref{BeXB}).

\subsection{Properties of OB+NS binary systems} \label{OBNSfid}

\begin{figure*}[t] 
	\centering
 	\includegraphics[width=0.8\linewidth]{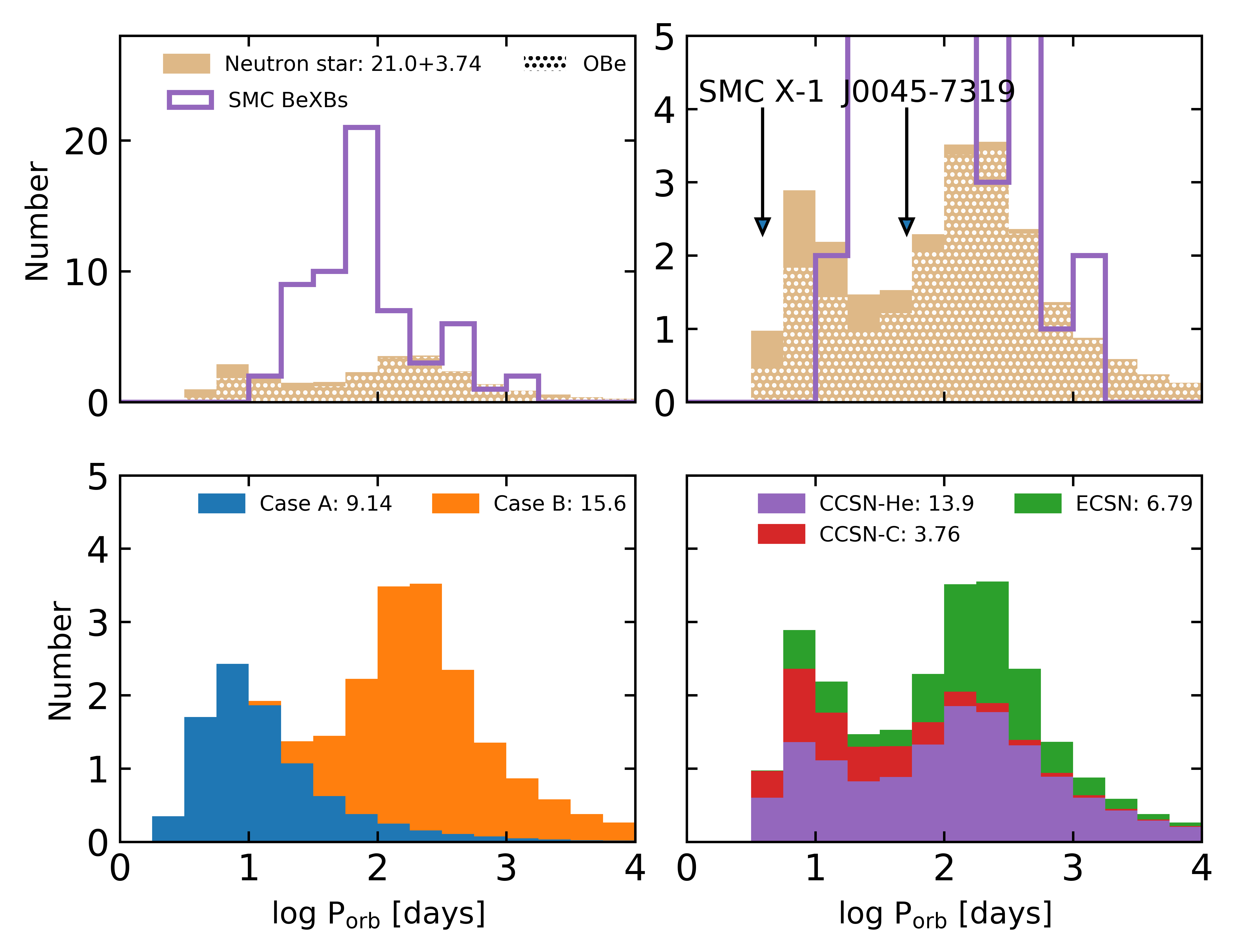}
	\caption{\label{porbNS} Distributions of the orbital periods of the $\sim$$25$ OB+NS binaries in our fiducial synthetic SMC population.
	Upper left: Predicted period distribution (yellow histogram), where the $\sim$$21$ OB+NS binaries in which the OB stars rotates faster than 95\% of their critical rotational velocity are indicated by shading (labelled `OBe'). 
	The period distribution of observed Be/-X-ray binaries in the SMC \citep{Haberl2016} is plotted as purple line. 
	Upper right: A zoom-in of the upper left panel. The B star + radio pulsar binary J0045-7319 \citep{Bell1995} and the supergiant X-ray binary SMC X-1 \citep{Rawls2011,Falanga2015} are indicated by arrows.
	Lower left: The same predicted distribution as in the top panels, with the colours indicating which of the systems formed through Case\,A and Case\,B mass transfer (blue and orange, respectively), with the corresponding total numbers given in the legend.
	Lower right: The same predicted distribution as in the top panels, with the colours indicating which type of SN, and therefore which NS kick distribution, was assumed for the different systems [green for electron-capture supernova; red for helium-envelope-stripped SN (CCSN-C); purple for hydrogen-envelope-stripped SN (CCSN-He); cf., Sect.\,\ref{natal_kick}]. The predicted total numbers related to different SN types are given in the legend.
    }
\end{figure*}

\begin{figure}[t]
    \includegraphics[width=\linewidth]{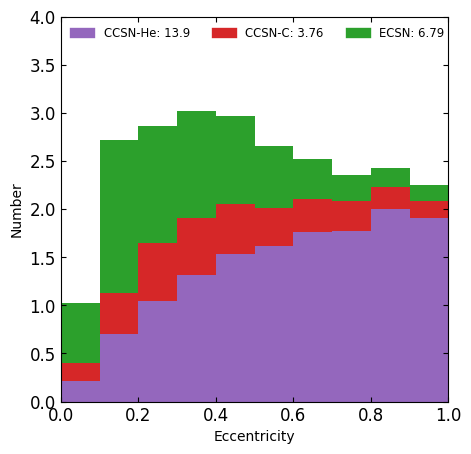}
    \caption{Distribution of the eccentricities of our OB+NS model binaries, with the colours indicating which type of SN was assumed for the different systems, as in the bottom right panel of Fig.\,\ref{porbNS}.  
    \label{ECC}}
\end{figure}    

\begin{figure}[t]
    \includegraphics[width=\linewidth]{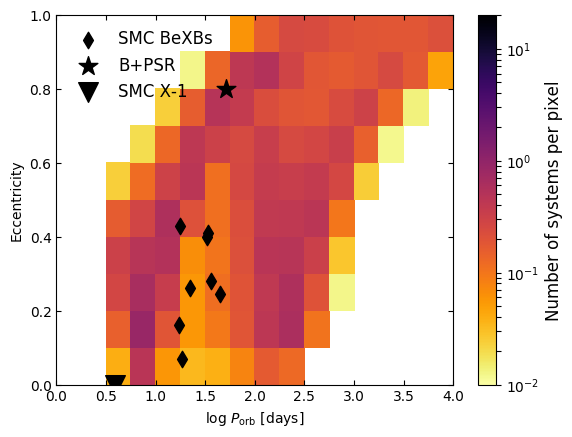} 
    \caption{2D-distribution of the $\sim$$24$ OB+NS binaries of our fiducial synthetic SMC population in the orbital period - eccentricity plane. The number in each pixel is colour-coded. The 7 SMC Be X-ray binaries with eccentricity measurements \citep{Townsend2011,Coe2015} are indicated by black diamonds. The B star + radio pulsar binary \citep{Kaspi1994,Bell1995} is marked by a black star, the supergiant X-ray binary SMC\,X-1  by a black triangle \citep{Falanga2015}. 
    \label{ECC2}
    }
\end{figure}

We find $\sim$25 OB+NS binaries in our fiducial synthetic population (cf., Tab.\,\ref{tab3}). We compare the synthetic with the observed population in Sect,\,\ref{BeXB} and discuss implications in Sect\,\ref{discuss}. Figure \ref{porbNS} presents their orbital period distribution. We find two distinct sub-populations with orbital period peaks near 10 and 150 days, which are associated with  the different modes of mass transfer. As seen in Figs.\,\ref{fig_2} and \ref{A1k}, OB+NS binaries are formed from two clearly-separated triangular  regions in the Case A and Case B regimes due to the upper limit on mass transfer rate set by Eq.\,\eqref{qmin}. In addition, some OB+NS binaries have orbital periods exceeding the upper initial orbital period bound ($\simeq 3000\,$days), which are widened by the SN kicks. 

Depending on the final structure of our donor stars, we assume that neutron stars are formed through different types of SNe, with correspondingly different NS birth kick distributions (cf., Sect.\,\ref{natal_kick}) are distinguished. We find  $\sim$6.8 OB+NS binaries in which the NS was assumed to form via an electron capture SN (ECSN), 13.9 via hydrogen-stripped  (CCSN-He), and 3.8 vie helium-stripped SNe (CCSN-C). ECSNe contribute a considerable fraction of the population at orbital period around 150 days, where the second dredge up of the primary star, which would reduce the helium core mass in single stars, is avoided due to the mass transfer, making ECSNe to occur much more frequently than in single stars \citep{Podsiadlowski2004}. We see that systems with a He-envelope-stripped SN prefer narrow orbits since the binary has to be close enough to get the donor star deeply stripped \citep[also see][]{Ercolino2024b}.
Systems with H-envelope-stripped SNe do not show a strong orbital period preference.

Figure\,\ref{ECC} presents the predicted distribution of eccentricities of our OB+NS 
binary population. It peaks at around 0.3. The OB stars of highly 
eccentric binaries have the chance to fill their Roche lobe during 
periastron passage, which causes the number of systems to drop in the high-eccentricity region. Due to the difference in the magnitude of the adopted kicks, 
ECSN mainly contributes binaries with $e\sim0.3$, 
while stripped SN contributes most of high-$e$ binaries. 

We further show the distribution of OB+NS binaries in the 
orbital period-eccentricity plane in Fig.\,\ref{ECC2}. We see that
wide binaries tend to have higher eccentricities since they are more easily 
disrupted than close binaries. Further, this plot confirms the prediction of two distinct populations with orbital periods of $\sim$10\,d and $\sim$150\,d, which have similar eccentricity distributions.
The upper left edge is populated by systems which undergo 
Roche Lobe overflow during periastron passage.

\subsection{Properties of OB+BH binary systems\label{SMC_OBBH}}

\begin{figure*}[!htbp]
\vskip 1.0cm
\includegraphics[width=0.9\linewidth]{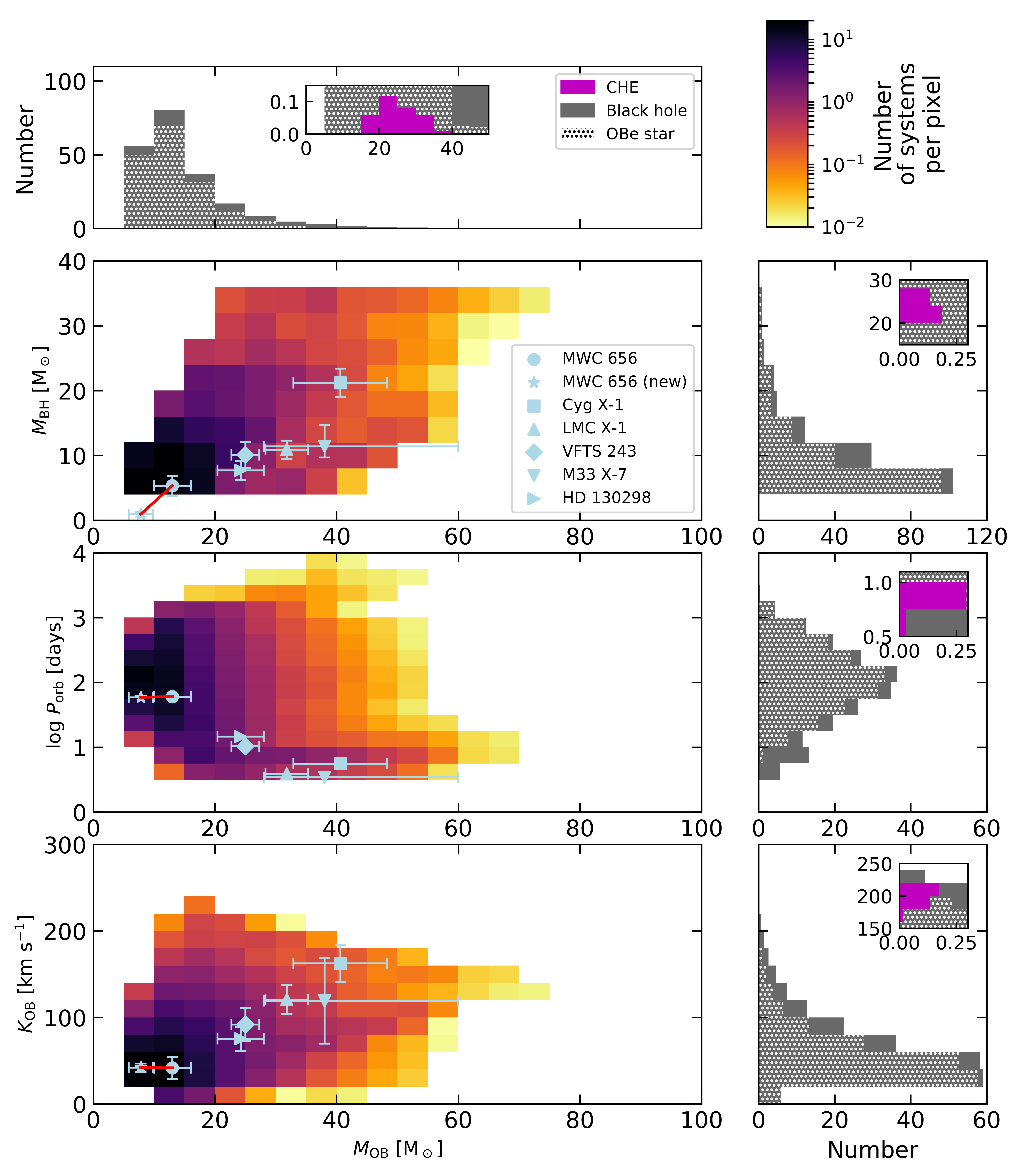}
\caption{Properties of the $\sim$211 OB+BH binaries in our fiducial synthetic SMC population (cf., Tab.\,\ref{tab3}), as function of the mass of the OB star. The three main plots show the 2D number density distributions of the systems, with the OB star mass on the X-axis, and black hole mass, orbital period, and OB\,star orbital velocity semi-amplitude on the Y-axis, respectively. The number density is colour coded (see colour bar on the top right). The known OB+BH binaries, irrespective of their host galaxy metallicity, are marked by blue symbols, and correspond to MWC 656 \citep{Casares2014}, Cyg X-1 \citep{Miller-Jones2021}, LMC X-1 \citep{Orosz2009}, M33 X-7 \citep{Ramachandran2022}, VFTS 243 \citep{Shenar2022}, and HD 130298 \citep{Mahy2022}. The BH nature of the secondary star in MWC 656 debated in the literature. The solution by \citet{Janssens2023} is labelled by "MWC 656 (new)" and connected to the solution by \citet{Casares2014} through a red line. The four histograms give the corresponding 1D projections, with shading identifying rapidly rotating OB stars (labelled `OBe'), and insets magnifying the small contributions from chemically homogeneous evolution (CHE). The orbital velocity semi-amplitudes for observed systems are calculated using Eq.\,\eqref{KOB} with the observed masses and orbital periods. The errors are calculated by error propagation.
}
 \label{f_bh}
\end{figure*}	

Figure\,\ref{f_bh} gives an overview of selected properties of the OB+BH binaries in our fiducial synthetic SMC population. The total number of predicted systems is $\sim$210, which means that $\sim$5\% of the O and early B stars are expected to have a BH companion (Sect.\,\ref{s_pie}). The top histogram of Fig.\,\ref{f_bh} implies that the majority of BHs are expected to have B\, star companions. We find $\sim$135 systems (64\%) with a main sequence star mass below 15$\mso$, Those systems are predicted to contain rather light BHs (5---10$\mso$ for most, up to $20\mso$ for $\sim$10\%) in orbits with typically 100\,d periods, in which the B\,star moves with $20$---$60\kms$. The top histogram shows that most of these B stars are expected to be Be stars, since the corresponding stellar model rotates faster than with 95\% of its critical rotation velocity. 

The O\,star systems are expected to contain a wider mass range of BHs, with 95\% of them in the range 
$5$---$25\mso$, and few with BHs up to 35$\mso$. There orbital periods and orbital velocities spread a bit wider than those of the B star systems, and in particular we expect O\,star systems with orbital periods below 10\,d, with O\,star orbital velocities of up to 200$\kms$. However, also many systems with wider orbits and orbital velocities well below 100$\kms$ are predicted. As in the B star regime, most O\,type companions are found to rotate very rapidly. 

The expected number of systems drops for higher main sequence star masses, mostly because of the adopted IMF. Below 10$\ms$, the chance that the companion star is massive enough to form a BH decreases. Our lowest-mass main sequence star with a BH companion has about 6$\mso$ (cf., Fig.\,\ref{Companions}). The BH mass distribution is also affected by the IMF. The lightest BH we predict has $\sim$4.9$\ms$, for which the $6.6\mso$ pre-collapse star ejected about 0.5$\ms$ of helium-rich envelope, and 1.2$\ms$ is lost due to the release of gravitational binding energy. The most massive BH in our model sample weighs about 35$\ms$, and the mass ejection from its progenitor is set by the pulsational-pair instability (PPI) mechanism (cf., Sect.\,\ref{CC_formation}). 

The majority of our OB+BH binaries have orbital periods around 100 days. Binaries that are widened by the PPI-induced mass loss occupy the period regime above 3000 days. The effect of tides, which can limit the spin-up of the accreting MS star, is relevant only in the closest binaries. Therefore we expect OBe companions to BHs in most of the wide binaries ($\porb>10\,$days), while slower rotators dominate in closer binaries. The most massive MS stars may rotate relatively slowly even in wide binaries because of wind braking \citep{Langer1998}. 

The models in our grid which in principle could produce the OB+BH systems with the most massive main sequence stars are those very massive systems which start with a mass ratio close to one. However, at higher initial primary masses, the minimum initial mass ratio above which the secondary star ends core hydrogen burning before the primary ends its life is decreasing (Fig.\,\ref{A2k})\footnote{This is expected, since the exponent $\alpha$ in the mass-luminosity relation for main sequence stars ($L\propto M^{\alpha}$) tends towards $\alpha=1$ for high mass \citep[cf., fig.\,17 in][]{Kohler2015}, at which point the timescale of core hydrogen burning becomes independent of mass.}. Therefore, those systems are assumed to merge when reverse mass transfer starts, such that they do not contribute to our synthetic SMC compact object binary population. Very massive OB stars are also not produced from the most massive systems with lower mass ratios because the accretion efficiency is quite limited, and stellar wind mass loss is also considerable at the highest masses. Therefore,
our population has no statistically significant contribution to the BH binary population at OB star masses above about 70$\ms$. 

Comparing with the orbital period distribution of our NS binaries (Fig.\,\ref{porbNS} and \ref{App_porb}), we see that the clear signature of the two different modes of mass transfer is absent in the BH binary period distribution (middle panel in Fig.\,\ref{f_bh}). The reason is that more binaries near the Case A/B boundary avoid merging in the BH-forming regime compared to the NS-forming regime (see Fig.\,\ref{A2k}), which fills the orbital period gap between the two modes.

The lowest panel of Fig.\,\ref{f_bh} presents the distribution in the 
OB star mass - orbital velocity semi-amplitude of OB star $\kob$ plane.
In contrast to the WR binaries, which are assumed to be circular as consequence of the RLOF, the assumed mass loss at BH formation does produce a kick, which makes the orbits mildly eccentric. Therefore, 
$\kob$ is defined here by
\begin{equation}
    K_\mathrm{ OB}=\frac{M_\mathrm{ cc}}{(M_\mathrm{ cc}+M_\mathrm{ OB})}\sqrt{\frac{G(M_\mathrm{ cc}+M_\mathrm{ OB})}{a(1-e^2)}},
    \label{KOB}
\end{equation}
where $M_\mathrm{cc}$ is the mass of compact object. Inclination effects are not included here.
The mass loss during the BH formation produces eccentricities of less than 0.1, which makes the deviation of the orbital velocities of OB stars from their semi-amplitudes within 10\%.
Corresponding to the peak of the orbital period distribution of about 100\,days seen in the middle panel, our OB+BH binaries have velocity semi-amplitudes
peaking near 40$\kms$. The highest velocity semi-amplitudes we find are about 250$\kms$ (Fig.\,\ref{App_porb}), which is near the initial orbital period boundary between stable mass transfer and L2 overflow (see Figs.\,\ref{fig_2}, \ref{A1k}, and \ref{A2k}). 

\section{Parameter variations\label{initial_parameters_var}}
\label{sect4}

\begin{table*}
\caption{Predicted populations derived with various different assumptions. The  predictions of fiducial model are listed for comparison purpose, which are computed by the Kroupa IMF (Eq. \ref{IMF}), the Sana distributions for initial mass ratios and orbital periods (Eq. \ref{f_qi} and Eq. \ref{f_logpi}) with our fiducial kick velocity distributions (Tab.\,\ref{tab2}). Be feature is assumed with $\vrot/\vc > 0.95$ and BH forms with He core mass above 6.6$\ms$ at the core He depletion. We vary these assumptions in different models (see Sect.\,\ref{initial_parameters_var} for the definitions). O stars are defined by effective temperatures above 31.6 kK, which are further divided into pre-interaction, merger products, and the main-sequence companions in post-interaction binaries. In the table, "$=$" means the same value as the fiducial model, and "+" indicates the number change compared with the fiducial model, for example, in the NS-limit model, the number of NS+OB binaries is "3.74+14.2", which means that the NS-limit predicts 14.2 more NS+OB binaries than the fiducial model.
}
\begin{center}
\begin{tabular}{l|c|c|c|c|c|c|c|c|c|c}\hline\hline 
\rule{0pt}{\dimexpr.7\normalbaselineskip+1mm}
         & Fiducial\tablefootmark{a} & Kick-265 & Kick-0 & Kick-BH & logPq-flat  & SFH-S & SFH-R & $\vc$-0.98 & $\vc$-0.8 &
         NS-limit\\[1pt]
         \hline
\rule{0pt}{\dimexpr.7\normalbaselineskip+1mm}
WR+O     & 2.75    & $=$  & $=$  & $=$   &  2.62   &  1.08    &  2.78 & 4.91 & 1.68 & $=$ \\[1pt]
WR+Oe    & 3.97    & $=$  & $=$  & $=$   &  5.19   &  1.82    &  4.01 & 1.80 & 5.03 & $=$\\[1pt]
He+OB   & 12.1     & $=$  & $=$  & $=$   & 8.63   &   10.2   &   15.1  &   15.6  &   8.71 & $=$\\[1pt]
He+OBe  & 223   & $=$  & $=$  & $=$   & 200 &   221 &   347 &   220 &   226 & $=$\\[6pt]
BH+OB    & 40.3   & $=$  & $=$  & 29.5    &  42.1  &  30.5   & 52.3  & 64.0 &17.1 & 0 \\[1pt]
BH+OBe   & 170  & $=$  & $=$  & 96.4    &  226 &  165  & 261 & 147 &194 & 0\\[1pt]
NS+OB    & 3.74    &  1.41   & 8.02    & $=$   &  2.79 & $=$   &  6.75  & 4.39  & 1.84 & 3.74+14.2\\[1pt]
NS+OBe   & 21.0   &  3.24   & 95.1   & $=$   &  21.8 &  $=$   & 46.3  & 20.4  &23.0  & 21.0+28.1\\[6pt]
OB disr. & 2.83    & 5.38   & 0    & 2.83+7.45  &  1.73   &  $=$  & 4.74   & 3.47  & 1.36 & 2.83+24.7\\[1pt]
OBe disr. & 71.5   &  91.0  & 0    & 71.5+74.9 &  93.5  &  $=$  &  153 & 70.8  & 73.0 & 71.5+138 \\[6pt]
OBe   & 489   &  $=$   & $=$  & $=$   &  546 &  502  & 811  & 460 & 521 & $=$ \\[1pt]
$-$ Oe   & 22.6   &  $=$   & $=$  & $=$   & 29.5 & 17.3  & 23.0  & 16.1 & 27.4 & $=$ \\[1pt]

\hline 
\rule{0pt}{\dimexpr.7\normalbaselineskip+1mm}
O stars & 1072 & $=$  & $=$  & $=$ & 1080 & 444 & 1083 & $=$ & $=$  & $=$ \\[1pt]
$-$ pre-int.  & 979 & $=$  & $=$  & $=$   &  1010  & 367  & $=$ & $=$ & $=$  & $=$ \\[1pt]   
$-$ merger  &  43.1 & $=$  & $=$  & $=$   &  20.0  & 37.4  & 52.5 & $=$ & $=$  & $=$ \\[1pt]   
$-$ post-int.  & 49.3 & $=$  & $=$  & $=$   &  49.2  & 38.6  & 51.0 & $=$ & $=$  & $=$ \\[1pt]   
\hline\hline
\end{tabular}\label{DIC}
\end{center}
\end{table*}

\begin{figure*}
    \centering
    \begin{tabular}{cc}
    \includegraphics[width=0.46\linewidth]{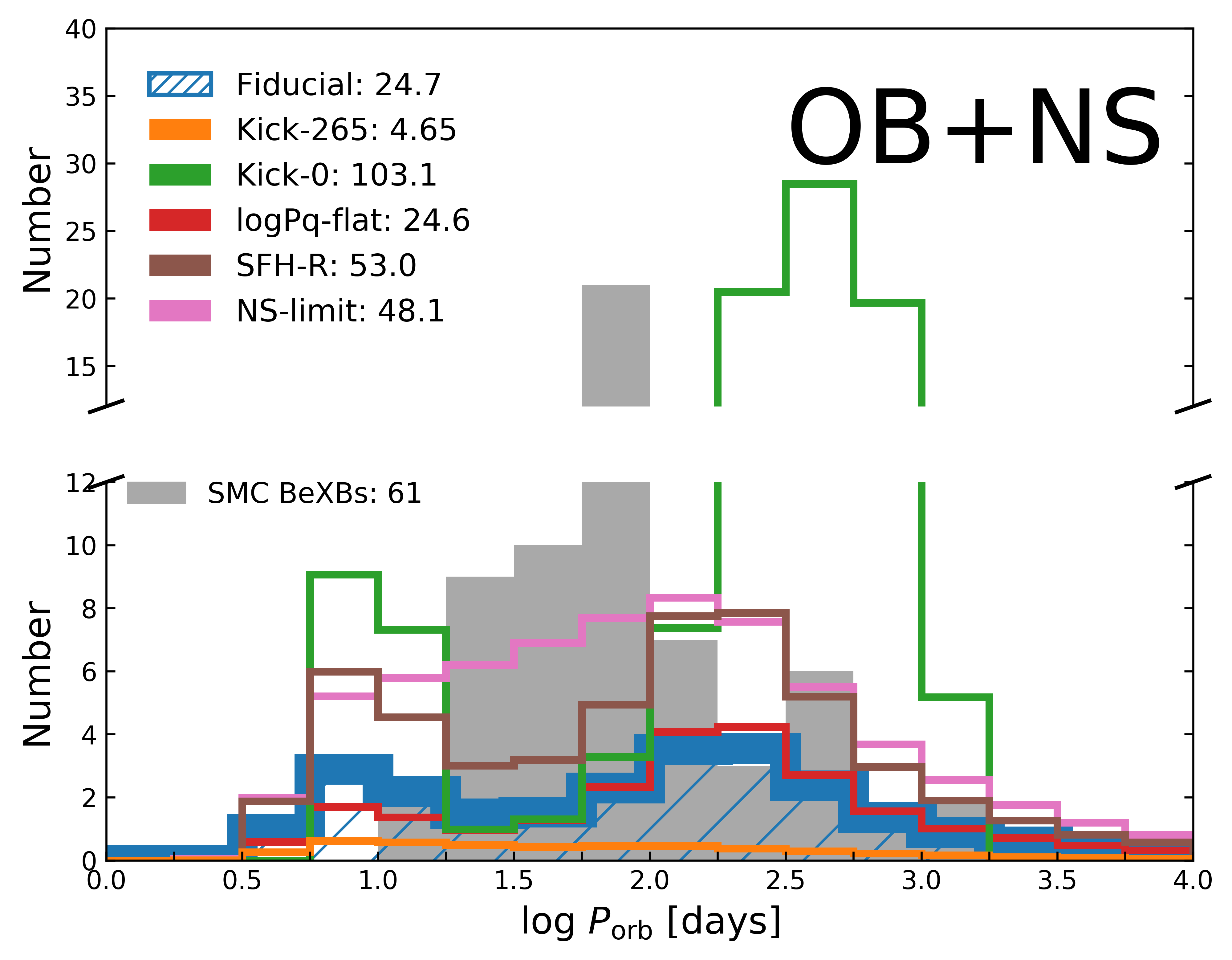} &
    \includegraphics[width=0.46\linewidth]{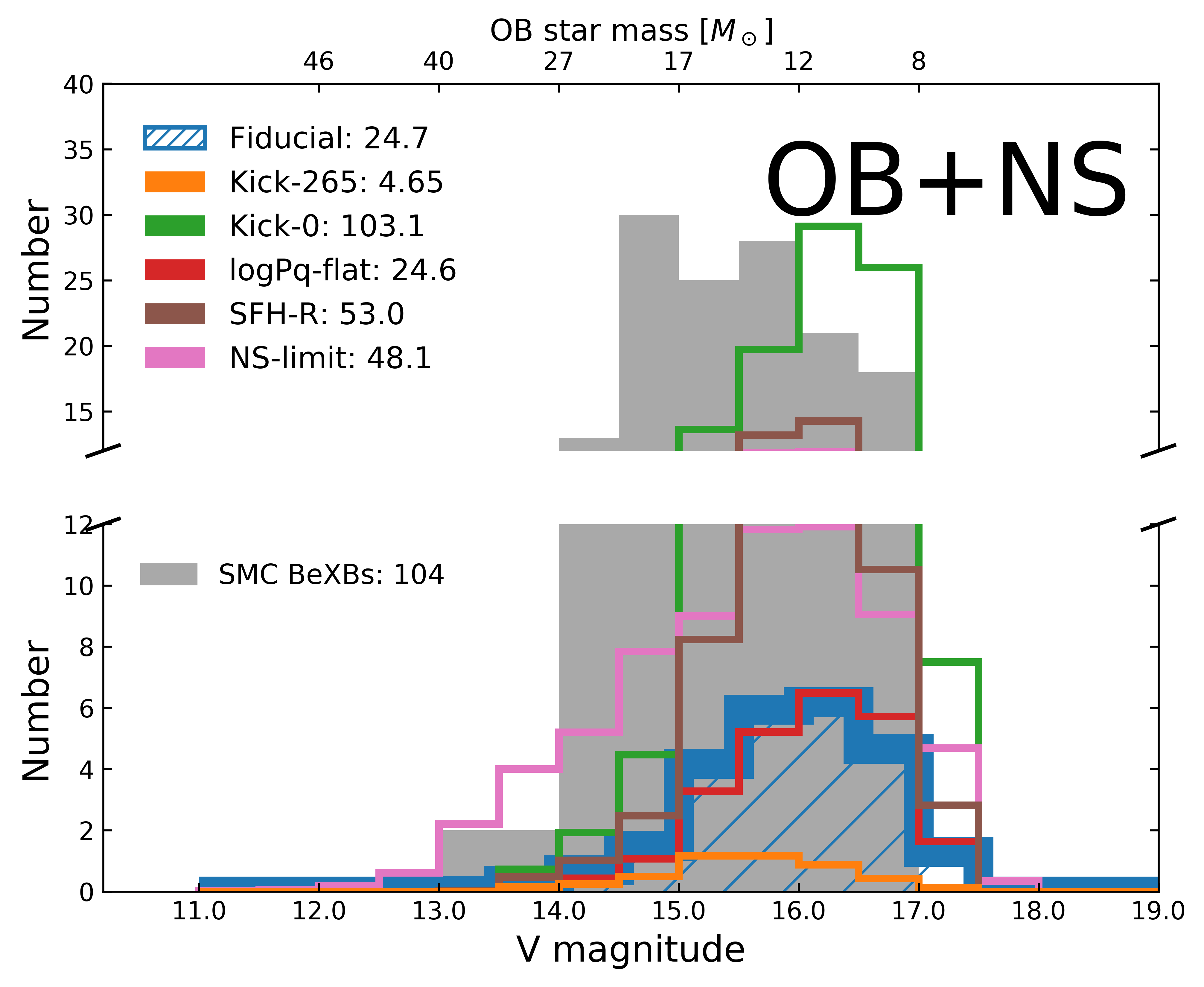} \\
    \includegraphics[width=0.46\linewidth]{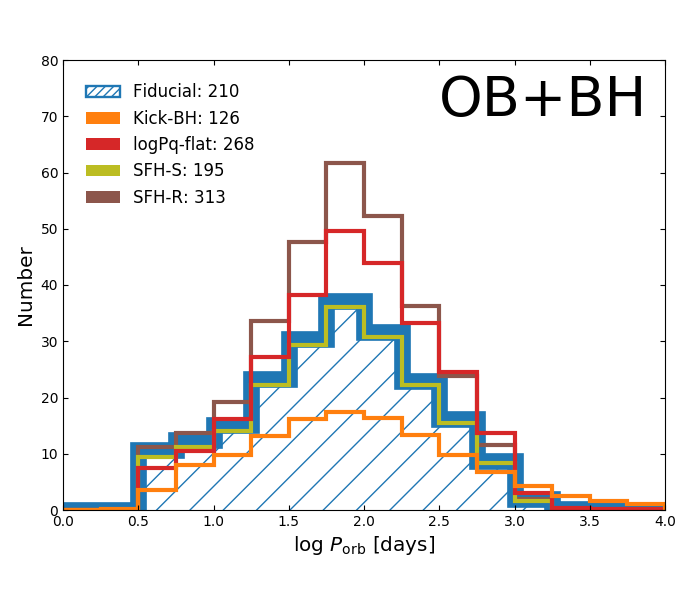} &
    \includegraphics[width=0.46\linewidth]{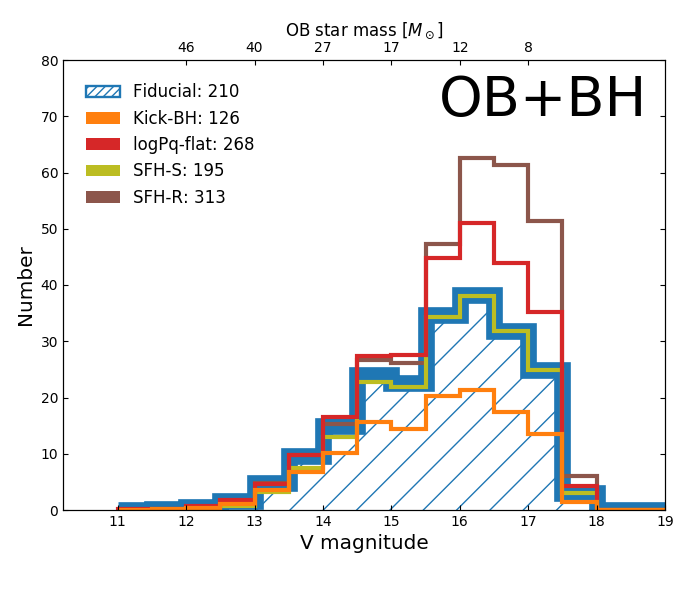}
    \end{tabular}
    \caption{Orbital period (left plots) and V\,band magnitude or mass (right) distributions of the OB+NS (top) and OB+BH (bottom) binaries resulting from our parameter variations. The predictions from the different population models are shown with differently coloured lines (see legend), where the predictions from the fiducial model are indicated by the hatched histograms with blue thick lines. 
    The population models which produce the same distributions as the fiducial model are not shown in the corresponding plots. In the bottom plots, the SFH-R distributions are very close to the fiducial ones. 
    We show the distributions of the observed SMC BeXBs in the upper panels with gray histograms. The observed orbital periods and V-band magnitudes of the SMC BeXBs are from \citet{Haberl2016}. 
    }
    \label{DIC_BH_NS}
\end{figure*}

\begin{figure}[!htbp]
    \centering
    \includegraphics[width=\linewidth]{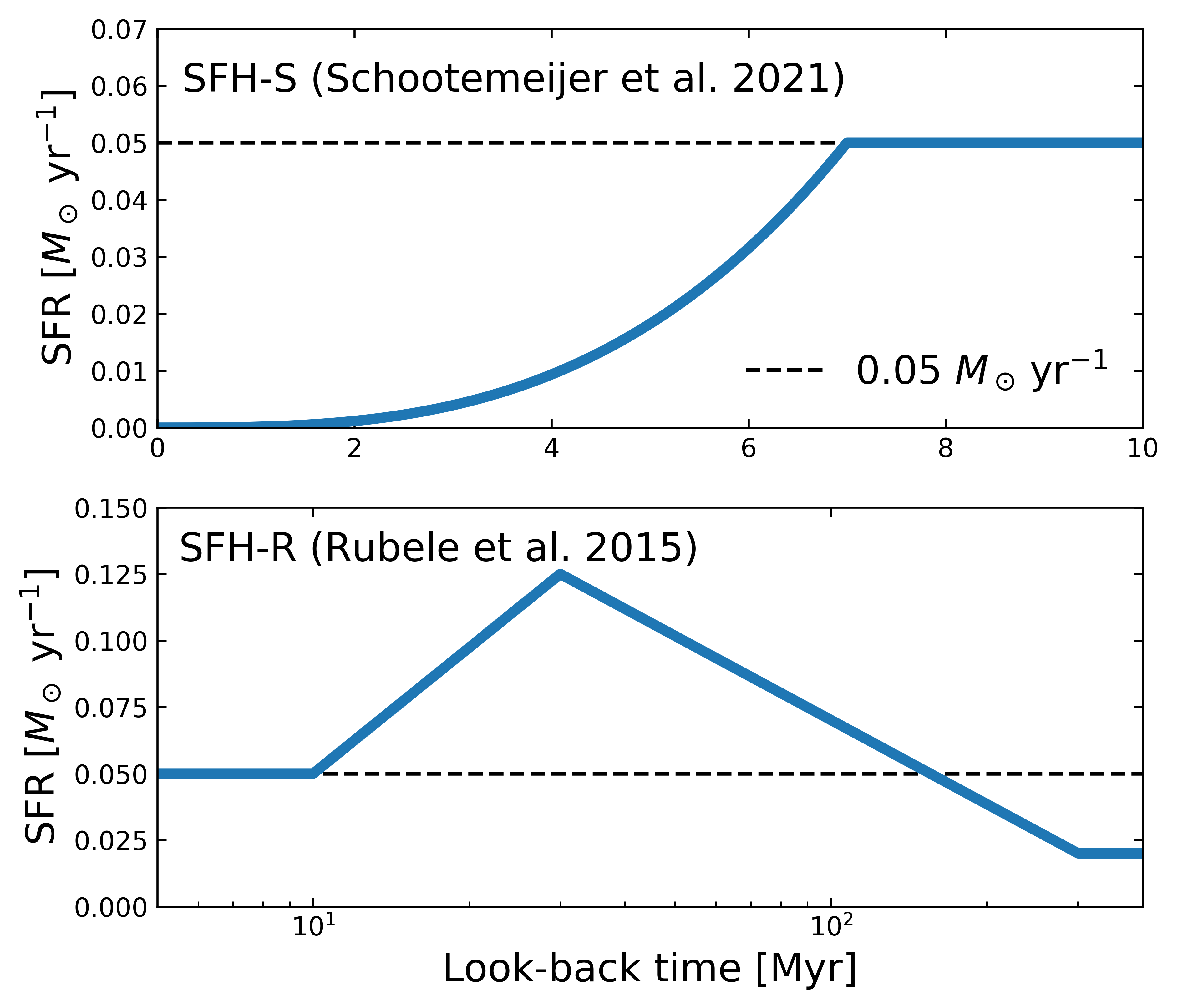}
    \caption{The star formation history adopted to compute the population models SFH-S
    \citep[upper panel,][]{Schootemeijer2021} and SFH-R \citep[lower panel,][]{Rubele2015}, where the dashed line indicates the constant star formation rate of 0.05$\msoy$ used in the fiducial model.}
    \label{SFH}
\end{figure}

For our SMC population synthesis, we cannot change the underlying stellar evolution models. Therefore, uncertain assumptions in those, in particular concerning the efficiency of mass transfer or the criterion for unstable mass transfer and merger, cannot be varied here. However, we can vary the parameters that enter directly into the population synthesis calculations. Those relate to assumptions on compact object birth kicks, on the time dependence of the star formation rate, on the initial binary parameter distributions, on the lowest rotation velocity of OBe star, and on the upper core mass limit for NS formation.  With this idea, we introduce the following population synthesis models, in which only the mentioned parameter differs from the parameters adopted in our fiducial population synthesis calculation.
\begin{itemize}
    \item Kick-265: the distribution of NS kick velocities is taken to be the Maxwellian distribution with $\sigma=265\kms$ for all types of SNe.
    \item Kick-0: All kick velocities are fixed to zero.
    \item Kick-BH:  BHs produced by fallback of previously ejected material may produce a supernova and receive a momentum kick at birth \citep{Janka2012}. We adopt a flat distribution in the range [0, 200]$\kms$ for natal kicks imparted to newborn BHs \citep[][and Paper II]{Kruckow2018};
    \item logPq-flat: we use flat distributions for the initial mass ratios and logarithms of the orbital periods of our binaries.
    \item SFH-S: we assume a star formation history with no star-formation at the present time that
increases up to 7 Myrs ago, before which it stays constant (see the upper panel in Fig. \ref{SFH}). 
    \item SFH-R: we assume a star formation rate with a recent peak $\sim$$20 - 40$ Myrs ago
    \citep[][; see the lower panel in Fig.\,\ref{SFH}]{Rubele2015}.    
    \item $\vc$-0.98: the threshold value of $\upsilon_\mathrm{ rot}/\upsilon_\mathrm{ crit}$ for defining OBe stars is taken to be  0.98.
    \item $\vc$-0.8: the threshold value of $\upsilon_\mathrm{ rot}/\upsilon_\mathrm{ crit}$ for defining OBe stars is taken to be  0.8.
    \item NS-limit: we assume all collapsing stellar cores form NSs regardless of their helium core mass.
\end{itemize}
Below, we discuss the impact of the parameter changes on the different types of stars in the 10 simulated populations. Table\,\ref{DIC} lists the numbers for the various types of post-interaction binaries appearing in the different model populations, as well as the number of O\,type stars and O\,type stars in pre-interaction binaries. Figure\,\ref{DIC_BH_NS} compares the distributions of orbital periods and the intrinsic V-band magnitudes\footnote{To estimate the intrinsic V-band magnitudes, We adopt the same method as in \citet{Schootemeijer2021}. The distance modulus of the SMC is taken to be 18.91 \citep{Hilditch2005}, and we calculate the bolometric correction by using a polynomial fit to MIST values \citep{Dotter2016,Choi2016}. The contribution of the Be disc in the V band is not accounted for, which is thought to be significant only in redder bands \citep[see Fig. A.16 in][]{Schootemeijer2022}} of OB stars in OB+cc binaries. 

\subsection{O stars} \label{Ostars}
Except for Model\,SFH-S, our synthetic populations always contain just over 1000 O\,stars. Model\,SFH-S contains only 444, because in this model the assumed star formation rate in the SMC dropped starting 7\,Myr ago. This corresponds roughly to the time spent by a 20$\mso$ model in the O star phase. In Model\,SFH-R, the star formation rate is equal to that in our fiducial model for the first 10\,Myr, such that the number of O\,stars remains essentially unaffected. A similar inference holds for the number of O\,stars in pre-interaction binaries.

The number of O stars produced from mergers or accretion is also reduced in Model\,SFH-S although by a smaller factor, since they originate from smaller initial masses. We obtain somewhat more O\,star merger and accretion products in Model\,SFH-R compared to our fiducial population because the star formation rate is elevated for ages above 10\,Myrs.

\subsection{OBe threshold} \label{OBe-stars }

\citet{Townsend2004} proposes that most of Be stars are very close to their critical rotation. Following this idea, we assume in our fiducial population model that a star shows the Be phenomenon when it reaches 0.95$\vc$. 
However, sub-critically rotating Be stars are observed \citep[][and references therein]{Rivinius2013}. A recent study shows that the mean rotational velocity of Be stars may be only about 0.68$\vc$ \citep{Dufton2022}. To assess these uncertainties in defining OBe stars, we set the threshold value of $\vrot/\vc$ to be 0.98 and 0.80 in Models\,$\vc$-0.98 and $\vc$-0.8, respectively. This threshold value has no repercussion in the model populations other than for the count of OBe versus ordinary OB stars. However, assuming that an OBe disc is required to make an OB star appear as Be/X-ray binary, the expected number of these systems is influenced. 

Our results show that the effects of changing the $\vrot/\vc$ threshold value are only considerable near the high-mass end, where wind braking becomes significant. When the $\vrot/\vc$ threshold is increased from 0.80 to 0.98, the expected number of BH+OBe binaries is decreasing from 194 to 147, while that of predicted NS+OBe binaries changes from 23 to 20. The predicted $\vrot$ and $\vrot/\vc$ distributions of the OB stars in OB+cc binaries are provided in Fig.\,\ref{App_v_rot}.

\subsection{Wolf-Rayet and helium stars} \label{WR+He}

Since to be counted as Wolf-Rayet stars, our stripped stars need to have luminosities above $\llso > 5.6$ (Sect.\,\ref{He_WR}), they originate from stars with initial masses above $\sim$$40\mso$ (cf., Fig.\,\ref{smc_single}). The lifetime of 40$\mso$ stars is only 4.5\,Myr. Therefore, the number of WR+OB binaries is more than halved in Model\,SFR-S.  It remains unaffected in Model\,SFR-R.

The adopted initial mass ratio and period distributions in our fiducial model prefer close binaries and high mass ratios (see Eqs. \eqref{f_logpi} and \eqref{f_qi}) in Sect.\,\ref{PopSyn}, compared to distributions which are flat in $\log P_{\rm i}$ and $q$. 
Changing to a flat distribution reduces the number of OB+He-star binaries by $\sim$10\%. This is because the He-star population is dominated by low-mass systems (Tab.\,\ref{tab3}), which are formed with relatively low initial orbital periods and high initial mass ratios (Fig.\,\ref{A1k}). Wolf-Rayet stars formed in wide binaries and low-mass-ratio binaries are boosted by a flat distribution, increasing the WR+Oe number from 4.0 (fiducial) to 5.2 (Model\,logPq-flat).   

\subsection{BH+OB binaries} \label{BH-binaries}

We see strong changes in the number of predicted BH+OB systems when we change assumptions related to the formation of BHs. When we introduce a BH kick as described above, about 82.3 systems break up, such that their predicted number in the SMC drops by 37\%, from 210 to 126 (Model\,Kick-BH). While this reduces the number of BH+OB binaries with orbital periods of $\sim$100\,d, and that with B\,star companions ($M_{\rm OB} \simle 15\mso$), the peaks of the respective distributions remain largely unchanged (Fig. \ref{DIC_BH_NS}). BH kicks produce some eccentric BH binaries with orbital periods of up to $\sim$40\,yr. Assuming that no BHs form of course produces zero BH binaries (Model\,NS-limit). 

In Model\,SFH-S, the number of BH+OB binaries is reduced from 210 (in the fiducial model) to 195 (by only 7.6\%). This reflects the fact that most of our BH binaries stem from progenitors with lifetimes of more than 7\,Myr (i.e., from stars below $\sim$$40\mso$). This is in contrast to our WR stars, which are formed from more massive primaries (Sect.\,\ref{WR+He}). Accordingly, Model\,SFH-R, with a peak in star formation at an age of 30 Myr, enhances the OB+BH population by $\sim$100, to a total number of 313.

Also the initial binary parameter distributions affect the predicted number of OB+BH binaries. Comparing with the Sana distributions \citep{Sana2012}, a flat distribution in $\log P$ has a larger fraction of wide-orbit binaries. Hence, Model\,logPq-flat predicts more OB+cc binaries in period range of 100 - 300 days, and less below 20 days,
compared to the fiducial model. The total number of BH systems in Model\,logPq-flat is boosted from 210 to 268.

\subsection{NS+OB binaries} \label{NS-binaries}

In our parameter study, we explore the maximum possible effects. Our Model\,Kick-265 shows that a Hobbs-kick applied to all NSs destroys almost all NS binaries; only 4.65 would remain in our SMC population. Assuming NSs were born without a birth momentum kick (Model\,Kick-0), on the other hand, increases their number by more than a factor\,4, with respect to our fiducial model. While both assumptions are not realistic, these experiments show the strong dependence of the expected numbers on the adopted kicks. Fig.\,\ref{DIC_BH_NS} shows that the kicks also affect the orbital period distribution. Larger kicks shift the peak of the distribution to larger periods, while smaller kicks shift it to shorter periods. 

The peak in the OB\,star brightness (or mass) distribution is not much affected. The upper right panel of Fig.\,\ref{DIC_BH_NS} shows that the most probable OB star mass in OB+NS binaries is about 8-15$\ms$ for all different kick models, while the predicted number is largely different. In addition, different kick velocities can change the age when the OB star fills its Roche Lobe during the OB+NS phase, which slightly affects the predicted total number of NS systems (NS binaries + disrupted systems).

In the fiducial model, if a star has a helium core mass $M_\mathrm{He,c}$ larger than 6.6$\ms$ at the core helium depletion, we assume it to produce a BH. This expectation is based on the detailed simulation on the compactness of pre-SN stars \citep{Sukhbold2018}. In order to examine how this assumption affects our NS population, we consider the extreme case (Model\,NS-limit) that all stars with $M_\mathrm{He,c}$ exceeding 6.6$\ms$ produce NSs. While our fiducial model predicts over 200 BH systems, Model\,NS-limit only predicts 67 NS binaries in total, because momentum kicks are included, and because more mass is ejected by  the SN than in the case of BH formation. Still, the number of NS+OB systems with orbital periods between 10 and 300 days is largely enhanced, smoothing out the bi-modality found in the fiducial model (Figs.\,\ref{ECC2} and\,\ref{DIC_BH_NS}). Furthermore, the number of massive and bright OB\,companions to NSs is increased. Their V-band magnitude can reach up to 13\,mag (Fig.\,\ref{DIC_BH_NS}).

Finally, the number of OB+NS binaries is also affected by the adopted star formation history. There may be evidence suggesting a non-constant star formation rate in the recent past.
The population of high-mass X-ray binaries in the SMC could infer
a peak in star formation rate tens of million years ago 
\citep{Antoniou2010,Antoniou2019,Ramachandran2019}, which may be questionable \citep[][and Paper II]{Schootemeijer2021}. \citet{Rubele2015} identified 
two peaks at $\sim$$30\,$Myrs and $\sim$$ 5\,$Gyrs.
\citet{Rubele2018} confirmed the existence of the bi-modality, while the 
recent peak was shifted to recent 10 Myrs. On the other hand,
\citet{Schootemeijer2021} found that one needs star formation rate decreases to
zero within 7 Myrs to explain the dearth of young massive stars in the SMC.
While Model\,SFH-S has little effect, the star formation history in Model\,SFH-R was designed to explain the large number of Be/X-ray binaries found in the SMC. As shown by the numbers in Tab.\,\ref{DIC}, it does its job, and boosts the predicted number of OB+NS binaries by more than a factor of\,2. It affects the period and mass distributions very little. 

\subsection{Runaway stars} \label{runaways }

In our fiducial model, $\sim$$74$ OB runaway stars are formed, due to the disruption of the binary when the compact object forms. Table\,\ref{DIC} shows that this number can increase substantially, depending on our assumptions. Larger NS kicks have a relatively small effect, because the NS kicks in our fiducial model already break up $\sim$$75\%$ of the systems. However, adding a BH kick (Model\,Kick-BH) roughly doubles the number, and  producing only NSs and no BHs (Model\,NS-limit) quadruples it, and it can produce very massive runaway stars.

Notably,  each of our broken-up binaries also produces a runaway NS or BH. While those BHs will likely remain undetectable, the NSs may develop into radio pulsars. We refrain here from simulating their population, because this would require to include a significant number of new parameters, describing the radio emission, time evolution, and space motion of the pulsars \citep[cf.,][]{Titus2020}.

\section{Comparisons with observations}\label{s_obs}

\subsection{O stars and OBe stars\label{obs_OBe}}

Our fiducial model predicts 1072 O stars in the SMC, which include the O stars in pre- and post-interaction binaries and merger products. As discussed in Sect.\,\ref{no}, this number may exceed the number of observed O stars, because observed counterparts of their early hydrogen-burning evolution are lacking. While a recent drop in the SFR can address this issue (Sect.\,\ref{Ostars}), it would lead to a clear underproduction of WR stars (Sect.\,\ref{wrobs}). A more fine-tuned SFR may fit the numbers of WR star binaries and O stars simultaneously. 
The observed fractions of Oe stars and of O stars with evolved companions, compared to all O stars, should be considered as upper limits when compared to our predictions, because the latter include the unobserved, potentially embedded young O\,stars \citep{Schootemeijer2021}.

\begin{figure}[t]
    \centering
    \includegraphics[width=\linewidth]{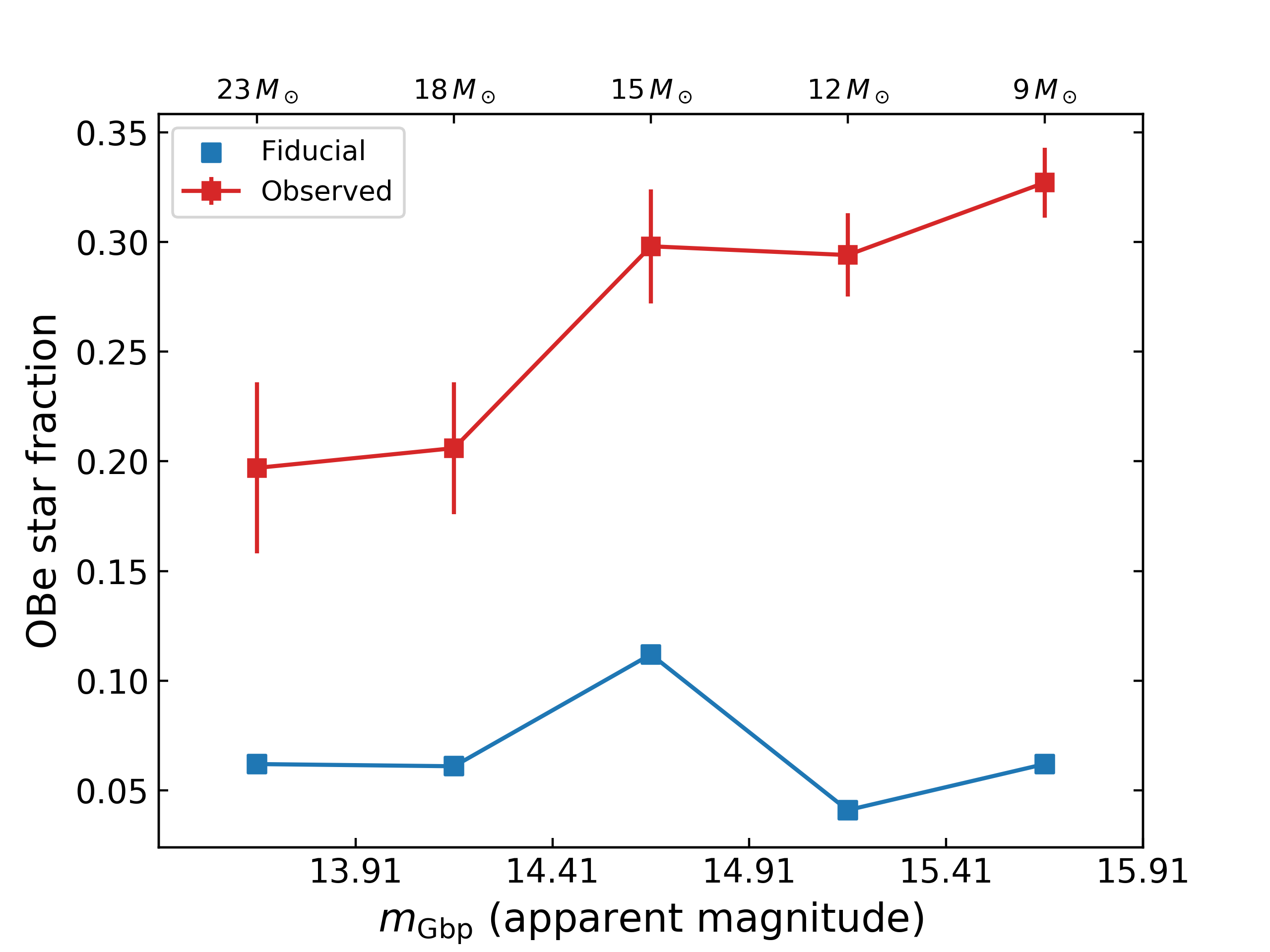}
    \caption{Predicted fraction of OBe stars in our fiducial synthetic SMC population as a function of apparent G$_\mathrm{bp}$ magnitude (blue). The OB stars in pre-interaction binaries are included. The observed OBe star fraction is plotted with red \citep{Schootemeijer2022}. On the top we show the averaged evolutionary mass in each bin \citep{Schootemeijer2022}. 
    }
    \label{OBe_frac}
\end{figure}

Our fiducial population predicts that about 7\% of the OB stars should appear as OBe stars. This is much less than the observed fraction \citep[$\sim$$20\%-30\%$,][]{Schootemeijer2022}, where both, the predicted and the empirical fractions are computed using the same $V$-magnitude threshold of 15.8\,mag and upper limit of 13.4\,mag (corresponding roughly to $9\mso$ and 23$\mso$ respectively). Most of our post-mass transfer binaries ($>$80\%) are expected to contain an OBe star (cf., Fig.\,\ref{Companions}), and it is insensitive to our adopted threshold in the fraction of critical rotation required for a model to be counted as OBe star (Sect.\,\ref{OBe-stars }). While correcting for embedded stars would reduce this discrepancy, it affects mostly O\,stars, while the OBe population is dominated by Be stars, so this correction would be small. On the other hand, the discrepancy is smallest for the most massive OBe stars (Fig.\,\ref{OBe_frac}).  
Including the O stars with masses above $23\mso$, our fiducial model shows about 2.1\% of O stars to be Oe stars, which implies 22.6 Oe stars out of 1072 O stars.  
The BLOeM survey has identified 20 Oe stars out of 159 O stars, giving a fraction of 13\% \citep{Shenar2024}, implying about 60 Oe stars exist in the SMC with a completeness of 35\%. Our parameter variations give a range for the predicted number Oe stars of 16-30 (Tab.\,\ref{DIC}).   

Notably, the wind induced spin-down of rotating O\,stars is uncertain and could be too strong in our models \citep{Nathaniel2025}. Furthermore, 
it would be an effective way to boost the number of predicted OBe stars by assuming that the product of main sequence mergers would be rotating rapidly enough to satisfy our OBe criterion.  However, as discussed in Sect.\,\ref{binary_grid}, it appears more likely that such mergers lose most of their angular momentum in the process, and appear as slow rotators after their thermal relaxation. 

Our underprediction of the number of OBe stars is accompanied by a similar underprediction of the number of Be/X-ray binaries (see below). To account for the large number of observed Be/X-ray binaries in the SMC, a peak in the SF rate some 30\,Myr ago has been proposed, which we explore in our ``SFH-R'' population model (cf., Sect.\,\ref{NS-binaries}). In this scenario, also the number of predicted OBe stars is boosted by a factor of\,1.7 (Tab.\,\ref{DIC}). While this result implies that the past SF history in the SMC may indeed play a role, its deviation from a constant SF rate would need to be more extreme to explain the high number of OBe stars within our fiducial physics assumptions. 

It is therefore mandatory to consider variations in the key physics assumptions used in our binary evolution models. The most relevant ones concern the mass transfer efficiency during RLOF, and the conditions applied for assuming a binary model will merge during mass transfer. 
A higher mass transfer efficiency would allow more IMF-preferred low-mass stars to form OBe stars, and a relaxed merger criterion would help that more OB binaries survive the first mass transfer phase and produce OBe stars.
Since we cannot vary these assumptions in retrospect in our detailed binary evolution models, we discuss potential consequences qualitatively in Sect.\,\ref{discuss}. For a more quantitative discussion, we refer to Paper II.

\subsection{Be/X-ray binaries\label{BeXB}} 

There are about 150 high-mass X-ray binary candidates found in the SMC, of which $\sim$100 are identified to be Be X-ray binaries. In 63 Be X-ray binaries, the NS spin period has ben measured \citep{Haberl2016,Treiber2025}\footnote{Online up-to-date catalogue: https://www.mpe.mpg.de/heg/SMC}. However, our fiducial model only predicts $\sim$25 OBe+NS binaries (Sect. \ref{initial_parameters_var}). This discrepancy is similar to that in the number of OBe stars (see above). It may partly be remedied in a similar way.

As for the OBe stars, the SF history may play a role. In our ``SFH-R'' population model (cf., Sect.\,\ref{NS-binaries}), the number of predicted Be/X-ray binaries is more than doubled. Again, we conclude that a mini-starburst $\sim$20\,Myr ago may help to reduce the discrepancy, but that other mechanisms may be worth to consider. Different to the case of the OBe stars, the number of Be/X-ray binaries cannot be increased through the merger channel.

As shown in Sect.\,\ref{NS-binaries}, the number of predicted Be/X-ray binaries can also be boosted by reducing the adopted NS birth kick (by up to a factor of 4.5), or the NS/BH-formation core mass threshold (by about a factor\,2 assuming our fiducial NS kicks; see Tab.\,\ref{DIC}). However, these factors are derived from extreme assumptions, which are unlikely to be realized in nature. NS containing Be/X-ray binaries could also be produced from WD binaries through accretion-indued collapse. However, this evolutionary channel, which requires a common envelope stage and is not included in our synthetic population, works only with main
sequence components in the mass range $2\dots 4\mso$ \citep{WangBo2020}. As such, these systems could only add a minor contribution to the low mass end of the observed SMC
Be/X-ray binaries distribution (Fig.\,\ref{DIC_BH_NS}).

On the other hand, our model predicts about 400 OBe stars with He-star and with BH companions, which have no clear observed counterparts (see below). For both cases, the production of X-rays has been suggested \citep{Casares2014,Langer2020a}. Since the nature of the companion stars in about half of the observed Be/X-ray binaries is undetermined,
alternatives to assuming  BHs for them could be considered (cf., Sects.\,\ref{Heobs} and\,\ref{BH-binaries-obs} below). In addition, the missing of BH companions may also be explained by a different BH formation prescription that makes a large fraction of our predicted BH progenitors undergo supernovae and produce NSs instead \citep[e.g.,][]{Schneider2021,Aguilera-Dena2023}.

While our fiducial model produces an insufficient number of Be/X-ray model binaries, their general properties agree rather well with observations.  Figure\,\ref{porbNS} shows that the predicted orbital period range  is similar to the observed one. While the predicted long-period tail is absent in the observed distribution,  this might be so because the periods of the longest period systems are hardest to measure.  Also our shortest-period systems with orbital periods
below 10 days have no observed counterparts. This overprediction in the short-period regime is also found in the prediction from the POSYDON code \citep{Rocha2024}.
This discrepancy could be related to tidally-induced disc truncation.
Short-period OB+NS binaries are less eccentric (Fig.\,\ref{ECC2}),
and in low-eccentricity binaries the 
gap between the circumstellar disc and the L1 point is so large that the NS can 
not generate strong X-ray emission \citep{Okazaki2001}.
The observed supergiant X-ray binary 
SMC X-1 is close to Roche-lobe filling and has an orbital period of 3.9 days 
\citep{Rawls2011,Falanga2015}, which may have
evolved from such short-period B(e)+NS binaries.

We also find the predicted mass distribution of the Be stars in Be/X-ray binaries broadly consistent with the observed distribution (Fig.\,\ref{DIC_BH_NS}). 
We expect most OB stars in OB+NS binaries to $V$-magnitudes brighter than 17.5\,mag (spectral types earlier than B3, or masses larger than $\sim8\mso$), 
which is consistent with the observed spectral types of 
the SMC BeXBs \citep{McBride2008}.  This latest spectral type reflects 
the minimum mass of a main-sequence star to have a NS companion. The observed distribution shows more stars in the high-mass end (14-15\,mag), which may imply a higher accretion efficiency is required to better reproduce the observed distribution.

We find a bi-modality in our predicted orbital period distribution, which is related to Case A and Case B binaries (Fig.\,\ref{porbNS}). This  bi-modality is also visible in the period-eccentricity distribution
(Fig.\,\ref{ECC2}). Observationally, \citet{Knigge2011} found two subpopulations of 
Be X-ray binaries, those in close orbits with short 
NS spin periods and wide orbits with long NS spin periods. We cannot compare directly to their analysis, because our model does not include 
the NS spin evolution. The observed bimodal feature 
in NS spin period distribution is suggested to relate to the underlying supernova mechanisms \citep{Knigge2011} or different accretion modes \citep{Cheng2014,Xu2019}. 


As shown in Fig.\,\ref{ECC2}, observed BeXBs have eccentricity $e$ below 0.4 
\citep{Townsend2004,Coe2015},
while  we expect half of the population to have eccentricity above that. 
These predicted high-$e$ binaries
also have wide orbit, making the periastron passages of NSs very fast and the corresponding systems
hard to be identified as BeXBs. 
Recently, an extreme case of BeXBs, A0538-66, is observed in the LMC, which has one of the shortest orbital period, 16.6\,d, but highest eccentricity, 0.72, of BeXBs \citep{Rajoelimanana2017,Ducci2022}, marginally covered by our predicted distribution (Fig.\,\ref{ECC2}). This system may have a relatively wide pre-SN orbit, where the mass gainer was spun up  by mass transfer but not affected by tides, and the supernova explosion has reduced its orbital period to 16.6\,d, which requires the kick velocity has a component directed opposite to the orbital velocity of the NS progenitor.

In the SMC, one B-type star + radio pulsar binary is observed \citep[PSR J0045-7319][]{Kaspi1994,Bell1995}. 
This binary is highly eccentric with an orbital period of 51 days (Fig.\,\ref{ECC2}) 
but it does not show X-ray emission. 
According to our models, it could be formed through Case A mass transfer, 
where tides spins down the B star, whereafter its orbit gets widened due to supernova explosion.

\subsection{OB+helium star binaries} \label{Heobs}

Our fiducial population model predicts 235 OB+He star binaries to exist currently in the SMC, where the OB star is expected to rotate rapidly in most of them (223). With most of them in rather wide orbits ($\sim$ 100\,d), and the OB star rotating rapidly and dominating the optical spectra, they are difficult to identify observationally \citep{Wellstein2001,Gotberg2018}. About 25 candidates have been been found recently \citep{Drout2023}, but strong biases preclude a solid comparison with our results.  

Of the expected OB+He star binaries, we predict $\sim$32 to contain He stars more massive than $\sim$$3\mso$ (cf., Tab.\,\ref{tab3}). With luminosities exceeding $\sim$$10^4\lso$ \citep{Langer1989}, these stars can be expected to emanate a fast wind \citep{Vink2017,Sander2020}, which could generate X-rays through the collision with the companion star's wind or disc. Some of the observed Be/X-ray binaries in the SMC, in particular those with continuous X-ray emission, could thus contain He stars rather than NSs  \citep{Langer2020a}. In fact, if the number of OB+He star binaries with massive He stars in our fiducial population would be underestimated by a similar amount as the number of OBe stars or that of Be/X-ray binaries, we could expect more than 100 OB+He star binaries with He star masses above $3\mso$. A dearth of He\,stars due to the delayed expansion of massive stars at low metallicity, as proposed by \citet{2024arXiv241205356H} appears unlikely, as this would also make it harder to explain the large number of OBe stars in the SMC, and the drop of the observed binary fraction in evolved massive SMC stars \citep{Patrick2025}. 

\subsection{OB+Wolf-Rayet star binaries} \label{wrobs}

There are 12 WR stars observed in the SMC, of which 4 have O\,star companions and one contains a H-deficient WR-type star \citep{Foellmi2003,Foellmi2004,Koenigsberger2014,Hainich2015,Shenar2016,Shenar2018,Neugent2018,Schootemeijer2018}. 
From our fiducial population model, we expect 6.7 WR+O binaries in the SMC. 
All of the observed SMC WR star binaries have orbital periods below 20\,d, where we predict 2.6 systems. Considering the Poisson error of the observed sample, our model well reproduces the observed number of WR+O binaries.
While our models predict half of them to contain Oe star companions (cf., Tab.\,\ref{DIC}), all observed O star companions 
appear to be moderate rotators \citep[projected velocities comparable or lower than $\sim$$200\,\kms$,][]{Shenar2016,Shenar2018}. This confirms a known discrepancy between the observed rotation velocities of O\, stars in WR binaries and predictions from evolutionary models \citep{Shara2020}, which may suggest that circumstellar discs are easily disrupted by O star winds and radiation. Alternatively, the wind induced spin-down in the evolutionary models is underestimated \citep[also see][]{Nathaniel2025}, or the massive accretors never spin up to critical rotation.

The location of the WR components of the WR binaries in the HR diagram are well reproduced (Fig.\,\ref{SMC_WR}). The hydrogen-free one, the WO\,star SMC\,AB8 \citep{Shenar2016}, is covered by our H-free models with $\log\,T_\mathrm{eff}/\mathrm{K} > 5.1$ \citep[also see][]{Wang2019}. Notably, a large temperature correction due to the optical depth of the WR wind is not expected for the SMC WR stars 
\citep{Aguilera-Dena2022,Sen2023}. Particularly, SMC AB7 is near the H-free region, and observationally this WR stars has the lowest surface hydrogen abundance compared to SMC AB6 and SMC AB3. The WR+WR binary SMC AB5 may have formed through the tidally-induced chemically homogeneous evolution in equal-mass binaries \citep{dM2016,Marchant2016}. For SMC AB2 and AB4, their cool surface temperatures suggests that they are core-hydrogen burning \citep{Schootemeijer2018}, 
core-helium-burning WR stars formed from the initial secondary stars \citep{Pauli2023}.

The agreement of the properties of the four WR+O star binaries with our models is particularly satisfying. 
\citet{Schootemeijer2018} have measured the slope of the H/He-gradient in the envelope, which affects their effective temperature, in a model-independent way. The result implied a H/He-gradient coincident with that left by the receding convective core during core hydrogen burning, which is indeed what we obtain in our short period binary models, which have orbital periods in the WR+O star binary phase of less than 20\,d.   

On the other hand, our fiducial population model predicts distributions of the binary properties of WR+O binaries which are much broader than the observed ones. Specifically, we expect 4.2 WR+O binaries with orbital periods above 20\,d, while the observed ones all have shorter periods. This is reflected in the distribution of the predicted orbital velocities (bottom panel of Fig.\,\ref{SMC_WR}). Furthermore, we expect a larger range in WR/O-mass ratios (0.5-1.5; Fig.\,\ref{WR_Pq}), while the three objects where this is measured show values near 0.5. Notably, both, WR+O star binaries with long orbital periods, and with less massive companions (possibly B stars) are harder to detect as such due to their orbital velocities (Fig.\,\ref{WR_Vorb}). However extensive searches in the seven apparently single SMC WR stars have excluded the presence of companions in the long-period or large-mass-ratio part of the parameter space (\citealt {Foellmi2003}, \citealt{Schootemeijer2024}; see also \citealt{Deshmukh2024} and \citealt{Dsilva2020,Dsilva2022,Dsilva2023} for the Milky Way WR stars).  \citet{Almeida2017} have revealed a flatter orbital period distribution in the 30 Doradus region than that assumed in our fiducial model, which would enlarge the discrepancy in the long-period regime between model predictions and observations (cf., Model logPq-flat in Sect.\,\ref{initial_parameters_var}).

Even though, with 12 observed WR stars in the SMC, we have to face large statistical errors,
their presence strongly disfavours the SFH adopted in our ``SFH-S''-population model. As shown in Tab.\,\ref{DIC}, this model produces 2.9\,WR binaries, of which only one have an orbital period below 20\,d, while four such systems are observed.

\subsection{OB+BH binaries} \label{BH-binaries-obs}

Our fiducial model predicts 210 OB+BH binaries to reside in the SMC. While this number may appear surprisingly large at first, it is not in relation to the number of OB+NS binaries. Based on the large number of Be/X-ray binaries, the order of magnitude of the OB+NS binaries in the SMC is surely at least 100. Given that, based on the Salpeter-IMF, the birthrate of NSs and BHs is comparable, differences in the number of binaries may come from differences in the lifetime of the OB companion, and from differences in the binary disruption rate. The former favours OB+NS binaries, but only slightly so, since, in our prediction,  also the OB+BH binaries are dominated by B star companions (Fig.\,\ref{DIC_BH_NS}). Contrarily, the disruption rate strongly favours the OB+BH binaries, because 80\% of the OB+NS systems are disrupted
(see Tab.\,\ref{DIC} for varying the kick-assumptions for BHs). Therefore, independent of strong assumptions, we may expect that the order of magnitudes of the number of OB+NS and of OB+BH binaries in the SMC are similar. 

In our fiducial SMC population, we find about 1000 O\,stars and 40 O+BH binaries. 
Given that a fraction of the O stars may still be embedded (see Sect.\,\ref{obs_OBe}),
we would expect 4.0-6.0\% of the observed SMC O\,stars to have a BH companion. A very similar value has been found from a comparable binary population synthesis study for the LMC \citep{Langer2020}. So far, no massive BH binarity has been discovered in the SMC. We will therefore, when assessing the observable parameters of OB+BH binaries, compare with the known OB+BH systems found in other galaxies, wary of their metallicity difference.

Five O+BH binaries have been detected, 
Cyg\,X-1 in the Milky Way \citep{Miller-Jones2021}, 
LMC\,X-1 in the Large Magellanic Cloud \citep{Orosz2009}, 
M33\,X-7 in the M33 \citep{Ramachandran2022}, 
HD\,130298 in the Milky Way \citep{Mahy2022}, 
and VFTS\,243 in the Large Magellanic Cloud \citep{Shenar2022}. 
The first three sources are wind-fed X-ray binaries. The recently discovered VFTS\,243 and HD\,130298 are X-ray quiet and have the relatively long orbital periods (10.1 days and 14.6 days). 
The Be+BH nature 
of MWC\,656 \citep{Casares2014} is challenged by new spectral data \citep{Rivinius2022,Janssens2023}. 
We include both solutions in Fig.\,\ref{f_bh}.
While none of these systems are found at SMC metallicity, 
the orbital properties of the OB+BH binaries appear to be insensitive to the variation of metallicity
\citep[Fig. B.1 in][]{Janssens2022}.

Figure\,\ref{f_bh} shows that the parameters of the observed BH binaries fall into the regime that is well populated in our fiducial model, and follow the trends with the mass of the OB star.  
On the other hand, the orbital period distribution of our predicted OB+BH binaries peaks near 100\,d, while --- in particular when  MWC\,656 is removed --- there are no observed massive BH binaries with orbital periods above 15\,d. This could in part be due to an observational bias. Most massive BH binaries have so far been detected due to their X-ray emission, which is expected to be much weaker or absent in wind accreting binaries with orbital periods above $\sim$10\,d \citep{Sen2021, Sen2024}. Accretion from an OBe disc might produce weak X-ray emission if the main sequence companion is rapidly rotating and emanates a disc (which occurs less often in the OB+WR progenitor binaries than we expect; cf., Sect.\,\ref{wrobs}), as suggested for MWC\,656. In fact, it cannot be exclude that some of the observed Be/X-ray binaries in the SMC host BHs.

Optical spectroscopy, however, can be expected to detect long period OB+BH binaries. In a circular $15\mso+10\mso$ OB+BH binary with an orbital period of 100\,d, the orbital velocity of the OB\,star is $\sim$$50\kms$, which should be easily detectable even with a significant inclination of the orbit. 
Two long-period OB+BH candidates have been identified in the LMC, whose BH nature needs verification by follow-up observations \citep{Shenar2022b}. Still, recent and ongoing campaigns targeting the Milky Way \citep{Mahy2022} and the SMC \citep{Bodensteiner2025} fail to find candidate systems.

Therefore, it is possible that the predicted long-period BH binaries do not exist, or are much rarer than expected. In fact, this would be quite consistent with the apparent absence of long-period WR+OB binaries in the Magellanic clouds (see  Sect.\,\ref{wrobs}), which are progenitors of OB+BH binaries. It would be an immediate consequence of this scenario that the currently favoured channel to produce merging massive black holes, which involves a common envelope evolution of a massive star with a black hole in a wide orbit \citep{Belczynski2016}, would hardly occur in nature. We will discuss implications for the  mass transfer physics in massive binaries in Sect.\,\ref{discuss}.

\section{Inferences for mass transfer physics\label{discuss}}

\subsection{Lower mass regime}

In Sect.\,\ref{s_obs}, several discrepancies between our model population and the observed populations of massive stars in the SMC have been uncovered. Each of these discrepancies can potentially allow us to suggest improvements to our model assumptions, and may thus be helpful in future binary population synthesis calculations.

Our fiducial model predicts too few OBe stars above $\sim$$9\mso$. As the OBe star populations is dominated by the lowest mass binaries considered in our study, we may seek to change assumptions made for this mass group. Considering Fig.\,\ref{A1k}, we see that it is a prominent feature of the lower mass binary models in our grid that the merger fraction is very high ($\sim$$80$\% at $M_{\rm 1,i}=10\mso$). A straightforward cure to obtain more OBe stars would be less systems merging. According to Paper II, this could about double the predicted number of OBe stars. A further increase can be achieved by increasing the mass transfer efficiency, as then more OBe stars are obtained from the more numerous lower mass stars. 

In a similar realm, one might obtain more Be/X-ray binaries. However, for those, we have to match the lowest Be star mass in Be/X-ray binaries. For the lowest primary mass to produce NSs of $\sim$$10\mso$ (Fig.\,\ref{A1k}), the lowest Be star mass in Be/X-ray binaries of $\sim$$8\mso$ (Fig.\,\ref{DIC_BH_NS}) implies that for low accretion efficiency (as in our binary models) the lowest secondary mass surviving the mass transfer event is $\sim$$8\mso$, and binaries with $10\mso$ primaries and initial mass ratios below 0.8 would merge \citep[also see][]{Rocha2024}. For fully conservative mass transfer, this critical mass ratio could be as low as 0.1. 

\citet{Schurmann2024} and \citet{Zhao2024} showed that the minimum mass ratio for stable mass transfer depends strongly on the mass transfer efficiency. This is not surprising, since the swelling of the secondary upon accretion is a function of its accretion rate (rather than of the mass transfer rate).  
Sch\"urmann et al. (Paper II) find that this interdependance fixes the accretion efficiency to $\sim$50\%, leading here to a critical mass ratio of about 0.5. This result is consistent with the recent study by \citet{Vinciguerra2020} based on rapid binary population synthesis of Be/X-ray binaries. 
Sch\"urmann et al. (Paper II) find that this can also help to explain the large number of observed OBe stars in the SMC.

While it is satisfactory to find consistent constraints on the mass transfer physics in the considered mass range ($M_1\simeq 10\mso$), the mass transfer physics itself remains not well understood. The large merger fraction and high critical mass ratio in our binary models (Fig.\,\ref{A1k}) occur because we deem all systems to merge in which the energy input to achieve the required rate of mass loss from the binary exceeds the available photon energy (Sect.\,\ref{MESA_input}). The required energy input is particularly high for our models, because we assume that the mass gainer stops accreting once it is spun up, which mostly occurs after very little accretion. Clearly, this assumption cannot hold in the binaries considered here \citep[cf.,][]{Langer2012}. However, even with an accretion efficiency of 50\%, about $3.5\mso$ need to be removed from $\sim$$10\mso$ binaries. When the scenario that this matter is ejected by a photon-driven wind fails, the only other available energy source may be the orbital energy. If this is taped for ejecting the matter, we would expect significant effects on the orbital separations.       

\subsection{High mass regime}

In the high mass regime, the main discrepancies concern not so much the number of objects --- though for the OB+BH binaries it could, as none are observed in the SMC (see below). To predict as many as 7 OB+WR binaries requires a low merger fraction (cf., Fig.\,\ref{A2k}) which is only achieved with a low accretion efficiency \citep{Schurmann2024}. In this respect, out fiducial model appears acceptable.

At high mass, the main concerns are the orbital period distributions. Many massive O\,star binaries with orbital periods above $\sim$$20\,$d are found in the LMC \citep{Sana2013,Mahy2020b}
and SMC (Sana et al. 2025, submitted), which is reflected in our initial binary period distribution (Eq.\,\ref{f_logpi} in Sect.\,\ref{PopSyn}). However, there are essentially no OB+WR binaries (with He-burning WR stars) found with such periods (Sect.\,\ref{wrobs}). Similarly, even though spectroscopic searches could have identified them, there are so far not any undisputed OB+BH binaries with periods above 20\,d found (Sect.\,\ref{BH-binaries-obs}). This situation is in contrast to that in the lower mass regime, where the orbital period distribution of the observed lower-mass counterparts, the OBe/X-ray binaries, peaks near $100\,$d, in agreement with our predictions (Fig.\,\ref{DIC_BH_NS}). Also, the observed Be+subdwarf binaries often have orbital periods on the order of 50-200\,d \citep{Peters2008,Peters2013,Mourard2015,Peters2016,Wang2017,Wang2023}.

To explain this, one might assume that long-period O\,star binaries merge during their mass transfer phase. The stellar type of the merger product is uncertain, and depends on the fraction of the H-rich envelope which is lost in the process. In contrast to the stars in the lower mass regime, the proximity to the Eddington limit of the high mass stars facilitates the ejection of matter from the binary system.  Due to the same reason, however, a merger might be prevented. 

Like in the lower mass regime, the physics of mass transfer is not well understood here. Possibly, the progenitors of the apparently single SMC WR stars, which are stars initially more massive than $\sim$$35\mso$ \citep{Hainich2015,Schootemeijer2024}, do not require a binary companion to remove their H-rich envelope \citep{Grassitelli2021,Schootemeijer2024}. However, the absence of long-period evolved massive binaries argues for some of the apparently single WR stars to be the result of a binary merger \citep[cf.,][]{Shenar2023} While long-period WR and BH binaries are harder to detect than short-period ones (except perhaps with Gaia), the possibility that future surveys find such systems cannot be excluded. However, at this time, the chance to obtain merging double-BHs from the Common Envelope channel appears rather small (Sect.\,\ref{BH-binaries-obs}). 

Notably, if we would assume that long-period binaries which would have produced OB+BH systems would merge, the predicted number of OB+BH binaries in the SMC would drop. By which fraction depends on the mass transfer physics in the mass range $20$---$50\mso$, which is not well constrained by the observed populations of post-interaction binaries. In this mass range, also the fate of massive stars is as yet unclear. While we assume in our fiducial model, that single stars in this mass range form BHs, a fraction of them may explode as SN and form NSs instead (Sect.\,\ref{CC_formation}). 

\section{Conclusions\label{conclusion}}

Using a large grid of detailed massive binary evolution models, we have constructed a synthetic massive star population of the Small Magellanic Cloud. Through comparing the number and property distributions of specific subpopulations of post-mass transfer binaries with corresponding observed SMC populations, we were able to obtain strong constraints on uncertain models assumptions, in particular those used to describe the first mass transfer which occurs in these binaries. 

To reproduce the observed OBe stars and OBe/X-ray binaries, the majority of the binary stars need to avoid merging during the mass transfer. At the same time, the average mass transfer efficiency needs to be relatively high in the lower mass regime, and low in the high mass regime of the studied binaries. A lack of observed long-period (> 20\,d) OB+WR and possibly of OB+BH binaries may imply that the observed massive long-period O\,star binaries merge, which would disfavor the Common Envelope channel for the production of merging massive BHs. The physical processes which determine the mass budget and the stability or instability of the first mass transfer remain largely undetermined. 

The observed populations of evolved massive binaries are still small, in particular in the upper mass range of the investigated binaries. It is therefore desirable to enhance these samples, and thereby obtain tighter constraints, not only to the first phase of mass transfer in these systems, but also to achieve more reliable predictions of the contribution of the isolated binary channel to the observed realm of stellar explosions and gravitational wave sources.

\begin{acknowledgements} The authors thank the referee Bo Wang for  the valuable feedback on the manuscript. 
    CW is thankful for the financial support from the CSC scholarship. TS acknowledges support from the Israel Science Foundation (ISF) under grant number 0603225041 and from the European Research Council (ERC) under the European Union's Horizon 2020 research and innovation programme (grant agreement 101164755/METAL). PM acknowledges support from the FWO senior fellowship number 12ZY523N. 
\end{acknowledgements}

\bibliographystyle{aa}
\bibliography{Xu_SMC.bib}

\begin{appendix}
\section{Outcomes of our model grid}
\label{Pq}

\begin{figure*}\label{fig_g1}
\includegraphics[width=\linewidth]{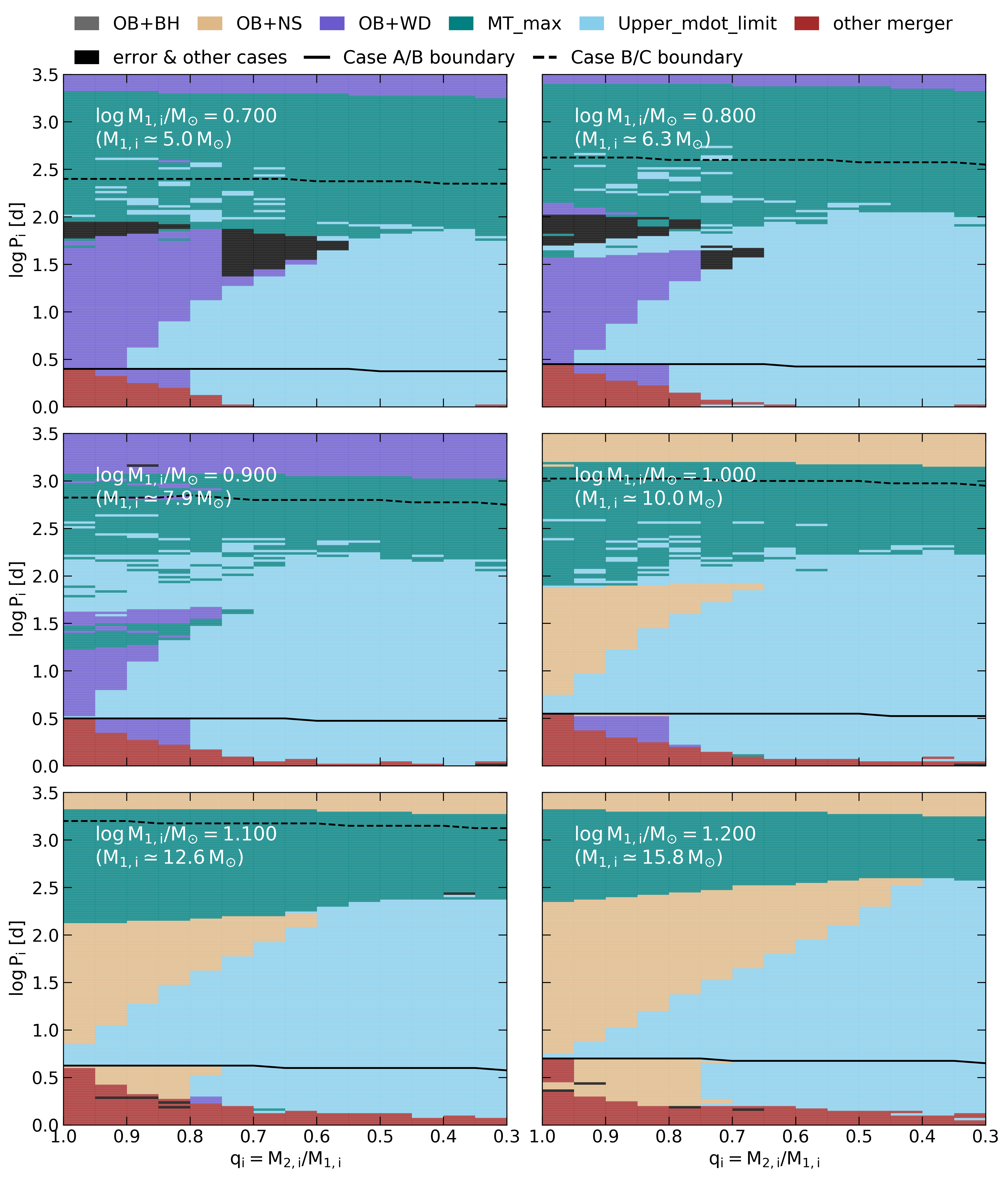}
			\caption{The outcomes of our detailed binary evolution models. The same as Fig.\ref{summary} but for initial primary mass 5.0$\mso$, 
 6.3$\mso$, 7.9$\mso$, 10.0$\mso$, 12.6$\mso$, and 15.8$\mso$ \label{A1k}}
	\end{figure*}

\begin{figure*}
 \includegraphics[width=\linewidth]{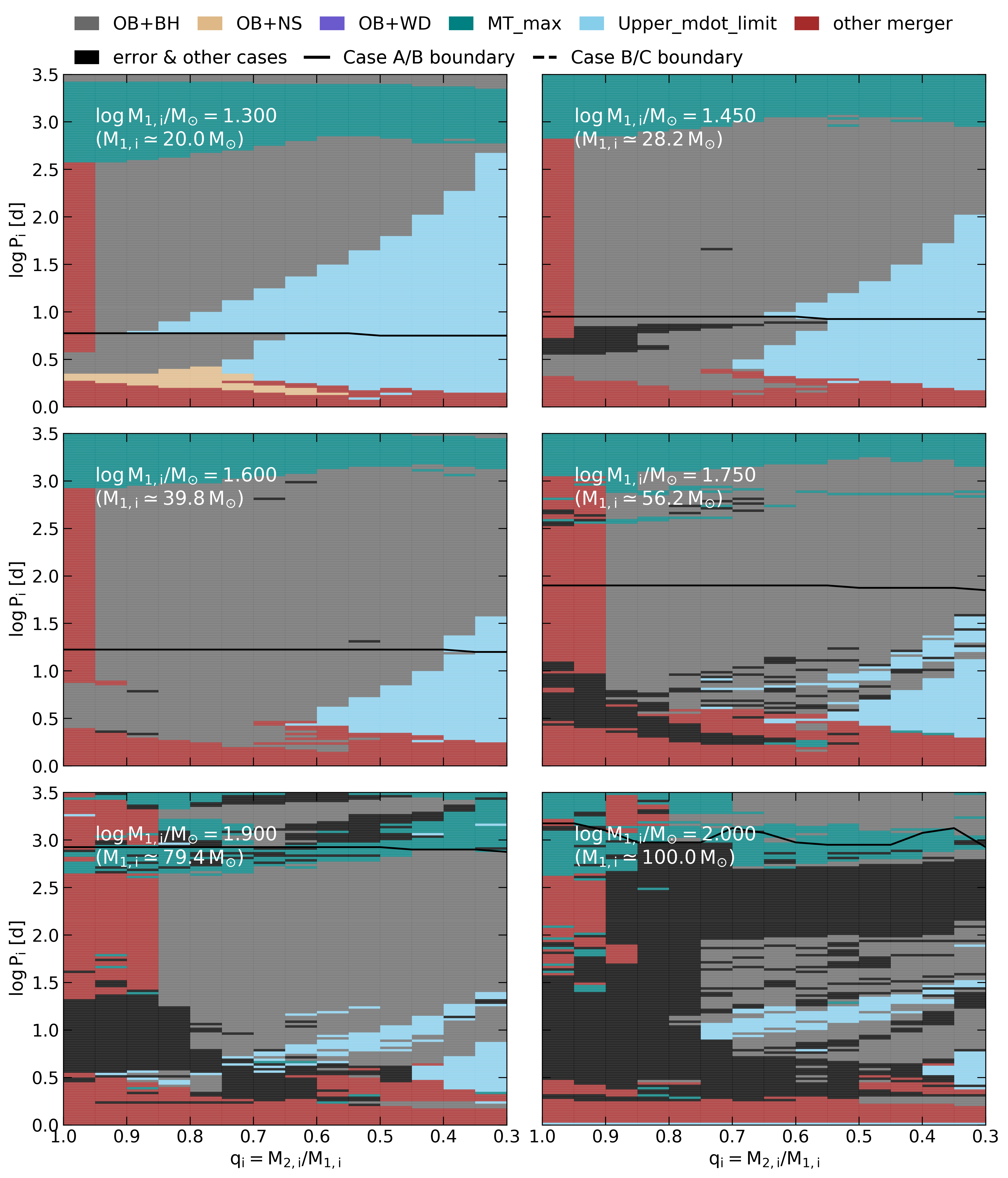}
			\caption{The outcomes of our detailed binary evolution models. The same as Fig. \ref{summary} but for initial primary mass 20.0$\mso$, 
 28.2$\mso$, 39.8$\mso$, 56.2$\mso$, 79.4$\mso$, and 100.0$\mso$. 
 \label{A2k}}
	\end{figure*}

The outcomes of our detailed binary evolution models with initial primary mass from 5.0$\mso$ to 15.8$\mso$ (Fig.\,\ref{A1k}) and from 20$\mso$ to 100$\mso$ (Fig.\,\ref{A2k}), where each 
pixel represents one detailed MESA binary model, and the related evolutionary 
outcome is coded in colour. 
For the sake of presentation, a subset of initial primary masses is shown.


\section{Supernova window\label{SN_window}}

The supernova windows, the initial parameter space allowing supernova to occur, play important roles by determining the magnitude of kick velocities.
However, the resolution of our SMC model grid is not good enough to distinguish different
types of SNe and we therefore adopt the SN windows computed by the ComBinE code \citep[][and Paper II]{Kruckow2018} 
Taking ECSNe as an example,
while ECSNe may happen in a very narrow mass range in single star case \citep{Poelarends2008,Janka2012}, 
the ECSN window can be broadened by binary interaction
\citep{Podsiadlowski2004,Langer2012,Shao2014,Poelarends2017,Siess2018}.
However, detailed simulations show that even including mass transfer, 
it is still narrower than 1$\ms$ \citep{Poelarends2017,Siess2018}. 
The ZAMS mass window computed by the ComBinE code 
is about $[9.5,\,10.2]\ms$. However, around this mass range, the SMC model grid only 
have three mass slices of 8.9$\ms$, 10$\ms$, and 11.22$\ms$. Instead of interpolating,
we simply use our grid points to calculate the systems undergoing ECSNe with a factor accounting for the fraction of ECSNe in each pixel of our model grid. While this could cast uncertainties on our NS populations,
it should be a minor effect comparing with merger criterion.

The fraction factor is calculated through the following approach. 
We use the ComBinE code simulates binaries with flat distribution 
for initial primary mass $M_\mathrm{ 1,i}$, initial mass ratio $q_\mathrm{ i}$, 
and initial logarithmic orbital period log$P_\mathrm{ orb,i}$. Then  we calculate the following statistical weight for all ComBinE models,
\begin{equation}
    W=M_\mathrm{1,i}^{-\alpha}q_\mathrm{ i}^{-\beta}(\log\,P_\mathrm{orb,i})^{-\gamma},
\end{equation}
where $(\alpha,\,\beta,\,\gamma)=(2.3,\,0.1,\,0.55)$ for our fiducial model (the Kroupa IMF and the Sana distribution), and $(\alpha,\,\beta,\,\gamma)=(2.3,\,0,\,0)$ for the logPq-flat model. With this statistical weight, we calculate the fraction of different types of SNe in each pixel of our SMC model grid. Taking ECSN as an example,
the ECSN fraction $f_\mathrm{ ECSN}$ is evaluated as
\begin{equation}
        f_\mathrm{ ECSN} = \frac{\sum_{ j = 1}^{N} \delta_\mathrm{ ECSN}\,W_{j}}{\sum_{ j = 1}^{N} W_{j}},
\end{equation}
where $N$ is the total number of ComBinE models inside a given pixel from our MESA model grid, and
\begin{equation}
\delta_\mathrm{ ECSN} = 
\begin{cases}
1\,\,\mathrm{ ~if~the~ComBinE~model~undergoes~ECSN}\\
0\,\,\mathrm{ ~the~other~cases}
\end{cases}.
\end{equation}
The fractions of other types of SNe are calculated in the same way. 
In our population synthesis calculations, we use the Monte Carlo method to generate a sample of kick velocities with a size of $n$ for each pre-SN systems,
where $n\times f_\mathrm{ ECSN}$ of the sample are draw from the kick distribution
corresponding to ECSN.

\begin{figure*}
    \centering
    \includegraphics[width=0.49\linewidth]{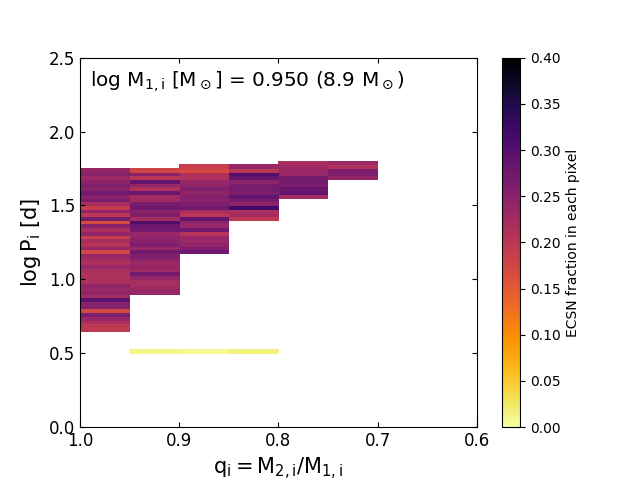}
    \includegraphics[width=0.49\linewidth]{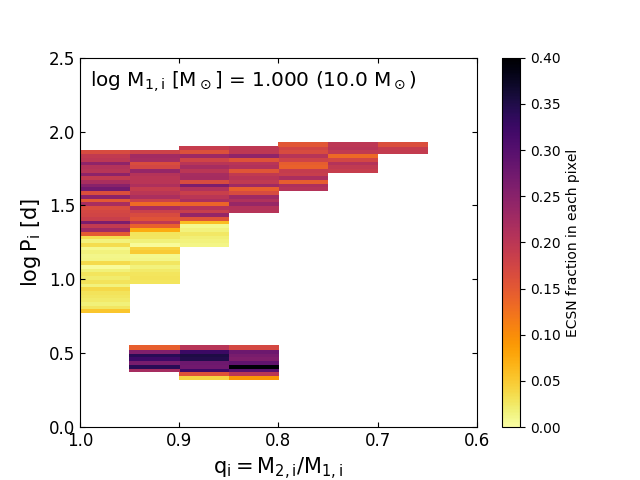}
    \caption{ECSN fraction on the $\qi-\logpi$ plane with initial primary mass 8.9 M$_\odot$ and 10 M$_\odot$.}
    \label{f_ecsn}
\end{figure*}

Figure \ref{f_ecsn} shows the ECSN fraction for each pixel in our model grid.
We see that the ECSN window 
behaves differently in Case A/B systems. In Case B systems, mass transfer 
helps the donor star avoid the second dredge-up, which makes ECSNe become possible
with relatively low stellar masses. In Case A systems, mass transfer happens when
the primary stars are still on the main sequence, which limits the growth of 
the inner core in the post-MS evolution, and consequently the 
ZAMS mass window of ECSNe is shifted towards the high-mass end.

The ComBinE code like other rapid codes assumes that 
the envelope of donor star is completely stripped by mass transfer.
With this assumption, Case BB or Case BC mass transfer only takes place in tight binaries.
However, there could be a considerable fraction of envelope left after 
mass transfer at low metallicity \citep{Laplace2020}, 
which allows Case\,BB or Case\,BC mass transfer occur in wide binaries \citep{Ercolino2024,Ercolino2024b}. 
We find a similar feature in our model grid. 
The remaining H-rich outer layer helps the core keep growing \citep{Ercolino2024}, which
may shift the boundary between NS and WD. 
The H-rich outer layer can expand to very large radii and trigger the mass 
transfer from a partially stripped star. 
Usually, the partially stripped star is less massive than the accretor, making 
mass transfer widen the orbit. 
The parameter space of helium-envelope-stripped supernova could be narrowed by 
this process. In addition, due to the
remaining material, stars could explode when they still fill their Roche lobe 
\citep{Laplace2020,Ercolino2024,Ercolino2024b}. The asymmetric structure of pre-SN star may have 
effects on the kick velocities. We do not expect our predicted NS population to be largely affected by the difference in the physics in ComBinE and MESA.

\section{Envelope inflation}

\citet{Sanyal2015} have shown that in massive main-sequence stars the maximum
Eddington factor can exceed one inside the stars due to the Fe-bump of opacity. 
Radiation then pushes the envelope to a very large radius, leading to core-hydrogen burning 
supergiant models. Envelope inflation is sensitive to the treatment of 
convection and can cause convergence issues in calculation.
The LMC binary models adopted in \citet{Langer2020} cannot handle
the effects of envelope inflation well and therefore  do not have models 
 with initial primary mass larger than about 40$\ms$. 
While envelope inflation can be avoided by assuming a more efficient energy transport inside the envelope
\citep[the {\tt MLT++ } option in the MESA code][]{Paxton2013}, we do not see the reason to assume that the traditional 
mixing length theory becomes invalid \citep{Bohm1958}.

Since the inflated envelope is highly convective \citep{Sanyal2015}, 
\citet{Langer2020} assume that the mass transfer from an inflated star is unstable.
In our SMC models we do find the binaries undergoing envelope inflation 
reach the OB+BH phase. Above 50$\ms$ initial primary mass, the effect of
envelope inflation becomes more and more significant. 
As a consequence, the orbital period window of Case A systems significantly 
widens. With a more efficient convection, the inflated envelope will expend less
and fill the Roche Lobe at a later time, which could slightly increase the 
mass of the stripped star. Since there are only 11 OB+BH binaries having 
initial primary mass above 50$\ms$, we do not expect our result to be largely
affected by this uncertainty.

\section{Pair instability}

We adopt the mass range and mass ejection of pulsational pair instability (PPI) computed by \citet{Marchant2019}. None of our models are massive enough to produce pair-production supernovae. Our 100$\mso$ single star model develops a helium core $M_{\rm He,c}$ of $58.6\mso$ at the core helium depletion, which is still below the threshold for pair-production supernova \citep[$61.1\mso$][]{Marchant2019}. Our fiducial model predicts 3.95 OB+BH binaries formed through PPI (hereafter PPI OB+BH), of which only one is significantly affected by the mass ejection during the PPI ($M_{\rm He,c} > 45\mso$, Fig.\,\ref{ppisn_fitting}).
The recently updated $^{12}{\rm C}(\alpha,\gamma)^{16}{\rm O}$ rate shifts the PPI mass range to a higher values \citep{Farag2022}, which would reduce our predicted number of PPI OB+BH binaries due to the effect of the IMF. 

The PPI OB+BH binaries are also affected by the winds of WR stars, which are not well understood \citep{Grafener2017}. In our binary model, primary stars with initial masses above $\sim$$70\mso$ can show a hydrogen-free WR star phase, during which the stellar winds reduce the masses of the WR stars. If our model had a weaker WR star wind, the predicted PPI OB+BH binaries would be slightly more eccentric due to a stronger mass ejection, while the total number of the PPI OB+BH binaries would not change. Given that the PPI OB+BH binaries only contribute a very small fraction to our synthetic population, we do not expect our result to be significantly affected by these uncertainties.

\section{Calculations of statistical weights}
\label{formulas}

We assume the distribution of the initial mass of primary star is described by the 
initial mass function (IMF), which is $f_\mathrm{IMF}\,\propto \mi^{-\alpha}$, 
and the distributions of initial mass ratio and initial orbital period are 
$f_{\qi}\propto \qi^{-\beta}$ and 
$f_{\logpi}\propto (\logpi)^{-\gamma}$. 
The predicted number contributed by a binary model $N_\mathrm{b}$ with initial parameter 
$(\mi,\,\qi,\,\logpi)$ is 
\begin{equation}
    {\rm d\,}N_\mathrm{ b} \propto \mi^{-\alpha}\qi^{-\beta}(\logpi)^{-\gamma}\,\mathrm{ d\,log}M_\mathrm{ 1,i}\,\mathrm{ d}q_\mathrm{ i}\,\mathrm{ d\,log}P_\mathrm{ orb,i}.
    \label{A1}
\end{equation}
In order to take into account star formation rate, we rewrite Eq. \eqref{A1}
into mass fraction form,
\begin{equation}
    {\rm d\,}F_\mathrm{ b} \propto  (M_\mathrm{ 1,i}+ M_\mathrm{ 1,i}\, q_\mathrm{ i}) \times {\rm d\,}N_{\rm b}.
\end{equation}
Then from a constant star formation rate (SFR), the predicted number of a OB+cc binary is given by 
\begin{equation}
	\centering
        \begin{split}
	N_\mathrm{ b}=\mathrm{ SFR\times\,lifetime}\times\frac{\int_{V}{\rm d\,}F_{\rm b}}{\langle M_\mathrm{ b}\rangle},
        \end{split}
        \label{A3}
\end{equation} 
where $V$ is the parameter space enclosed by $[\mathrm{ log}P_\mathrm{ orb,i},
\,\mathrm{ log}P_\mathrm{ orb,i}+\Delta\,\mathrm{ log}P_\mathrm{ orb,i}]$, 
$[q_\mathrm{ i},\,q_\mathrm{ i}+\Delta q_\mathrm{ i}]$, and 
$[\mathrm{ log\,}M_\mathrm{ 1,i},\,\mathrm{ log\,}M_\mathrm{ 1,i}+\Delta \mathrm{ log\,}M_\mathrm{ 1,i}]$,
$(\Delta \mathrm{ log\,}M_\mathrm{ 1,i},\,\Delta q_\mathrm{ i},\,\Delta\,\mathrm{ log}P_\mathrm{ orb,i})$
are the intervals of our model grid, lifetime is the lifetime of the OB+cc phase, and
$\langle M_\mathrm{ b}\rangle$ is the averaged mass of the binary within the parameter space $V$
weighted by the initial distributions, which is
\begin{equation}
    \langle M_\mathrm{ b}\rangle = \frac{\int_{V} (\mi+q\mi) {\rm d\,}N_{\rm b}}{\int_{V}  {\rm d\,}N_{\rm b} }.
\end{equation}
Defining statistical weight $W$ as following, 
\begin{equation}
    W(\mi,\,\qi,\,\logpi) =\frac{\int_{V}{\rm d\,}F_\mathrm{ b}}{\langle M_\mathrm{ b}\rangle},
\end{equation}
Eq. \eqref{A3} has the following form
\begin{equation}
    N_\mathrm{ b}=\mathrm{ SFR\times\,}\mathrm{lifetime\times\,}W(\mi,\,\qi,\,\logpi).
\end{equation}

In order to include non-constant star formation rate, we introduce the factor SFH,
\begin{equation}
    \mathrm{SFH} = \frac{\int_{t_\mathrm{f,OB+cc}}^{t_\mathrm{i,OB+cc}}\, \mathrm{SFR}(t)\, \mathrm{d}t}{\mathrm{lifetime}},
\end{equation}
where the SFR is the function of lookback time $t$, which means that $t = 0$ marks the observed
status, $t_\mathrm{i/f,OB+cc}$ are the binary age of entering/ending OB+cc phase.
The observed OB+cc population at $t=0$ comes from the star formation starting at
$t=t_\mathrm{f,OB+cc}$ and ending at $t=t_\mathrm{i,OB+cc}$.
With this, the predicted number of a OB+cc binary with non-constant SFR is given by 
\begin{equation}
    N_\mathrm{ b}=\mathrm{ SFH\times\,}\mathrm{lifetime\times\,}W(\mi,\,\qi,\,\logpi).
\end{equation}

The number of O stars, He stars, and WR stars are computed with the same 
method except the lifetimes are determined by effective temperature hotter
than 31.6 kK, core helium burning, and core helium burning with logarithmic
luminosity higher than 5.6.

\section{Tides during OB+cc phase\label{sync_timescale}}

As mentioned in Sect. \ref{MESA_input}, after the formation of BH or NS, we evolve
the secondary as a single star. However, during the OB+cc phase, the OBe star can
be spun down by tides. Here we take into account tidal interaction by
considering the synchronization timescale at the beginning of OB+cc phase.
If the synchronization timescale is shorter than 10\% of the lifetime of 
OB+cc phase\footnote{While we take 10\%  as the 
threshold value for strong tide, we also perform experiments with 30\%, 50\%, 
and 100\% and the predicted OBe+cc binaries remain unchanged.}, we expect 
in the following OB+cc phase the OB star is rapidly spun down by tides
and cannot form OBe stars.

In order to take into account the effect of eccentricity on tides, 
we introduce a different definition of synchronization timescale 
$\tau_\mathrm{ sync}$ basing on \citet{Hut1981} and \citet{Hurley2002}.
The spin evolution of stars induced by tides is given by \citet{Hut1981}
\begin{equation}
\begin{split}
    \frac{\domegas}{\dd t}&=3\left(\frac{k}{T}\right)_\mathrm{rad}
    \left(\frac{q^2}{r_\mathrm{ g}^2}\right)\left(\frac{R}{a}\right)^6
    \frac{\omegaorb}{(1-e^2)^6}\\
    &\times\left[f_2(e^2) - (1-e^2)^{3/2}\,f_5(e^2)\frac{\omegas}{\omegaorb}\right],
\end{split}
\end{equation}
where $\omegaorb$ and $\omegas$ are the orbital angular velocities and 
spin angular velocities of the OB star, mass ratio $q$ is $M_\mathrm{ cc}/\mob$, 
$r_\mathrm{g}$ is the 
ratio of gyration radius to stellar radius $R$, $e$ is eccentricity,
\begin{equation}
    \left(\frac{k}{T}\right)_\mathrm{ rad} = 1.9782\times 10^4\left(\frac{M_\mathrm{OB}R^2}{\ms \rs^2}\frac{\rs^5}{a^5}\right)^{1/2} (1+q)^{5/6}E_2\mathrm{\,yr}^{-1},
\end{equation}
the numerical factor $E_2$ is
\begin{equation}
    E_2 = 1.592\times10^{-9}\left(\frac{\mob}{\ms}\right)^{2.84},
\end{equation}
$f_2(e^2)$ and $f_5(e^2)$ are defined by \citet{Hut1981},
\begin{equation}
    f_2(e^2)=1+\frac{15}{2}e^2+\frac{45}{8}e^4+\frac{5}{16}e^6
    \end{equation}
    and
    \begin{equation}
    f_5(e^2)=1+3e^2+\frac{3}{8}e^4.
\end{equation}
Basing on the above equations, we can define the synchronization timescale 
$\tau_\mathrm{ sync}$ as
\begin{equation}
\begin{split}
    \tau_\mathrm{ sync} & = \abs{\frac{\omegas - \omegaorb}{\dot{\Omega}_\mathrm{ spin}} } \\
    & = \left[ 3 \left(\frac{k}{T}\right)_\mathrm{ rad}\left(\frac{q^2}{r_\mathrm{ g}^2}\right)\left(\frac{R}{a}\right)^6\right]^{-1}\\
    &\times \abs{\frac{(1-e^2)^6(\omegas - \omegaorb)}{f_2(e^2)\omegaorb - (1-e^2)^{3/2}\,f_5(e^2)\omegas}}.
    \label{tsync}
\end{split}
\end{equation}
For circular orbit, $e=0$, Eq. \eqref{tsync} becomes the widely used form
\begin{equation}
    \tau_\mathrm{ sync} (e=0)  = \left[ 3 \left(\frac{k}{T}\right)_\mathrm{ rad}\left(\frac{q^2}{r_\mathrm{ g}^2}\right)\left(\frac{R}{a}\right)^6\right]^{-1}.
\end{equation}

With the above assumption, we find that the tidal interaction during the OB+BH phase is too
weak to spin down the OB stars. In our fiducial model, 170.401 OBe+BH binaries are predicted. 
Without tide, 170.403 OBe+BH binaries are predicted.

\section{Further model details\label{further_details}}

\subsection{Single star models}

\begin{figure*} 
    \centering
    \includegraphics[width=\linewidth]{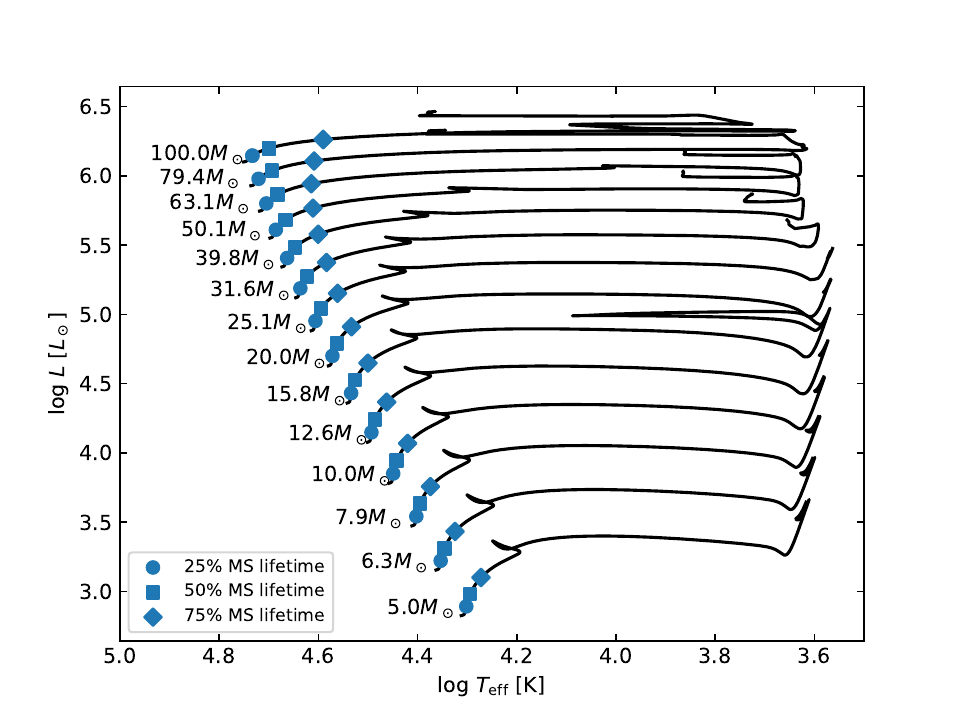}
    \caption{Evolutionary tracks of our non-interacting models in the Hertzsprung-Russell diagram, with the indicated initial masses. The tracks are terminated when the core helium abundance drops below $10^{-4}$. The circles, squares, and diamonds mark the ages of 25\%, 50\%, and 75\% of the MS lifetime.}
    \label{smc_single}
\end{figure*}

Figure\,\ref{smc_single} presents the evolution of our non-rotating single star models. The evolution begins at the zero-age main-sequence point. After the main-sequence phase, a hook appears due to the contraction before the ignition of hydrogen shell. About $50\mso$, envelope inflation becomes more and more significant, which allows Case A mass transfer to occurs in wide binaries. None of these single star models can produce WR stars through self-stripping.

\subsection{An example of chemically homogeneous evolution model}

\begin{figure*}
    \centering
    \includegraphics[width=\linewidth]{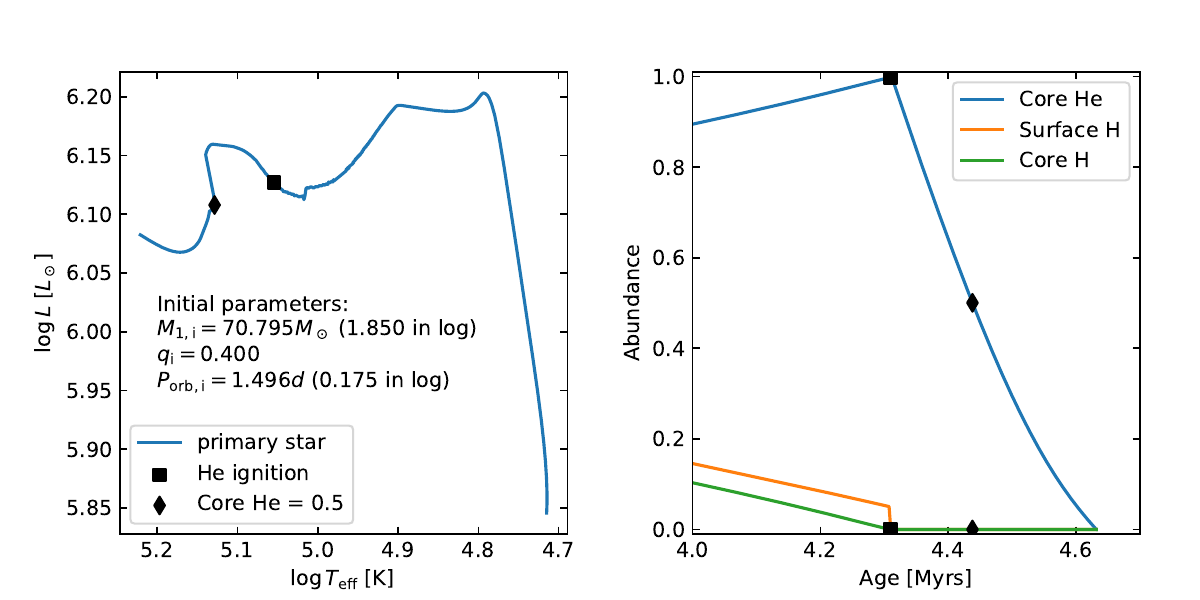}
    \caption{An example of our chemically homogeneous evolution model, with initial parameters indicated by text. The left panel presents the evolution of the chemically homogeneously evolving star on the Hertzsprung–Russell diagram, where the square and diamond mark helium ignition and the middle of helium burning. The right panel shows the chemical evolution of the chemically homogeneously evolving star as a function of its age. The blue, orange, and green correspond to the evolution of core helium, surface hydrogen, and core hydrogen. }
    \label{CHE_example}
\end{figure*}

All of our chemically homogeneous evolution models have a similar evolutionary history. 
Figure\,\ref{CHE_example} presents an example. The primary star is spun up by tide and evolves chemically homogeneously. As the primary star does not expand significantly, mass transfer is avoided.
During the late hydrogen-burning phase, the primary star reaches the Wolf-Rayet star regime, which is not considered in this work. When approaching the core helium ignition, the surface hydrogen rapidly drops to zero due to a enhanced stellar wind near the core hydrogen depletion.

\subsection{Properties of OB+WR binary systems}

\begin{figure*}
    \centering
    \includegraphics[width=0.9\linewidth]{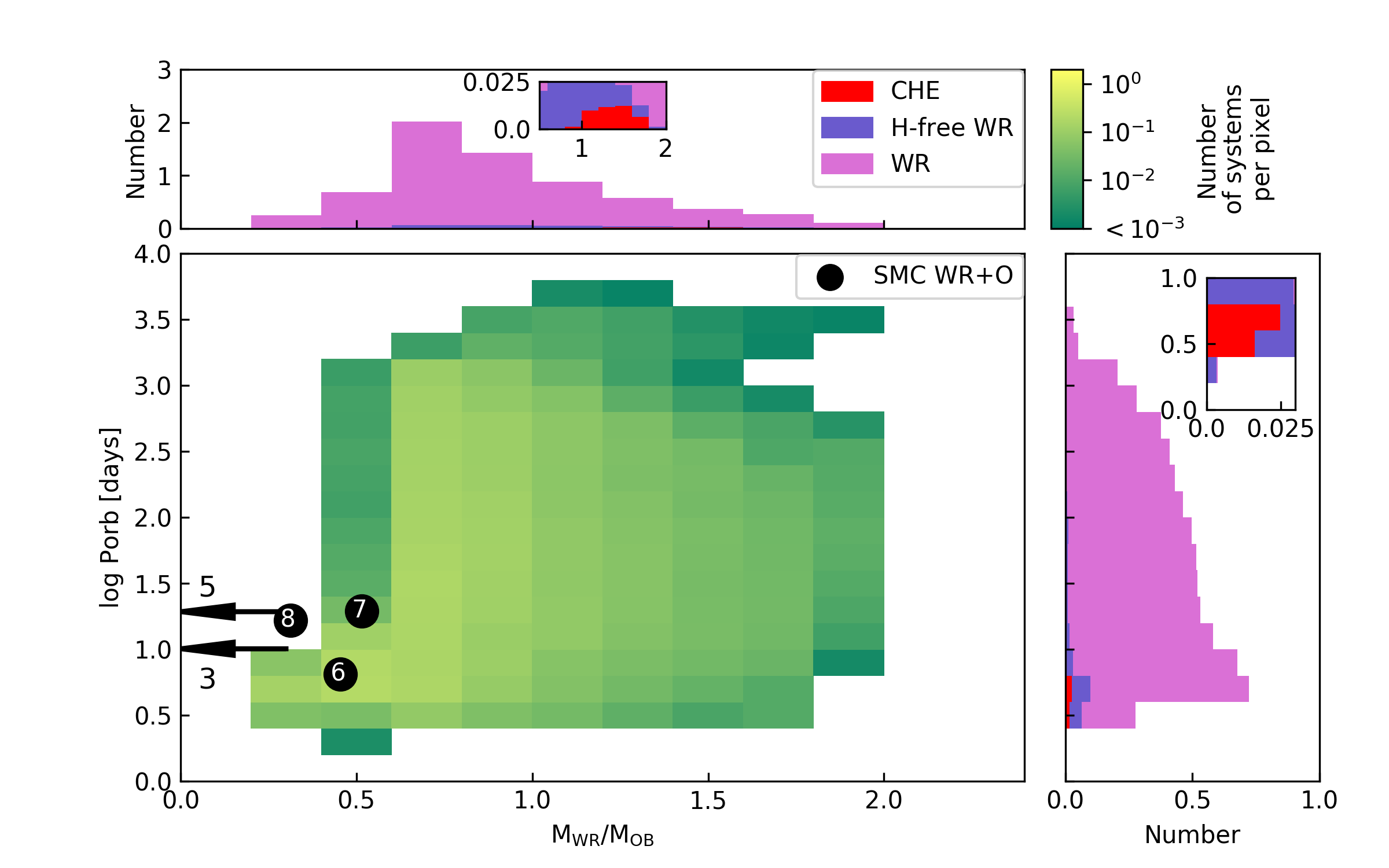}
    \caption{Predicted distribution of OB+WR binaries in the  mass ratio $M_\mathrm{WR}/M_\mathrm{OB}$ - logarithmic orbital period $\logp$ plane. The number in each pixel is coded in colour. The H-free and CHE WR stars are identified in 1D projection. The observed WR binaries \citep{Foellmi2003,Foellmi2004,Koenigsberger2014,Hainich2015,Shenar2016,Shenar2018} are plotted with black.}
    \label{WR_Pq}
\end{figure*}

Figure\,\ref{WR_Pq} presents the OB+WR binaries in the mass ratio $M_\mathrm{WR}/M_\mathrm{OB}$ - logarithmic orbital period $\logp$ plane. Most of our OB+WR binaries have a $M_\mathrm{WR}/M_\mathrm{OB}$ of 0.7. Below this value, binaries are formed with low initial orbital period and close-to-one initial mass ratio, where the mass gainer can accrete a large amount of mass.  The number drop towards high $\logp$ is related to the initial distribution (cf. Sect. \ref{SMC_He}). 

\begin{figure*}
    \centering
    \includegraphics[width=\linewidth]{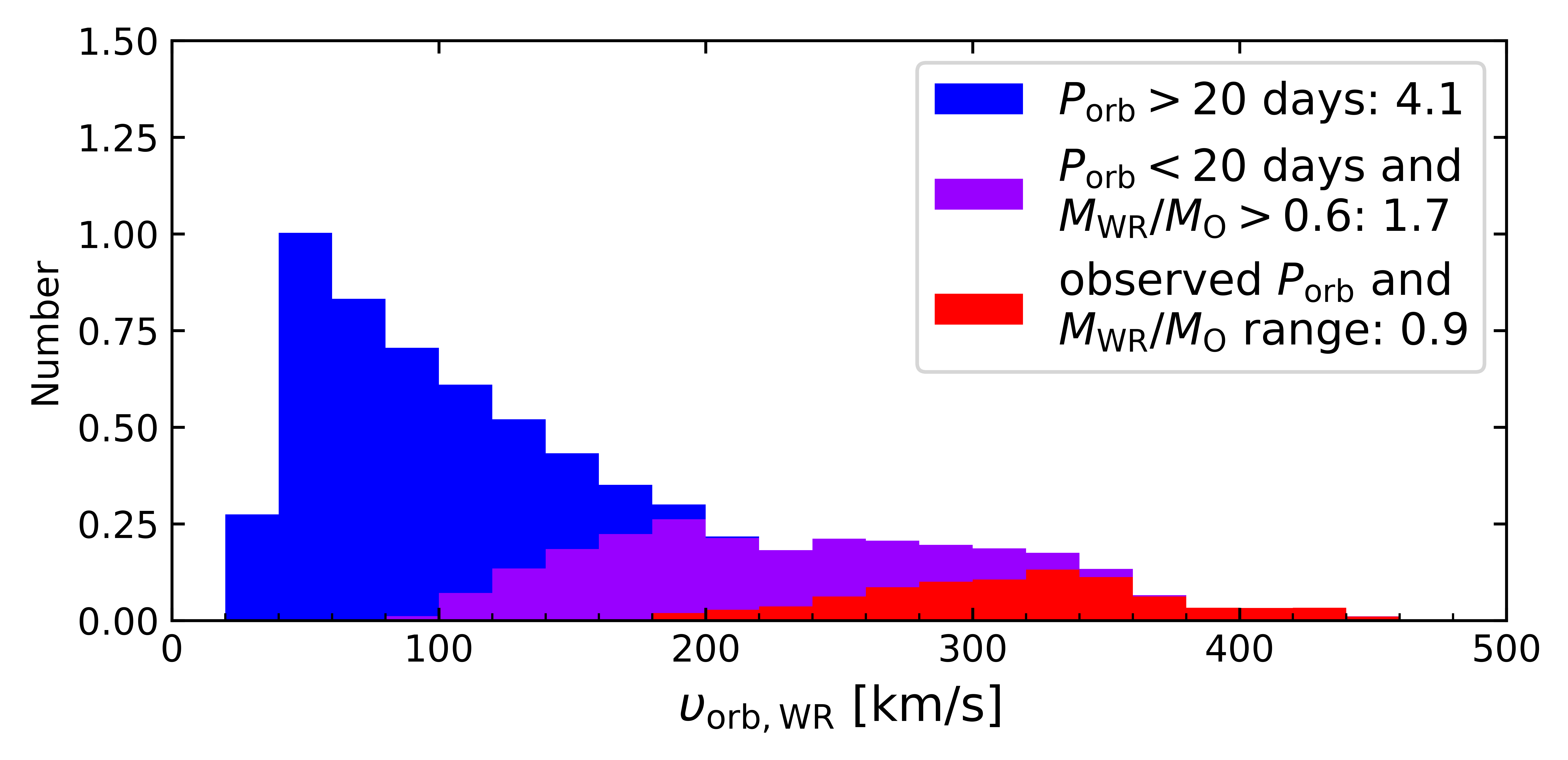}
    \caption{Predicted distribution of the orbital velocities $\upsilon_{\rm orb,WR}$ of the WR stars in the WR+O binaries. Three subpopulations are identified, which are featured by $P_{\rm orb}>20$ days (blue), $P_{\rm orb}<20$ days and $M_{\rm WR}/M_{\rm O}>0.6$ (purple), and $P_{\rm orb}<20$ days and $M_{\rm WR}/M_{\rm O}<0.6$ (observed region, red). The numbers of WR+O binaries correspond to each subpopulations are indicated by the text in the legend.}
    \label{WR_Vorb}
\end{figure*}

Figure\,\ref{WR_Vorb} presents the orbital velocities $\upsilon_{\rm orb,WR}$ of the WR stars in the WR+O binaries. The observed SMC WR+O binaries have orbital periods $P_{\rm orb}>20$ below 20 days and mass ratio $M_{\rm WR}/M_{\rm O}$ below 0.6, and we accordingly divide our predicted population into three subpopulations, which are featured by $P_{\rm orb}>20$ days, $P_{\rm orb}<20$ days and $M_{\rm WR}/M_{\rm O}>0.6$, and $P_{\rm orb}<20$ days and $M_{\rm WR}/M_{\rm O}<0.6$. The high-$\upsilon_{\rm orb,WR}$ regime is dominated by the low-$M_{\rm WR}/M_{\rm O}$ close WR+O binaries, which is the observed parameter space.  Due to the change in mass ratios, the high-$M_{\rm WR}/M_{\rm O}$ short WR+O binaries have lower $\upsilon_{\rm orb,WR}$, which is still above 100$\kms$.
Most of our WR stars have an orbital velocity of about 50$\kms$, corresponding to wide systems ($P_{\rm orb}>20$). 

\subsection{Properties of OB+cc binary systems}

We further distinguish  different modes of mass transfer (Case A or Case B).  Case A systems produce the most massive O stars
reaching 100$\ms$ and the slowest rotators, while Case B systems produce 
less massive OB stars and most of them are near-critically rotating (Fig.\,\ref{App_v_rot}).
Case A systems usually have tight orbits, resulting in strong 
tidal interaction. As a result they have relative high
accretion efficiency according to our rotation-dependent accretion efficiency.
The opposite takes place in Case B systems, which results in near-zero accretion efficiency.

\subsubsection{Masses and mass ratios}

\begin{figure*}
\centering
    \includegraphics[width=0.85\linewidth]{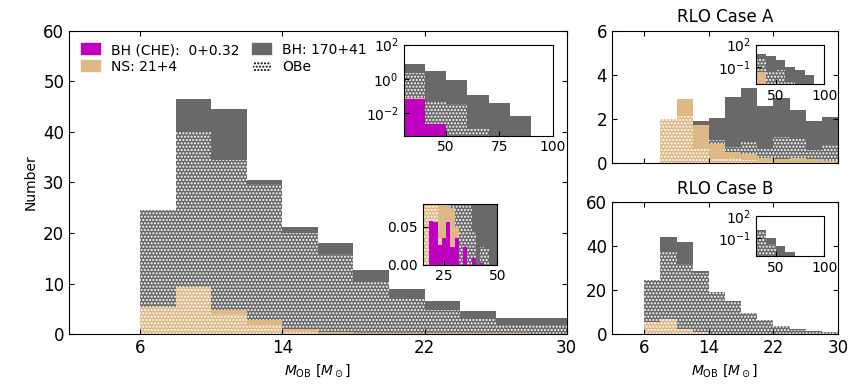}
    \includegraphics[width=0.85\linewidth]{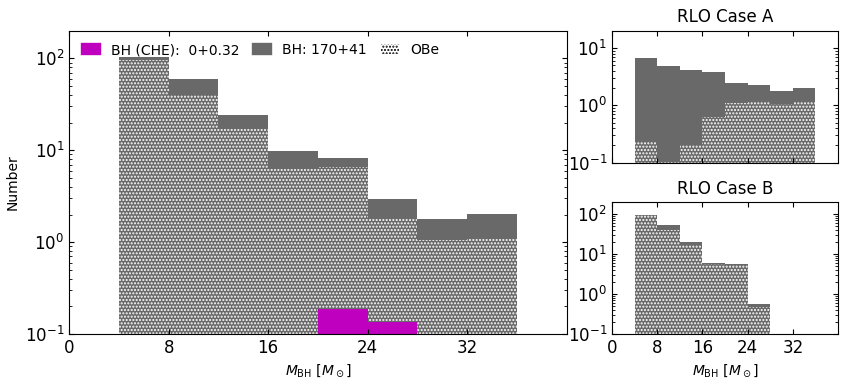}
 	\includegraphics[width=0.85\linewidth]{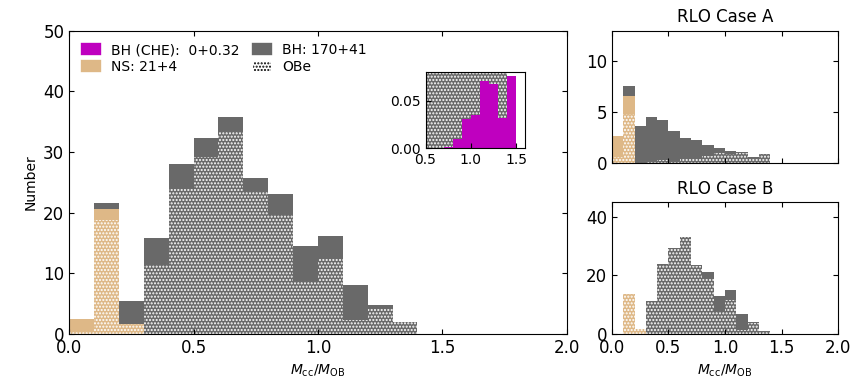}
	\caption{\label{App_Mass}Top panel: Distributions of OB star masses $\mob$ in OB+cc binaries. 
	The types of compact objected are coded in colour (BH: black, NS: brown), 
    and the shaded area is related to the OBe feature. The OB+BH binaries formed from CHE 
    are plotted with purple.
    The number in the legends is the predicted number of OBe stars
	and normal OB stars, e.g., "Black hole: 170+41" means 170 BH+OBe binaries and 41 BH+OB 
	binaries. The in-layer plot in the top panel shows the distribution in the range 
	30 - 100$\ms$ with bin width of 10$\ms$, while the main plot is produced 
	in 6 - 30$\ms$ with bin width of 2$\ms$. The left panel is the distribution
	of the total population, which is disentangled into Case A systems and Case B 
	systems in the right upper and lower panel respectively.
    Middle panel: Distributions of black hole masses.
    Bottom panel: Distributions of mass ratios of OB+cc binaries.
    }
\end{figure*}

The top panel of Fig.\,\ref{App_Mass} shows the predicted distribution of $\mob$. 
The distribution peaks at $\sim$$10 M_\odot$, below which stars have less
chance to form BHs or NSs. Above $10\ms$ the number drop is due to the 
effects of the IMF and lifetime.
Both BH systems and NS systems have a minimum companion mass about $6\ms$ (cf., Sect.\,\ref{s_numbers}). 

The middle panel of Fig.\,\ref{App_Mass} shows the predicted distribution of $\mbh$, which is mainly shaped by the IMF (cf. Sect. \ref{SMC_OBBH}). In Case A systems, mass transfer begins when primary stars are still on
the MS, which limits the growth of He core. 
Therefore, BH progenitors in Case A systems trend to have higher
initial primary masses than that in Case B systems.
Towards the high-mass end, the orbital period window of Case A binaries
becomes wider and wider due to the increasing importance 
of envelope inflation. These wide-orbit Case A systems produce 
the most massive BHs, which can have fast-rotating companions
because of weak tide.


The bottom panel of Fig.\,\ref{App_Mass} presents the distribution of mass ratio of OB+cc binaries 
($q = M_\mathrm{cc}/M_\mathrm{OB}$). 
In our calculations, the mass of NS is fixed to be 1.4$\ms$, 
while most of their companions have mass around 10$\ms$, 
leading to a peak in mass ratio at $\simeq 0.1$. Since our model
predicts all OB stars in OB(e)+NS binaries to be more massive than
6$\ms$, the highest mass ratio of NS systems is about 0.2-0.3. 

The mass ratio of OB+BH binaries peaks at $\simeq$0.6-0.7, clearly separated
from the NS systems. The drop in numbers towards high mass ratio is due
to the increasing fraction of the binaries undergoing unstable mass transfer.
The effects of decreasing lifetime cause the number drop towards low mass ratio. 
Case A systems contribute the
lowest mass ratio $\sim$0.2 and highest mass ratio $\sim$1.8, 
corresponding to close binaries with high accretion efficiency and wide 
binaries with inflated primary stars respectively. 

\subsubsection{Orbital properties}

\begin{figure*}
    \includegraphics[width=\linewidth]{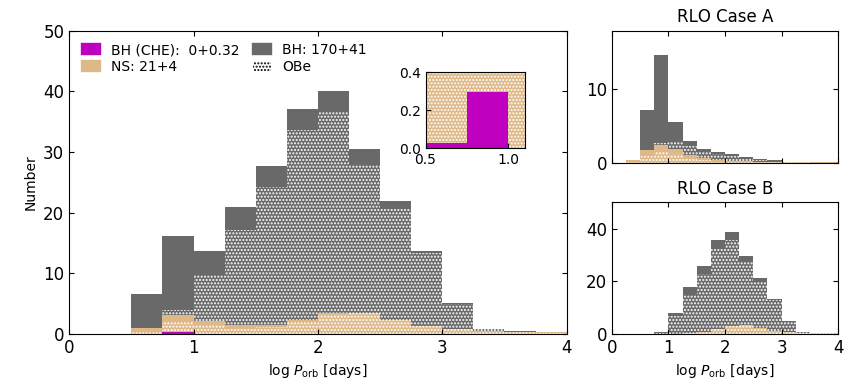}
    \includegraphics[width=\linewidth]{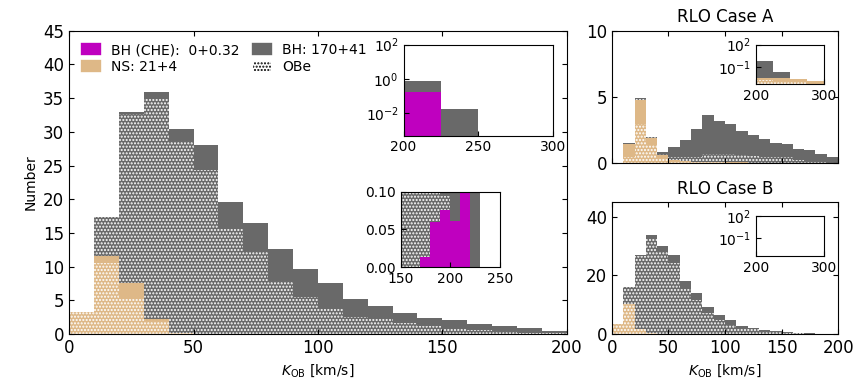}
	\caption{\label{App_porb}Top panel: Distribution of logarithmic orbital periods log$\,P_\mathrm{ orb}$ of OB+cc binaries. The colours and legends have the same meaning as Fig.\,\ref{App_Mass}.
    Lower panel: Semi-amplitude of orbital velocity of OB stars $\kob$.}
\end{figure*}

The top panel of Fig.\,\ref{App_porb} shows the distribution of orbital periods of OB+cc binaries $\porb$. Our merger criterion leads to a peak near $\porb=100\,$days (cf. Sect. \ref{SMC_OBBH}), which is dominated by Case B systems. Case A systems require close orbits, leading to a peak near 7\,days. A small fraction of Case A systems have orbital periods above 100 days due to envelope inflation (BH binaries) or SN kick (NS binaries). 

The bottom panel of Fig.\,\ref{App_porb} shows the distribution of the semi-amplitude of orbital velocities of OB stars $K_\mathrm{ OB}$.
For BH systems, the $\kob$ distribution peaks at 30 - 40$\kms$, 
corresponding to Case B systems. Case A systems peaks at 80 - 90$\kms$
since them have closer orbits. For NS systems, the OB stars are much 
more massive than the NSs, making $\kob$ less than 30$\kms$. A few NS 
binaries have $\kob > 200\kms$ due to their high eccentricity.

\subsubsection{Rotation of OB stars}

\begin{figure*}
\centering
\includegraphics[width=\linewidth]{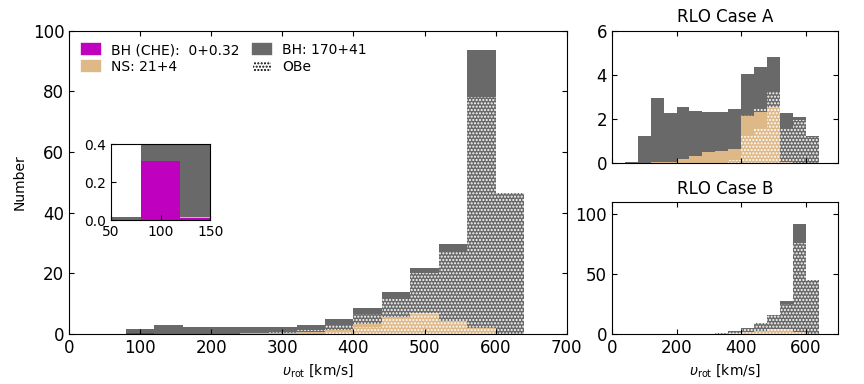}
\includegraphics[width=\linewidth]{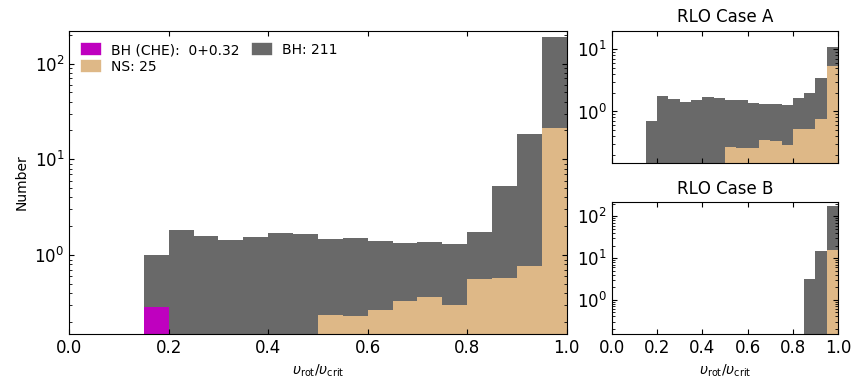}
	\caption{\label{App_v_rot}Distribution of rotation velocities of OB stars $\vrot$
 (top) and ratios of rotation velocity 
to critical velocity $\vrot/\vc$ of OB stars (bottom). The colours and legends have the same meaning as Fig. \ref{App_Mass}}
	\end{figure*}

Figure \ref{App_v_rot} presents the  distribution of rotational velocity
$\upsilon_\mathrm{ rot}$ of OB stars (top panel), which shows a fast-rotating peak 
around 600$\kms$ with a slow-rotating tail extended to 100$\kms$. 
The fast-rotating peak reflect the critical rotation velocities of stars
with mass around 10$\ms$. 
In Case A systems, tidal interaction plays an important role, which
makes $\vrot$ distributed in $100-600\kms$. We notice that some Case A
BH and NS binaries can rotate critically. For BH binaries, these 
critically rotating systems have inflated primary stars and wide orbit. 
For NS binaries,
the stripped star could not be massive enough to spin down the mass gainer.

We further present the distribution of the ratio of rotational velocity 
to critical velocity $\vrot/\vc$ in the bottom panel of Fig. \ref{App_v_rot}. Similar with the top panel, the $\vrot/\vc$ ratio shows a fast-rotating peak at 1 with a
slow-rotating tail extended to 0.2, corresponding to Case B and Case A systems.
In Case B systems, most of binaries have $\vrot/\vc > 0.95$ as expected,
while 18 of them with $\vrot/\vc < 0.95$ are braked by stellar wind.

\subsubsection{Surface abundance of OB stars\label{App_abundance_OB}}

\begin{figure*}
\centering
    \includegraphics[width=\linewidth]{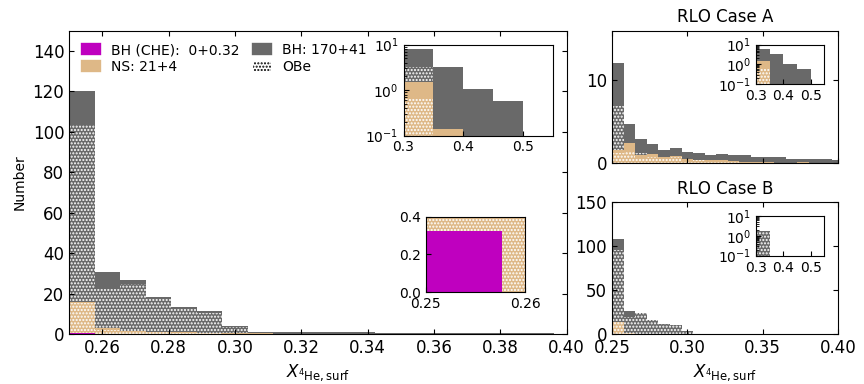}
	\includegraphics[width=\linewidth]{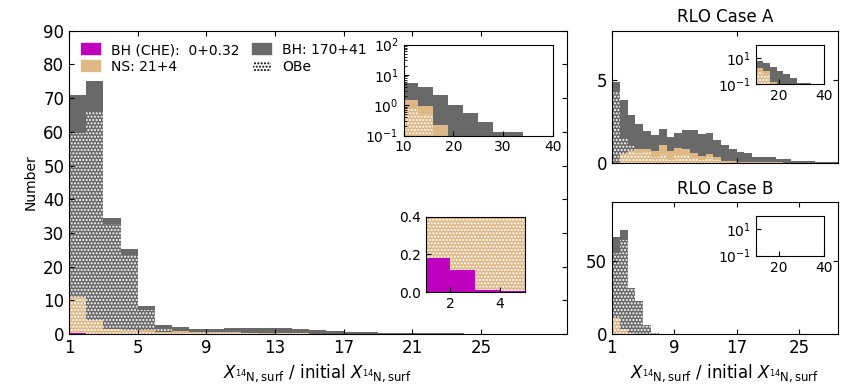}
	\caption{\label{App_abundance}The distributions of surface mass fraction of $^4$He $X_\mathrm{^4 He,surf}$ (top panel) and the enhancement factor of $^{14}$N $X_\mathrm{^{14}N,surf} /\,\mathrm{initial}\,X_\mathrm{^{14}N,surf}$ (bottom panel) of OB stars. The colours and legends have the same meaning as Fig. \ref{App_Mass}.}
\end{figure*}

We present the predicted distribution of surface abundance of OB stars in
Fig. \ref{App_abundance}. 
Surface abundance can be enriched through two ways, internal mixing
and mass transfer. For He, it is mainly enriched by mass transfer 
because the strong gradient in mean molecular weight between the core and 
the envelope prevents the transfer of He. Due to the near-zero accretion efficiency 
of wide binaries, most of OB stars have He unenriched.  
When the second mass transfer episode takes place, some accretors rotate 
sub-critically due to wind braking, allowing their surface He to be slightly enriched.
In Case A systems, due to the effect of tidal braking, accretion
efficiency can be up to 60\%. Consequently, the most enriched star has surface 
helium mass fraction about 0.5. The unenriched Case A binaries have inflated primary stars.

The distribution of surface nitrogen enhancement factor (surface nitrogen
mass fraction divided by initial surface nitrogen mass fraction)
is presented in the lower panel of Fig. \ref{App_abundance}.
Different from helium, surface nitrogen can be enriched by both mixing 
and mass transfer because CN-equilibrium is reached before the 
establishment of the strong gradient in mean molecular weight so that 
nitrogen in core can be transferred throughout envelope.
In Case B systems, surface nitrogen is mainly enriched by internal mixing,
resulting in an enrichment factor about 2-3. In Case A systems, 
the unenriched peak is related to the effect of envelope inflation and 
mass transfer leads to an enrichment factor of 10 to 15.
\citet{Hastings2020} has shown that the surface abundance of nitrogen
is sensitive with initial rotation velocity. An initially fast-rotating
star can have its surface nitrogen enriched by a factor of 30. In our 
models, all secondary stars are initially slow rotators. Hence our 
results give an lower limit on the nitrogen enhancement.

\end{appendix}

\end{document}